\documentclass[epj]{svjour}
\pdfoutput=1
\usepackage{amsmath}

\newcommand{\beq}{\begin{equation}}
\newcommand{\eeq}{\end{equation}}

\newcommand{\nub}{\bar{\nu}}

\newcommand{\nubar}[0]{$\overline{\nu}$}

\usepackage{array, graphicx, subfigure}
\usepackage{graphics}
\begin{document}
\title{Inelastic Axial and Vector Structure Functions for  Lepton-Nucleon Scattering 2021 Update}
\subtitle{Using Effective Leading order Parton Distribution Functions }
\author{Arie Bodek\inst{1}, Un Ki Yang\inst{2}, and Yang Xu\inst{1}}
\institute{Department of Physics and Astronomy, University of
Rochester, Rochester, NY  14627-0171 \\  \email {bodek@pas.rochester.edu}
\and Department of Physics and Astronomy, Seoul National University, Seoul 151-747, Korea \\
\email {ukyang@snu.ac.kr }}
\titlerunning{2021 update to Bodek-Yang model}
\authorrunning {Arie Bodek et al.}

\date{Received: date / Revised version:  V2.2   https://arxiv.org/abs/2108.09240 (2021) )}
%
\abstract{
We report on an update (2021) of a phenomenological model for inelastic neutrino- and electron-nucleon  scattering cross sections using effective leading order parton distribution functions with a new scaling variable $\xi_w$.  Non-perturbative effects  are  well described using the  $\xi_w$ scaling variable in combination with multiplicative $K$ factors at low $Q^2$. The model describes all inelastic charged-lepton-nucleon scattering data (HERA/NMC/BCDMS/SLAC/JLab) ranging from very high $Q^2$  to very low $Q^2$ and down to the $Q^2=0$ photo-production region. The model has been developed to be  used in analysis of  neutrino oscillation experiments  in the few GeV region.  The 2021 update  accounts for the difference between axial and vector structure function which brings it into much better agreement with  neutrino-nucleon total cross section measurements. The model has been developed primarily for  hadronic final state masses  $W$ above 1.8 GeV. However with additional parameters  the model  also describe the  $average$ neutrino cross sections in the resonance region down to $W$=1.4 GeV.
} 
\maketitle
\section{Introduction}
\label{intro}
The field of neutrino oscillation physics has progressed from the discovery of neutrino oscillation~\cite{ATM} to an era of precision measurements of mass splitting and  mixing angles. Uncertainties in modeling the  cross sections for neutrino interactions in the few GeV region result in systematic uncertainties in the extraction of mass splitting and  mixing parameters in neutrino oscillations experiments such as MINOS\cite{MINOS,MINOS2}, NO$\nu$A\cite{NOVA}, K2K \cite{K2K}, SuperK\cite{superK}, T2K\cite{T2K},  MiniBooNE\cite{MiniBooNE}, and DUNE\cite{DUNE}. A reliable  model of neutrino  inelastic cross sections at low energies is essential for precise neutrino oscillations experiments. 

The interest in neutrino interactions at low energies has resulted in the construction of  several near detectors  (e.g. MINOS\cite{MINOS2},  T2K\cite{T2K}) to measure low energy cross sections and fluxes,  as well as  experiments (e.g. SciBooNE \cite{sciboone}, MicroBooNE\cite{MicroBooNE}, ArgonNeu \cite{argoneut},  MINERvA\cite{minerva1}) and ICARUS\cite{ICARUS} at Fermilab  which are  designed  to measure neutrino cross sections at low energies. 

In this communication, we report on an update of  duality based model of neutrino interactions using effective leading order parton distribution functions (PDFs). Earlier  versions of the model\cite{nuint01-2,nuint04} have been incorporated into several Monte Carlo generators of neutrino interactions including NEUT\cite{NEUT},  GENIE\cite{GENIE}, NEUGEN\cite{NEUGEN} and  NUANCE\cite{NUANCE}.  The current version of GENIE  is using the NUINT04\cite{nuint04} version of the model. These early versions  assume that the axial structure functions are the same as the vector structure functions.

 In this 2021 update, we both further refine the model and $also$ account for the difference  between axial and vector structure functions at low values of $Q^2$. We refer to the version of the model which assumes that vector and axial structure functions are the same as "Type I (A=V)". The "Type I" version should be used to model electron and muon scattering.  We refer to the updated version of the model that accounts for the difference in vector and axial structure functions as "Type II  (A$>$V)". The "Type II  (A$>$V)" model should be used to model neutrino scattering,
 
In the few GeV region there are contributions from several kinds of lepton-nucleon interaction processes as defined by the final state invariant mass $W$ and the invariant square of the momentum transfer $Q^2$.   These  include quasi-elastic reactions ($W<1.07~GeV$),  the  $\Delta(1232)$  region  ($1.1<W<1.4~GeV$ ),  higher mass resonances  ($1.4<W<1.8~GeV$), and the  inelastic continuum region ($W> 1.8~GeV$).  At low momentum transfer the inelastic continuum is sometimes referred to as "shallow inelastic", and at high momentum transfer it is referred to as "deep inelastic".  It is quite challenging to disentangle each of those contributions  separately, and in particular  the  contribution of resonance production and the  inelastic scattering continuum.  At low $Q^2$  there are large non-perturbative  contributions to the inelastic cross section. These include kinematic target mass corrections,  dynamic higher twist effects,  higher order Quantum Chromodynamic (QCD) terms, and nuclear effects in nuclear targets.   

In this paper we focus on the inelastic part of the cross sections above the region of the $\Delta$(1232) resonance (i.e. the  higher mass resonances, and the inelastic continuum).  Other models (e.g. vector and axial form factors) should be used  describe the quasielastic and $\Delta$(1232) resonance contributions.

In  previous studies~\cite{highx,nnlo,yangthesis}, we have  investigated  non-perturbative effects within Leading Order (LO),  Next-to-Leading Order (NLO) and  Next-to-Next Leading Order (NNLO) QCD   using charged-lepton-nucleon scattering experimental  data~\cite{slac,bcdms,nmcdata}. We found that  in NLO QCD, most of the empirical  higher-twist terms needed to obtain good agreement with the low energy data for $Q^2>$ 1 GeV$^2$  originate primarily from target mass effects and the missing NNLO terms  (i.e. not from interactions with spectator quarks).

 If  such  is the case, then these terms should be the same in charged-leptons ($e$, $\mu$) and neutrino ($\nu_\mu$) scattering. Therefore, the vector part of  low energy  $\nu_\mu$  inelastic cross sections can be described by effective Parton Distribution Functions (PDFs) which are fit to high $Q^2$ charged-lepton-nucleon scattering  data, but  modified to include target mass and higher-twist corrections that are extracted from  low energy $e/\mu$ scattering data.   For $Q^2<$ 1 GeV$^2$ additional corrections for non-perturbative effects from spectator quarks are required. These corrections can be parametrized as multiplicative $K$ factors.  In neutrino interaction the  $K$ factor terms should be the same as in  $e/\mu$ scattering for the vector (but not axial) part of the structure functions. 

A model that describes electron and muon scattering can then be used to model  the vector contribution to neutrino scattering.  For large $Q^2$ (e.g. $Q^2>1$ GeV$^2$) the vector and  and axial structure functions are expected to be equal.  However, at low $Q^2$ the vector and axial structure functions are not equal. The axial structure functions at low values of $Q^{2}$ are not constrained by muon and electron scattering data, because the vector (but not axial)  structure functions must go to zero at  $Q^{2}$=0. The modeling of the difference between the low $Q^2$ vector and axial structure functions requires additional parameters.

  In this paper we use CCFR neutrino structure function measurements at low $x$ and low $Q^2$ to constrain the  low $Q^2$ axial $K$ factors for sea quarks,  and  neutrino total  cross section measurements to constrain the low $Q^2$ axial  $K$ factors for valence quarks.
%
%
\section{Electron-nucleon and muon-nucleon scattering}
%
In this section we define the kinematic variables for the case  of charged-lepton scattering from neutrons and protons. The differential cross section for scattering of an unpolarized charged-lepton with an incident energy $E_0$, final energy $E^{\prime}$ and scattering angle $\theta$ can be written in terms of the structure functions ${\cal F}_1$ and ${\cal F}_2$ as:
\begin{tabbing}
$\frac{d^2\sigma}{d\Omega dE^\prime}(E_0,E^{\prime},\theta)  =
   \frac{4\alpha^2E^{\prime 2}}{Q^4} \cos^2(\theta/2)$  \\ \\
  $\times   \left[{\cal F}_2(x,Q^2)/\nu +  2 \tan^2(\theta/2) {\cal F}_1(x,Q^2)/M\right],$
\end{tabbing}
where $\alpha$ is the fine structure constant, $M$ is the nucleon mass, $\nu=E_0-E^{\prime}$ is energy of the virtual photon which
mediates the interaction, $Q^2=4E_0E^{\prime} \sin ^2 (\theta/2)$ is the invariant four-momentum transfer squared, and the Bjorken variable  $x=Q^2/2M\nu$ is a measure of the longitudinal momentum carried by the struck partons in a frame in which the proton has high momentum.   Here ${\cal F}_2=\nu {\cal W}_2$, ${\cal F}_1=M{\cal W} _1$  (and for neutrino scattering  ${\cal F}_3=\nu {\cal W}_3$).

Alternatively, one could view this scattering process  as virtual photon absorption.  Unlike the real photon, the virtual photon can have two modes of polarization.  In terms of the cross section for the absorption of transverse $(\sigma_T)$ and longitudinal $(\sigma_L)$ virtual photons, the differential cross section can be written as,
\begin{equation}
\frac{d^2\sigma}{d\Omega dE^\prime} =
   \Gamma \left[\sigma_T(x,Q^2) + \epsilon \sigma_L(x,Q^2) \right],
\end{equation}
where
\begin{eqnarray}
 \Gamma &=& \frac{\alpha K E^\prime}{ 4 \pi^2 Q^2 E_0}  \left( \frac{2}{1-\epsilon } \right) \\
K &=& \frac{Q^2(1-x)}{2Mx} = \frac{2M \nu - Q^2 }{2M} \\
\epsilon &=& \left[ 1+2(1+\frac{Q^2}{4 M^2 x^2} ) tan^2 \frac{\theta}{2} \right] ^{-1}.
\end{eqnarray}
The  quantities $\Gamma$ and $\epsilon$ represent the flux and the degree of longitudinal polarization of the virtual photons respectively.   Alternatively we can express $\epsilon$  in terms of the inelasticity $y=\nu/E_0$ as follows:
\begin{eqnarray}
\epsilon &=& \frac{1-y-Q^2/(4E_0^2)}{1-y+y^2/2 +Q^2/(4E_0^2)},
\end{eqnarray}
which in the limit of $Q^2<<E_0^2$  is approximately
\begin{eqnarray}
\epsilon &=& \frac {2(1-y)}{2(1-y)+y^2}.
\end{eqnarray}
Here,  $y$ close to zero corresponds to $\epsilon=1$ and $y$ close to one corresponds to   $\epsilon=0$.

The quantity $ {\cal R}$ is defined as the ratio $\sigma_L/\sigma_T$, and is related to the structure functions by,
\begin{equation}
 {\cal R}(x,Q^2)
   = \frac {\sigma_L }{ \sigma_T}
   = \frac{{\cal F}_2 }{ 2x{\cal F}_1}(1+\frac{4M^2x^2 }{Q^2})-1
   = \frac{{\cal F}_L }{ 2x{\cal F}_1},
\end{equation}
where ${\cal F}_L$ is called the longitudinal structure function. The
structure functions are expressed in terms of $\sigma_L$ and
$\sigma_T$ as follows:
\begin{eqnarray}
 {\cal F}_1 &=& \frac{M K }{ 4 \pi^2 \alpha} \sigma_T \\
 {\cal F}_2 &=& \frac{\nu K (\sigma_L + \sigma_T)}{4 \pi^2 \alpha (1 + 
 \frac{Q^2 }{4 M^2 x^2} )} \\
 {\cal F}_L(x,Q^2) &=& {\cal F}_2 \left(1 + \frac{4 M^2 x^2 }{ Q^2}\right) - 2x{\cal F}_1,
\end{eqnarray}
or
\begin{equation}
2x{\cal F}_1 = {\cal F}_2 \left(1 + \frac{4 M^2 x^2 }{ Q^2}\right) -  {\cal F}_L(x,Q^2).
\label{eq:fl-rel}
\end{equation}

In addition, $2x{\cal F}_1$ is given by
\begin{eqnarray}
\label{eq2xF1}
2x{\cal F}_1 (x,Q^{2}) &=& {\cal F}_2 (x,Q^{2}) 
\frac{1+4M^2x^2/Q^2}{1+ {\cal R}(x,Q^{2})}.
\end{eqnarray}


Standard  PDFs are extracted  from global fits to various sets of deep inelastic (DIS) scattering data  at  high energies and high $Q^2$, where non-perturbative QCD effects are small. PDF fits are performed within the framework of QCD in either  LO, NLO or  NNLO. Here, using a new scaling variable ($\xi_w$)  we construct  effective LO PDFs  that  account for the contributions from  target mass corrections,  non-perturbative QCD effects,  and higher order QCD terms.  

We use LO PDFs because in the low Q$^2$ region, effective PDFs at NLO or NNLO  $cannot$  be constructed because the QCD NLO and NNLO corrections blow up and are not valid very low Q$^2$ (e.g  $Q^2<1.5$ GeV$^2$).
\section{The basic model: First iteration with GRV98 PDFs.}
\label{section3}
Our proposed scaling variable, $\xi_w$ is derived as follows. Using energy momentum conservation, the factional momentum, $\xi$  carried by a quark in a nucleon target of mass $M$ is 
\begin{eqnarray}
 \xi &=& \frac{2xQ^{'2}}{Q^{2}(1+\sqrt{1+4M^2x^2/Q^2})},
\end{eqnarray}
where
\begin{eqnarray}
2Q^{'2}  &=& [Q^2+M_f{^2}-M_i{^2}]  \nonumber \\
         &+& \sqrt{(Q^2 + M_f{^2}-M_i{^2})^2+4Q^2(M_i{^2}+P_{T}^{2})}.\nonumber 
\end{eqnarray}
Here $M_i$ is the initial quark mass with average initial transverse momentum $P_T$,  and $M_f$ is the mass of the final state  quark.  This expression for $\xi$  was previously derived~\cite{barb} for the case of quark  $P_T=0$. 

Assuming $M_i=0$ we construct  following scaling variable
\begin{eqnarray}
\label{eq:xiw}
\xi_w &=& \frac{2x(Q^2+M_f{^2}+B)}
        {Q^{2} [1+\sqrt{1+4M^2x^2/Q^2}]+2Ax},
\end{eqnarray}
or alternatively
\begin{eqnarray}
\label{eq:xiw2}
\xi_w &=& \frac{Q^2+M_f{^2}+B}
        {M\nu [1+\sqrt{1+Q^2/\nu^2}]+A},
\end{eqnarray}
where in general $M_f =0$, except  for the case of charm-production in neutrino scattering  for which we use  $M_f=1.32~GeV^2$.

If $A=0$ and $B=0$ and $M_f=0$   then  $ \xi_w$  is  equal to the target mass (or Nachtman\cite{Nachtman}) scaling variable 
 $ \xi_{TM}$ where, 
\begin{eqnarray}
\label{eq:xitm}
\xi_{TM}  &=& \frac{Q^2}
        {M\nu [1+\sqrt{1+Q^2/\nu^2}]}.
\end{eqnarray}

The parameters $A$ and $B$ are  enhanced target mass terms (the effects of the proton target  mass is already taken into account  in the denominator of $\xi_w$).
They (on average)  for the higher order QCD terms, dynamic higher twist, initial state quark transverse momentum ($P_T$),  and also for the effective  mass of the  initial state and final state quarks 
originating from multi-gluon interactions at low Q$^2$.   These two parameters  also allow us to describe data in the photoproduction limit (all the way down to $Q^{2}$=0).  At  Q$^2$=0, $\xi_{TM}$=0 for all $\nu$,  while $\xi_w$ at  Q$^2$=0  varies with  $\nu$.

In leading order QCD (e.g. GRV98 PDFs), ${\cal F}_{2,LO}$ for the scattering of electrons and muons on proton (or neutron) targets is given by the sum of quark and anti-quark distributions (where each is  weighted by the square of the quark charges):
\begin{eqnarray}
{\cal F}_{2, LO}^{e/\mu}(x,Q^{2}) = \Sigma_i e_i^2 \left [xq_i(x,Q^{2})+x\overline{q}_i(x,Q^{2}) \right].
\end{eqnarray}
Our proposed effective LO PDFs  GRV98 model includes the following: 
\begin{enumerate}
 \item The GRV98~\cite{grv98}  LO Parton Distribution Functions (PDFs) are used to describe  ${\cal F}_{2, LO}^{e/\mu}(x,Q^{2})$.  The minimum $Q^2$ value for these PDFs is 0.8 GeV$^2$.
 \item  In order to better describe neutrino and antineutrino cross sections, we increase the up and down quark sea by 5\%, and decrease the up and down valence quarks such that the sum of quark and antiquark distributions remain the same. i.e.
 \begin{eqnarray}
 \label{sea_part}
 d_{sea} &= &1.05 ~d_{sea}^{grv98} \nonumber \\
 \bar d_{sea}& = &1.05~ \bar d_{sea}^{grv98} \nonumber\\
 u_{sea} &=& 1.05 ~u_{sea}^{grv98} \nonumber \\
 \bar u_{sea} &= &1.05~ \bar u_{sea}^{grv98} \nonumber\\
  d_{valence} &= & d_{valence}^{grv98} -0.05~ (d_{sea}^{grv98}+\bar  d_{sea}^{grv98})   \nonumber \\
    u_{valence} &=&  u_{valence}^{grv98} -0.05~(u_{sea}^{grv98}+\bar  u_{sea}^{grv98})   
 \end{eqnarray}
 \item  The scaling variable $x$ is replaced with the  scaling variable $\xi_w$ as defined in Eq.~\ref{eq:xiw}. Here,
\begin{eqnarray}
 {\cal F}_{2, LO}^{e/\mu}(x,Q^{2})  =  \Sigma_i e_i^2   \nonumber  \\
  \times   \left [\xi_wq_i(\xi_w,Q^{2})+\xi_w\overline{q}_i(\xi_w,Q^{2}) \right].
 \end{eqnarray}
\item  As done in earlier non-QCD based fits~\cite{DL,bonnie,omegaw,bodek} to low energy charged-lepton scattering  data,  we multiply all PDFs by vector $K$ factors such that they have the correct form in the low $Q^2$ photo-production limit.  Here we use different forms for the sea and valence quarks separately;
\begin{eqnarray}	
\label{eq:kfac}
K_{sea}^{vector}(Q^2) &=& \frac{Q^2}{Q^2 +C_s} \nonumber  \\
K_{valence}^{vector}(Q^2) &=&[1-G_D^2(Q^2)] 
 \left(\frac{Q^2+C_{v2}}   {Q^{2} +C_{v1}}\right), 
\end{eqnarray}
where $G_D$ = $1/(1+Q^2/0.71)^2$ is the  proton elastic form factor. This form of the $K$ factor for valence quarks is motivated  by the closure arguments~\cite{close} and the Adler~\cite{adler,adler2} sum rule.  At low $Q^2$, $[1-G_D^2(Q^{2})]$ is approximately $Q^2/(Q^2 +0.178)$,  which is close to our earlier  (NUINT01)  fit result~\cite{nuint01-2}. These modifications are included  in order to describe low $Q^2$  data in the photoproduction limit ($Q^2$=0), where  ${\cal F}_{2}^{e/\mu}(x,Q^2)$ is related to the photoproduction cross section  according to
\begin{eqnarray}
 \sigma(\gamma p)& =& \frac{4\pi^{2}\alpha}{Q^{2}}{\cal F}_{2}^{e/\mu}(x,Q^2)\nonumber  \\
 &= &\frac{0.112~mb}{Q^2} {\cal F}_{2}^{e/\mu}(x,Q^2).
\label{eq:photo} 
\end{eqnarray}
 \item We freeze the evolution of the GRV98 PDFs at a value of $Q^2=0.80$ GeV$^2$. Below this $Q^2$, ${\cal F}_2$ is given by
\begin{eqnarray}
{\cal F}_2^{e/\mu}(x,Q^2<0.8) =\nonumber \\
  K^{vector}_{valence}(Q^2) {\cal F}_{2,LO}^{valence}(\xi_{w},Q^2=0.8)\nonumber \\ 
   +  K^{vector}_{sea}(Q^2) {\cal F}_{2,LO}^{sea}(\xi_{w},Q^2=0.8). 
\end{eqnarray}
 \item Finally, we fit for  the parameters  of the modified  effective GRV98 LO PDFs (e.g. $\xi_w$)   to  inelastic   charged-lepton scattering  data  on hydrogen and deuterium targets    (SLAC\cite{slac}/BCDMS\cite{bcdms}/NMC\cite{nmcdata}/H1\cite{h1data}.  In this first iteration, only data with an invariant final state mass $W>2$ $GeV$ are included,  where  $W^{2}=M^{2}+2M\nu-Q^{2}$.
\end{enumerate}
 In iteration 1 we obtain an excellent fit with the following initial parameters:   $A$=0.419, $B$=0.223, and  $C_{v1}$=0.544, $C_{v2}$=0.431,  and $C_{sea}$=0.380, with  $\chi^{2}/DOF=$ 1235/1200. Because of these additional  $K$ factors, we find that  the GRV98 PDFs need to be scaled up by a normalization factor  $N$=1.011.   
Here the parameters are in units of GeV$^2$.   These parameters are summerized in Table~\ref{iteration1}.
 In summary in iteration 1  we modify the  GRV98 ${\cal F}_2$ to describe low energy data down to photo-production limit as follows:
\begin{eqnarray}
{\cal F}_2^{e/\mu}(x,Q^2) =  \frac{Q^2}{Q^2+0.380} (1.011) {\cal F}_{2, LO}^{sea}(\xi_w,Q^2) \nonumber \\
       + (1-G_D^2)\frac{Q^2+0.431}{Q^2+0.544}(1.011) {\cal F}_{2, LO}^{valence}(\xi_w,Q^2),
\end{eqnarray}
where $\xi_w=\frac{2x(Q^2+0.223)}{Q^2[1+\sqrt{1+4M^2x^2/Q^2}]+2*0.419x}$.          

In fitting for the effective LO PDFs, the structure functions data are corrected for the relative normalizations   between the SLAC, BCDMS, NMC and H1 data (which  are allowed to float within the quoted normalization errors).     A systematic error shift is applied to the BCDMS data  to account for the uncertainty in their magnetic field, as described in the BCDMS publication\cite{bcdms}.  Only hydrogen and deuterium data are used in the fit. All deuterium data are corrected with a small correction for nuclear binding effects~\cite{highx,nnlo,yangthesis} as described in section \ref{nuclear}. We also include a separate additional  charm production contribution  using the photon-gluon fusion model in order to fit the very high energy HERA data. This  contribution is not necessary for any of the low energy comparisons, but is necessary  to describe the very high energy low  $Q^{2}$ HERA ${\cal F}_2$ and photoproduction data.  The charm contribution must be added separately because the GRV98 PDFs do not include  a charm sea.  Alternatively,  one may use a charm sea parametrization from another PDF.

The first iteration fit  successfully  describes all   inelastic  electron and muon   scattering data in the continuum region ($W>2~GeV$)   including the  very high and very  low $Q^2$ regions. 
 
 We find that although  photo-production data were  not included in the first iteration fit,   the predictions of the model in the continuum region  for  the  photo-production cross sections on protons and deuterons ($Q^2=0$ limit) are also in  good agreement with photoproduction measurements\cite{photo}.

\subsection {Quark-hadron duality in the resonance region}
The assumption of quark-hadron duality is that the basic cross section in the resonance region originate from the PDFs of the initial state quark, and bumps and valleys of resonances originate from final state interaction.   Therefore, if quark-hadron duality holds, PDFs can be used to describe the  $average$ cross sections in the resonance region.

We find that  quark-hadron duality holds, and  although no resonance data were included  iteration 1, the fit  also provides a reasonable description  of the $average$ value of ${\cal F}_2$ for SLAC and Jefferson data in the resonance region~\cite{jlab} (down to $Q^{2}$= 1.5  GeV$^2$).   For quark-hadron duality to work in the resonance region at lower values of  $Q^2$ (down to  $Q^{2}$=0) an additional K factor ($K^{LW}(\nu$)) is required as discussed in iteration 2 
%
%
\begin{table}[h]
    \begin{center}
\begin{tabular}{|l|l|l|l|l|}
\hline            
$A$ & $B$ & $C_{v1}$ & $C_{v2}$ & $\chi^2/ndf$ \\
$0.419$ & $0.223$ & $0.554$ & $0.431$  &  $1235/1200$ \\
\hline
\hline
$C_{sea}$& & &$N$ &  ${\cal F}_{valence}$ \\
$0.380$ & &  & $1.011$  &  $[1-G_D^2(Q^2)] $ \\
\hline
\end{tabular}
\caption{ First iteration with  GRV98  PDFs: vector parameters. Only inelastic electron and muon scattering on hydrogen and deuterium (in the continuum region $W>2~GeV$) are used in the fit ($\chi^{2}/DOF=$ 1235/1200). Here the parameters are in units of GeV$^2$. }
\label{iteration1}
\end{center}
\end{table}

\section{Second iteration with GRV98: Including photo-production data, resonances,  and additional parameters}
%
We now describe the second iteration of the fit \cite{nuint04}. Theoretically, the $K_{i}$ factors in Eq.~\ref{eq:kfac} are not required to be the same for the $u$ and $d$  valence quarks or for the $u$,  $d$,  $s$,  sea quarks and antiquarks. In order to allow flexibility in  the  effective LO model, we treat the  $K_{i}$ factors  for $u$ and $d$ valence  and for sea quarks and antiquarks separately.

\begin{figure}[ht]
\includegraphics[width=3.3in,height=3.0in]{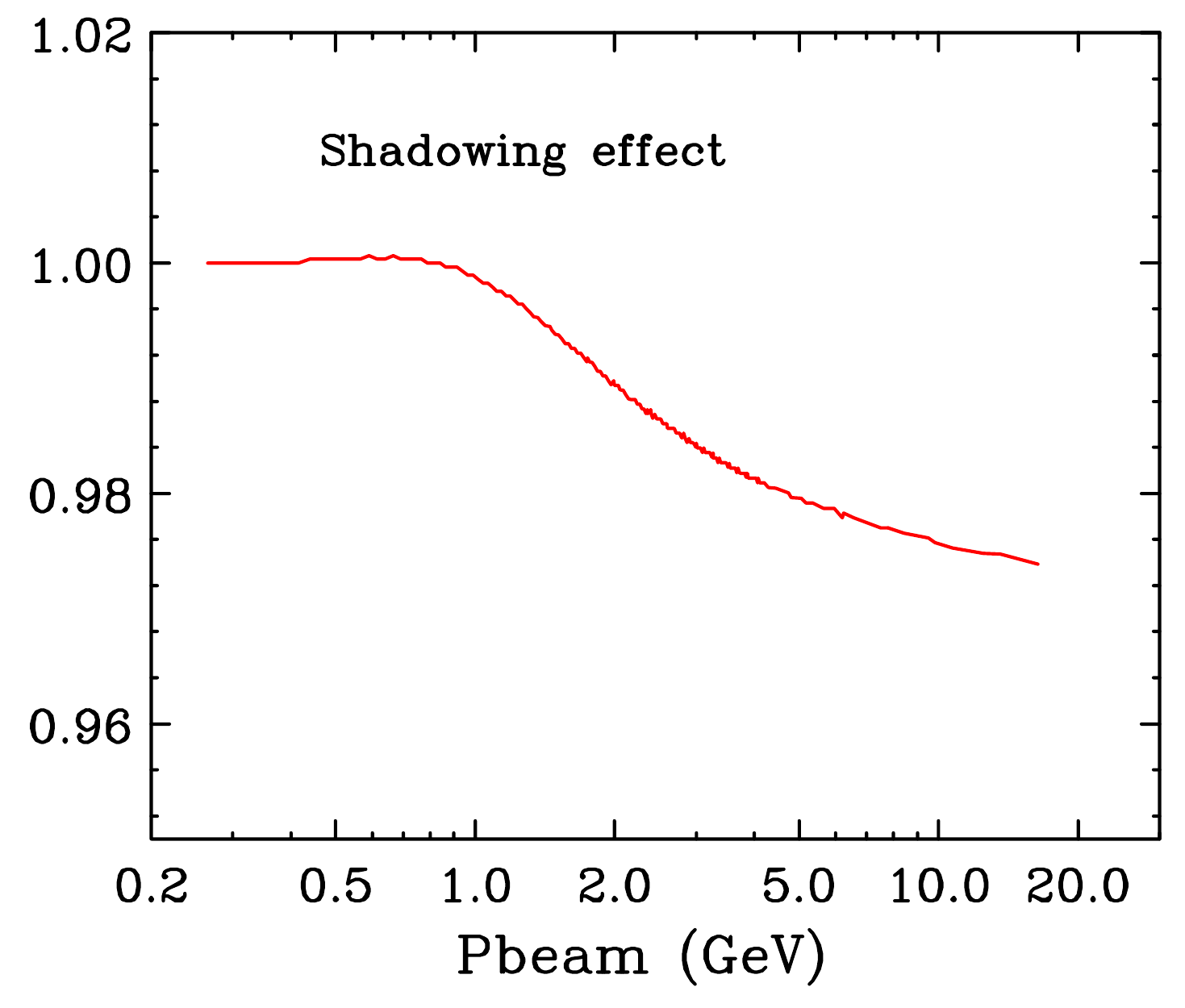}
\caption{ The ratio of photoproduction cross sections on deuterium to the sum
of the photoproduction cross sections on unbound  protons and neutrons.  This shadowing
correction\cite{yangthesis} is used to extract the photorproduction cross section on free neutrons and
protons.}
\label{fig:shadow}
\end{figure}

In this second iteration,  in order to get  additional constraints on the different  $K_{i}$ factors for up
and down quarks separately, we  include photo-production data  above the  $\Delta(1232)$  ($\nu>1~GeV$) for both hydrogen and deuterium. We  do not include  electron scattering data in the  resonance region  (on hydrogen and deuterium)  in the fit.   In order to extract neutron cross section  from photproduction cross sections on deuterium, we apply a small shadowing correction\cite{yangthesis} as shown in Fig.~\ref{fig:shadow}.  The small nuclear binding corrections  for the  inelastic lepton scattering data on deuterium is described in section~\ref{nuclear}. 
%

\begin{eqnarray}	
\label{eq:kfac2}
K^{LW}_{vector}(\nu)  &=& \frac{\nu^2+ C^{low-\nu}_{vector}  } {\nu^2}  \nonumber  \\
 K_{sea-strange}^{vector}(Q^2) &=& \frac{Q^2}{Q^2 +C_{sea-strange}}\nonumber  \\
 K_{sea-up}^{vector}(Q^2) &=& \frac{Q^2}{Q^2 +C_{sea}^{up}} \nonumber  \\
 K_{sea-down}^{vector}(Q^2) &=& \frac{Q^2}{Q^2 +C_{sea}^{down}}\nonumber  \\
 K_{valence-up}^{vector}(Q^2) &=&K^{LW}[1-G_D^2(Q^2)] \nonumber  \\
 &	\times & \left( \frac{Q^2+C_{v2u}}  {Q^{2} +C_{v1u}} \right)\nonumber \\
  K_{valence-down}^{vector}(Q^2) &=&K^{LW}([1-G_D^2(Q^2)] \nonumber  \\
  &	\times & \left(\frac{Q^2+C_{v2d}}  {Q^{2} +C_{v1d}}\right). 
\end{eqnarray}	
 The best fit iteration 2 parameters are   $A=0.621 \pm 0.009$, $B=0.380 \pm 0.004$, $C_{v1d}=0.341 \pm 0.007$, $C_{v1u}=0.417  \pm 0.024$, $C_{v2d}=0.323 \pm 0.051$,  $C_{v2u}=0.264 \pm 0.015$, and  $C^{low-\nu}=0.218\pm0.015$ for both down and up quarks.  The sea vector parameters  for iteration 2 are  $C_{sea}^{down}$=0.561, $C_{sea}^{up}$=0.369,  and $C_{sea}^{strange}$ is set to be the same as $C_{sea}^{down}$. Here,  the parameters are in units of GeV$^2$.   The  factor $K^{LW}_{vector}(\nu)$ with
 $C^{low-\nu}_{vector}$=0.218 is needed describe the resonance region for $Q^2<$ 1.5 GeV$^2$ as described below

The fit yields a  $\chi^{2}/DOF$ of $2357/1717$, and  $N=1.026 \pm 0.003$.  The photo-production resonance data (above the $\Delta(1232)$)  add  to the $\chi^2/ndf$  because the fit only provides a smooth $average$ over the higher resonances. No neutrino data are included in the fit. These parameters are summarized in Table~\ref{iteration2}.

The normalization of the various experiments are allowed to float within their errors with the normalization of the SLAC  proton data set to 1.0. The fit yields normalization factors of $0.986 \pm 0.002$,  $0.979 \pm 0.003$,  $0.998 \pm 0.003$,  $1.008 \pm 0.003$,  $1.001 \pm 0.004$,  and $0.987 \pm 0.005$ for the SLAC deuterium data,  BCDMS proton data,  BCDMS deuterium data, NMC proton data, NMC deuterium data, and H1 proton data, respectively. With these normalization, the GRV98 PDFs with our modifications  {\it should be multiplied by N=1.026 $\pm$ 0.003}.
%
%
\begin{table}[h]
    \begin{center}
\begin{tabular}{|l|l|l|l|l}
\hline            
$A$ & $B$ & $C_{v2d}$ & $C_{v2u}$  \\
$0.621$ & $0.380$ & $0.323$ & $0.264$   \\
\hline
\hline
 $C_{sea}^{down}$ & $C_{sea}^{up}$ &  $C_{v1d}$&  $C_{v1u}$ \\
$0.561$  &$0.369$ &   $0.341$  & $0.417$  \\
  \hline
  \hline
  $C_{sea}^{strange}$  & $C^{low-\nu}_{vector}$&  ${\cal F}_{valence}$   & $N$ \\
 $0.561$ & $0.218$   & $[1-G_D^2(Q^2)]$   & $1.026$   \\
 \hline
 \hline
 \end{tabular}
\caption{ Second iteration with  GRV98  PDFs: Vector Parameters. Here, we also include photoproduction data
on hydrogen and deuterium. No neutrino data are included in the fit.  When applicable, all parameters are in units of GeV$^2$.
 }
\label{iteration2}
   \end{center}
\end{table}
%
 %

As described in section \ref{dovu} we apply a small $d/u$ correction to the GRV98 PDFs. This correction increases the valence $d$ quark distribution at large $x$ and is extracted from NMC data for ${\cal F}_2^D/{\cal F}_2^P$.
%
\begin{figure}
\includegraphics[width=3.3in,height=3.5in]{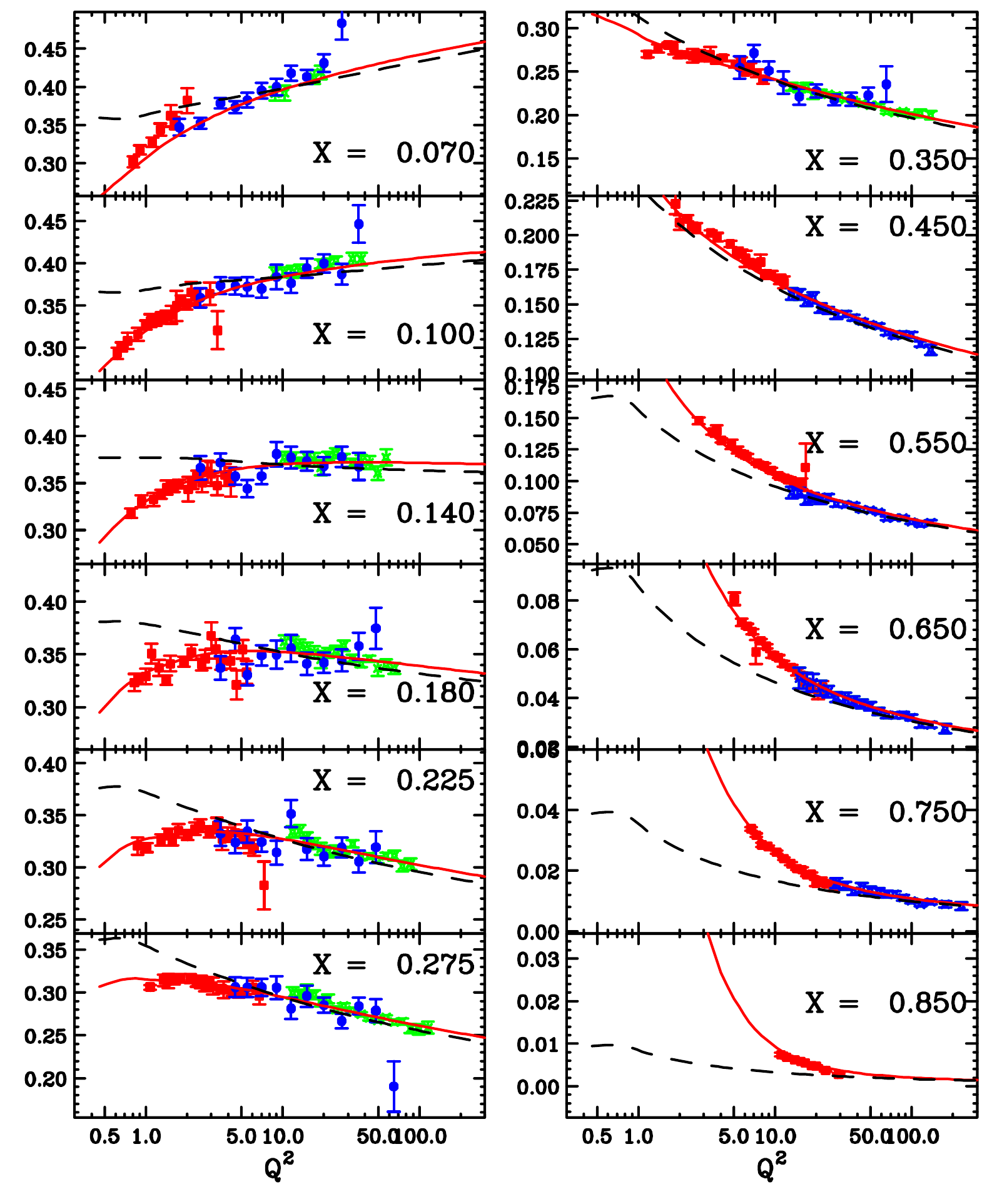}
\includegraphics[width=3.3in,height=3.5in]{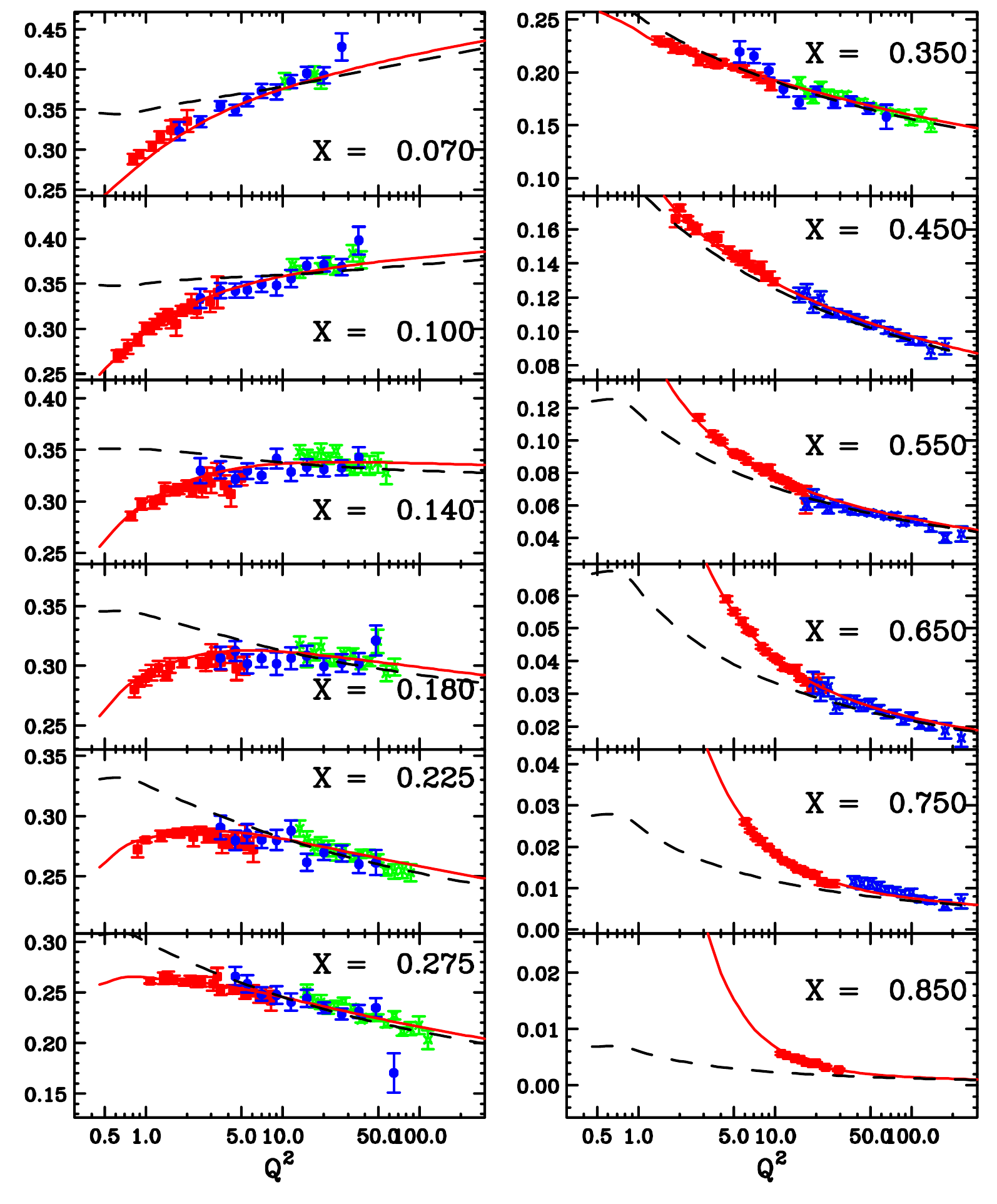}
\caption{The effective LO PDF model (iteration 2) compared to charged-lepton  ${\cal F}_2$ experimental data (SLAC, BCDMS, NMC)  at high $x$ (these data are included in the fit) :[top] ${\cal F}_2$ proton, [bot] ${\cal F}_2$ deuteron (per nucleon). The solid lines are the fit, and the  dashed lines are GRV98 .}
\label{fig:f2fit_highx}
\end{figure}
%
\begin{figure}[t]
\includegraphics[width=3.3in,height=3.5in]{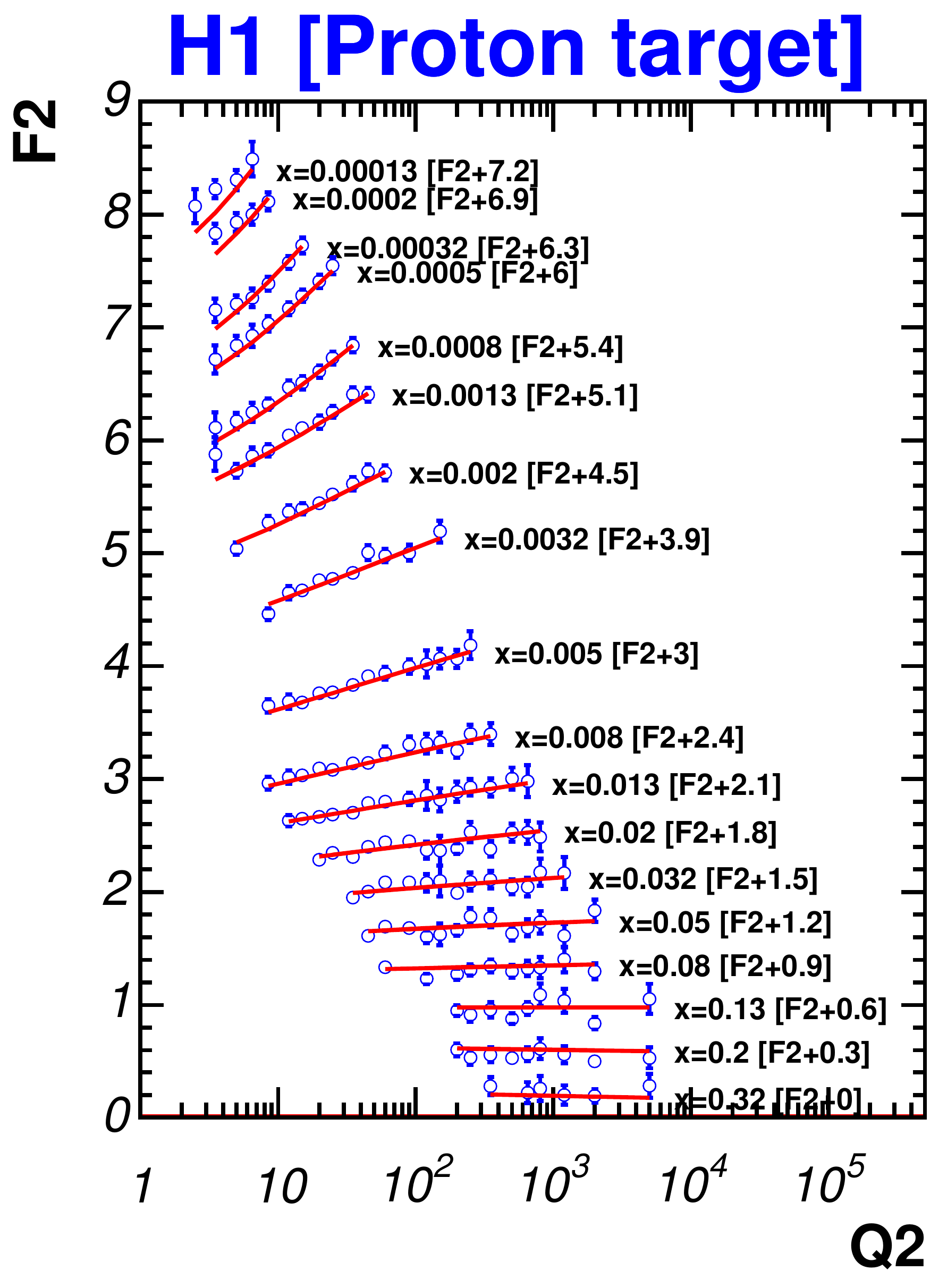}
\caption{The  effective LO PDF model (iteration 2)  compared to charged-lepton  ${\cal F}_2$ experimental data at low $x$ from H1 (these data are included in the fit).}
\label{fig:f2fit_lowx}
\end{figure}
		
\begin{figure}
\includegraphics[width=3.3in,height=3.3in]{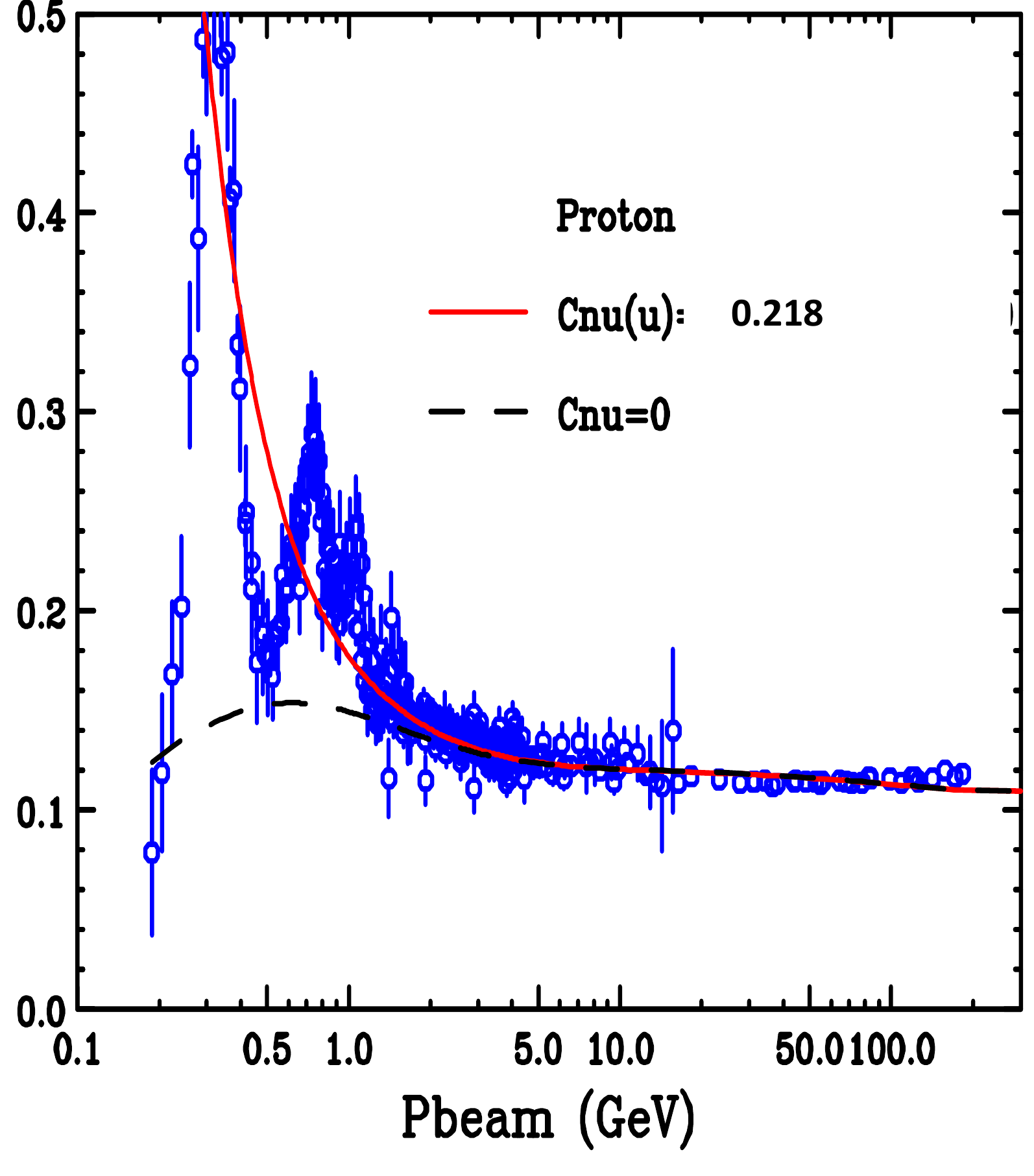}
\includegraphics[width=3.3in,height=3.3in]{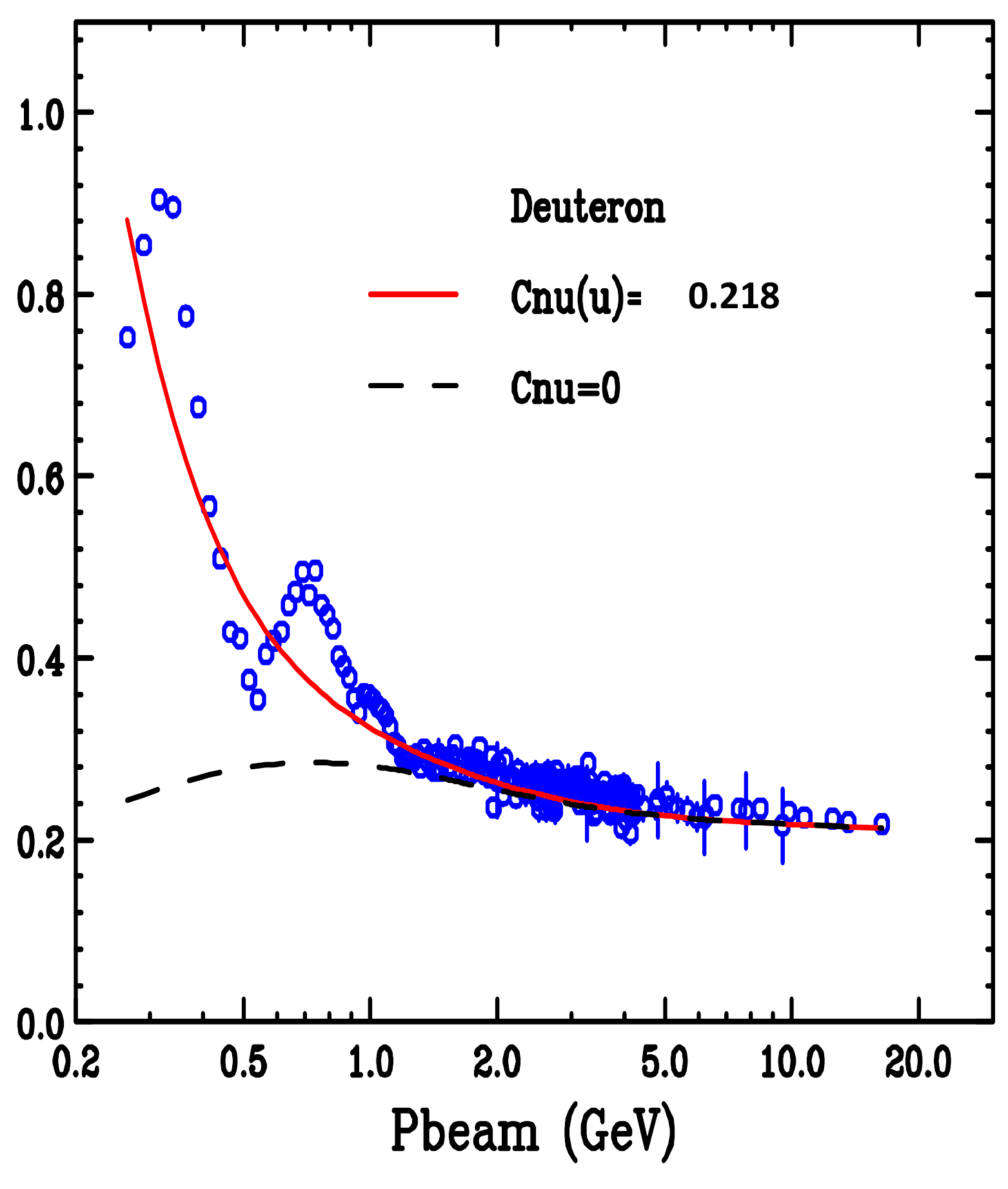}
\caption{The effective LO PDF model  (iteration 2)  compared to  photoproduction cross sections  ($Q^2=0$ limit)  at low and high energies (these data are included in the fit); [top] proton, [bot] deuteron (neutron plus proton). The cross sections are in units of mb.  At very high photon energy, we include charm contribution from gluon fusion process which is needed to describe the very high energy HERA data.  If we want to also describe the photoproduction data in the resonance region,  we need to multiply the $u$ and $d$ valence PDFs by  $K^{LW}=(\nu^2 + C^{low-\nu}_{vector})/\nu^2$ (where $C^{low-\nu}_{vector}$=0.218).  The red line includes include the  $K^{LW}$  factor, and the   dashed black  line does not includes the  $K^{LW}$ factor.   }
\label{fig:photo}
\end{figure}

Comparisons of the (iteration 2)  fit to various sets of  inelastic   electron and muon ${\cal F}_2$ data on proton and  deuteron  targets    are  shown in Fig.~\ref{fig:f2fit_highx}  (for SLAC, BCDMS and NMC).  Comparisons to H1(electron-proton) data at low values of $x$  are shown in Fig.~\ref{fig:f2fit_lowx}.  The effective LO model describes the inelastic charged-lepton  ${\cal F}_2$ data  both in  the  low $x$ as well as in  the high $x$ regions.  The model  also provides a very good description of both low energy and high energy  photoproduction cross sections\cite{photo}  on proton and deuteron targets for incident photon energies above $\nu=0.56$ GeV (which corresponds $W>1.4$ GeV)  as shown in Fig.~\ref{fig:photo}. 

%
%
%
\section{Comparison to resonance production data}
\label{resonance_section}
\begin{figure}
\includegraphics[width=3.3in,height=4.2in]{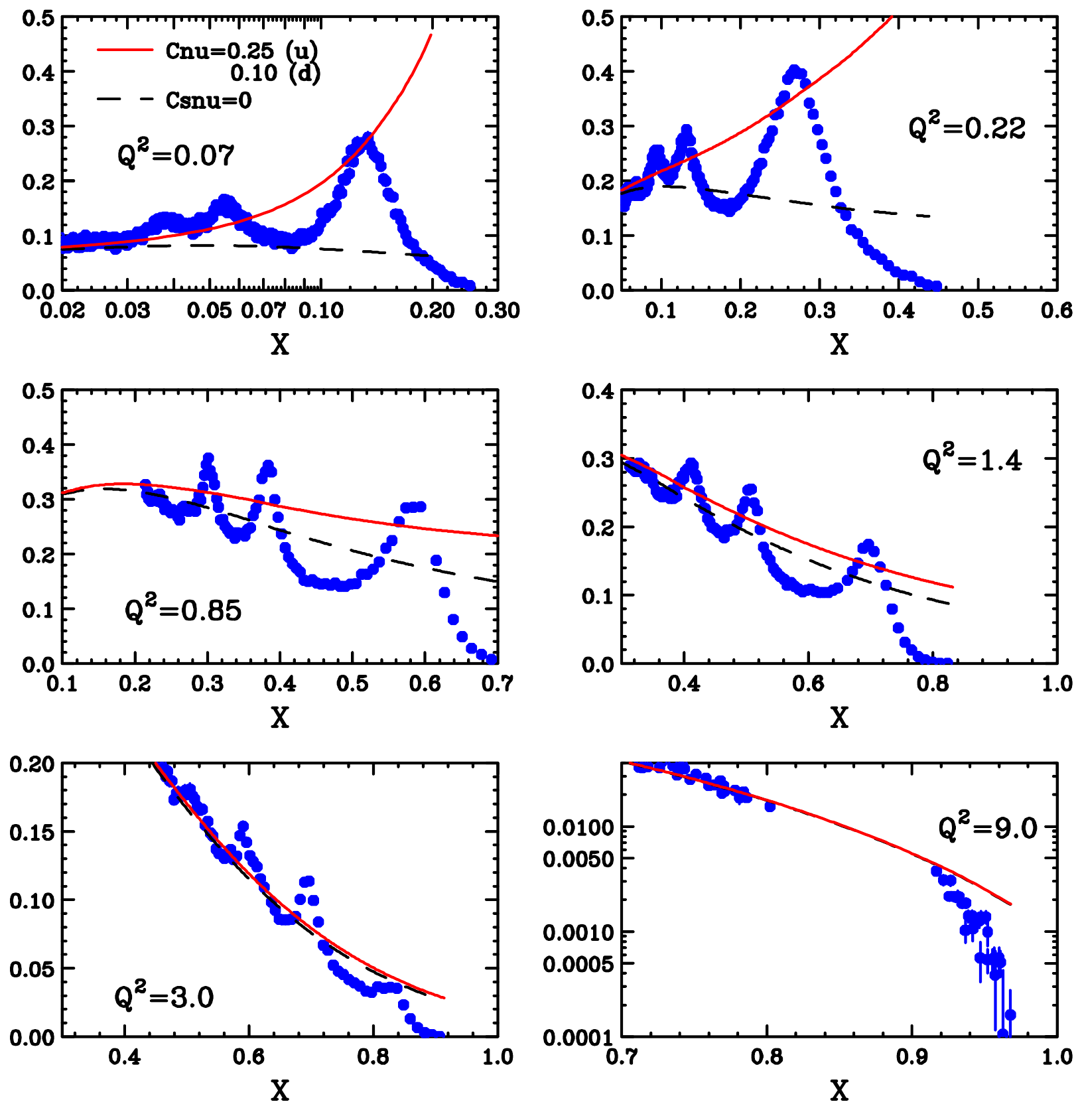}
\includegraphics[width=3.3in,height=3.0in]{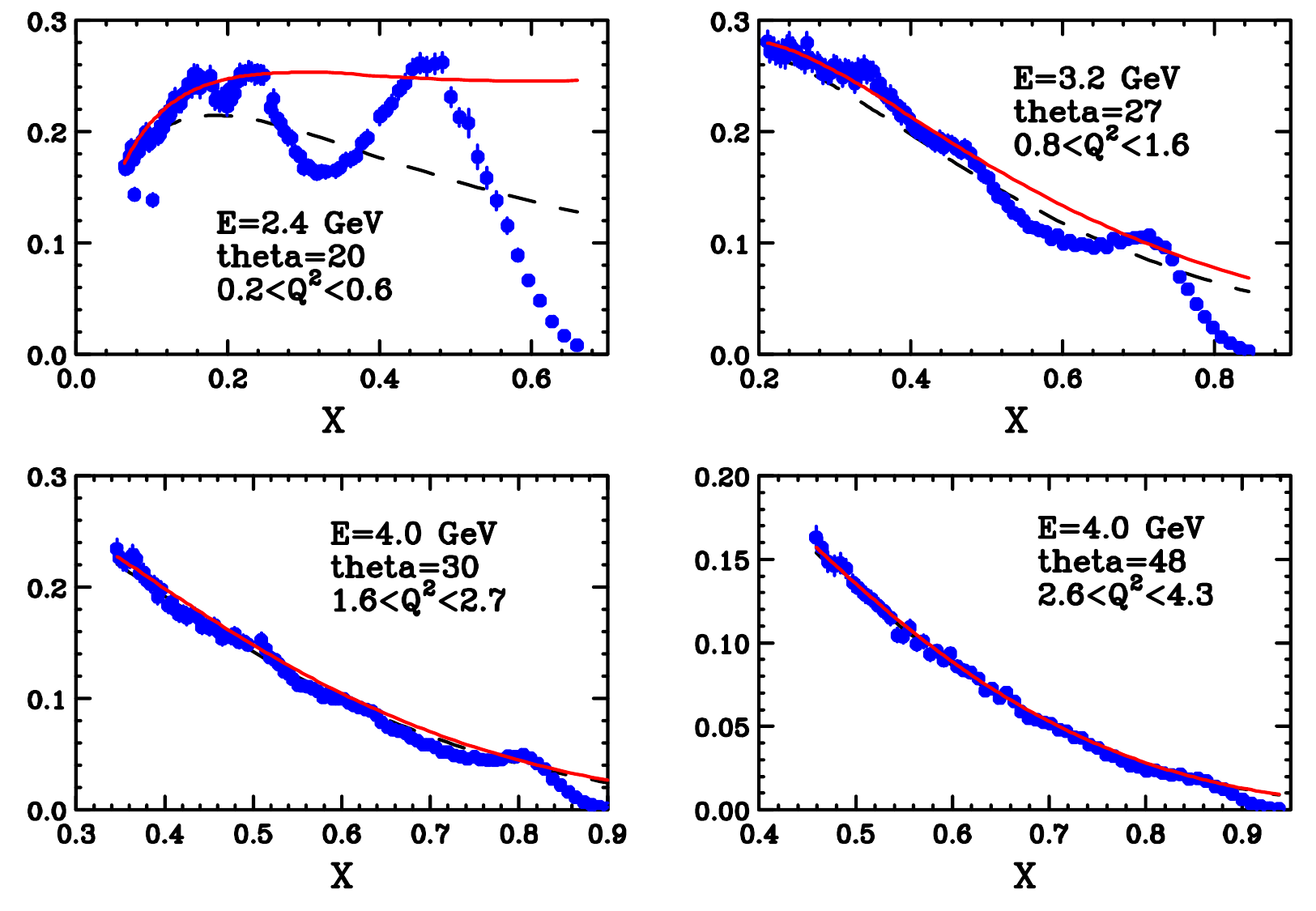}
\caption{ Comparisons of charged-lepton experimental data in the resonance region  to the  predictions of the iteration 2  effective  LO model. The top six plots are proton ${\cal F}_2$ data and the bottom four plots are deuteron  ${\cal F}_2$ data (per nucleon). The red line includes the  $K^{LW}$  factor and the  dashed black line does not include the  $K^{LW}$ factor.  Here, $K^{LW}_{vector}=(\nu^2 + C^{low-\nu}_{vector})/\nu^2$ (where $C^{low-\nu}_{vector}$=0.218).  }
\label{fig:res}
\end{figure}

\begin{figure}[t]
\includegraphics[width=3.3in,height=4.1in]{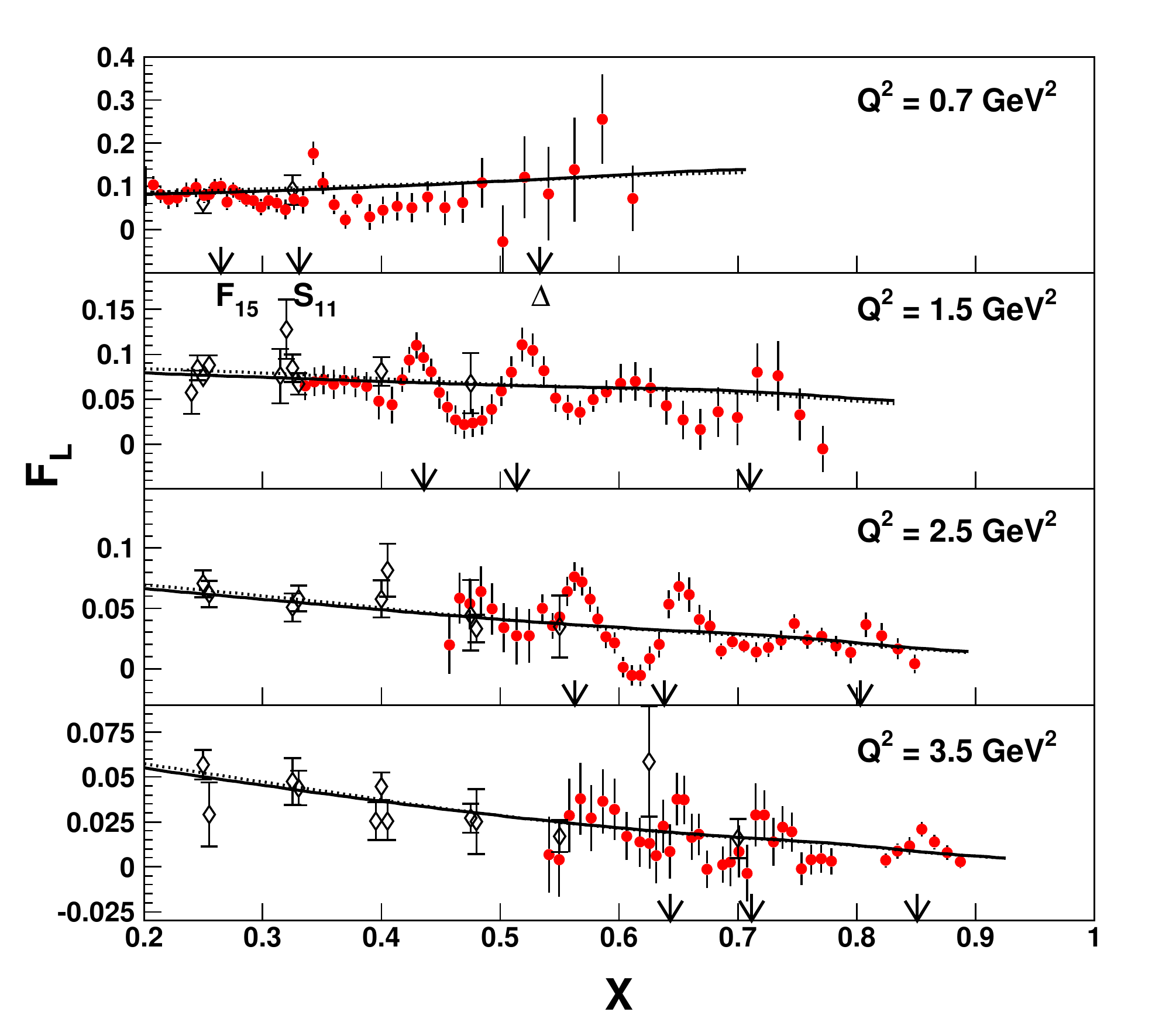}
\caption{ Comparisons of the predictions of the  iteration 2 model  to proton data for ${\cal F}_L$ (note that data for ${\cal F}_L$ are not included in the fit).}
 \label{fig:fL}
\end{figure}

Comparisons of the model predictionss  to  hydrogen and deuterium photoproduction cross sections ($Q^2=0$) including the resonance region are shown in Fig.~\ref{fig:photo}. The corresponding  electron scattering  data  in the resonance region~\cite{jlab} are shown in Fig.~\ref{fig:res}.  As expected from quark-hadron duality~\cite{bloom}, the model provides a reasonable description of both the inelastic region as well as the $average$ value of  the ${\cal F}_2$ data in the resonance region (down to $Q^{2}=0$), including the region of the first resonance ($W=1.23~GeV$).  We  also find good agreement with  recent ${\cal F}_L$  and ${\cal F}_2$ data in the resonance region from the E94-110, and JUPITER experiments~\cite{jlab,Liang:2004tj}  at Jlab, as shown in Fig.~\ref{fig:fL}. The predictions for  ${\cal F}_L$ are obtained  using the ${\cal F}_2$ model and the $R_{1998}$~\cite{R1998} parametrization (as discussed in section \ref{R-section}). 

 We find  good agreement with quark hadron duality down to very low $Q^{2}$ including the region of the $\Delta$(1238) resonance.  Other studies\cite{adler2} with unmodified GRV PDFs find large deviations from quark-hadron duality in the resonance region for electron and muon scattering.  This is because those studies do not include any low $Q^{2}$ $K$ factors and use the scaling variable $\xi$ (while we use the modified scaling variable $\xi_w$). We find that  quark hadron duality works at low $Q^2$  if we use the  modified scaling variable $\xi_w$,  and low $Q^{2}$ $K_{i} (Q^2)$  and $K^{LW}(\nu)$ factors.

%
In the  $Q^{2}=0$ photoproduction limit, the model  provides a good descriptions of the data for both the inelastic region as well as the  $average$ cross section in the resonance region   as shown in Fig.~\ref{fig:photo}.    

We recommend that the model should only be used for $W>1.4$ GeV. This is because the $\Delta$(1238) resonance has isospin 3/2 and  quarks have isospin of 1/2. Therefore,  effective leading PDFs and quark hadron duality should NOT be valid in the region of $\Delta$(1238) resonance
for neutrino scattering.   Both quasielastic scattering  and $\Delta$(1238) resonance production should be modeled in terms of vector and axial form factors.

%
%
\section{Application to neutrino scattering}
%
For very high energy neutrino scattering on $quarks$ and $antiquarks$  at high $Q^2$ ,  the vector and axial contributions are the same.   Therefore, at  high $Q^2$,  where the naive quark parton model is valid, both  the vector and axial  $K$ factors are  expected to be 1.0. The   neutrinos and antineutrino structure functions at high $Q^2$  are given by :
\begin{eqnarray}
{\cal F}_{2}^\nu(x,Q^{2}) &=& 2\Sigma_i \left [\xi_w q_i(\xi_w,Q^2) +\xi_w \overline{q}_i(\xi_w,Q^2)  \right].\nonumber 
\end{eqnarray}
and
\begin{eqnarray}
x{\cal F}_{3}^\nu(x,Q^{2}) &=&  2\Sigma_i \left [\xi_w q_i(\xi_w,Q^2) -\xi_w \overline{q}_i(\xi_w,Q^2) \right], \nonumber  
\end{eqnarray}
 where
 \begin{eqnarray}
 q^{\nu p}  &= & d+s;~~~~\bar{q}^{\nu p} = \bar{u} +\bar{c}\nonumber \\
 q^{\nu n}  &= & u+s;~~~~\bar{q}^{\nu n} = \bar{d} +\bar{c}\nonumber \\
 q^{\bar{\nu} p}  &= & u+c;~~~~\bar{q}^{\bar \nu p} = \bar{d} +\bar{s}\nonumber \\
 q^{\bar {\nu} n}  &= & d+c;~~~~\bar{q}^{\bar \nu n} = \bar{u} +\bar{c}.
 \end{eqnarray}
 
 Here,  ${\cal F}_2=\nu {\cal W}_2$, ${\cal F}_1=M{\cal W} _1$ and ${\cal F}_3=\nu {\cal W}_3$. Note that for the strangeness~conserving $(sc)$ part of the $u$ and $d$ quark distributions,    the PDFs  are multiplied by a factor of $cos^2\theta_c$=0.97417(21) where  $\theta_c$ is the Cabbibo angle.  For the  strangeness  non-conserving part the PDFs are are multiplied by a factor of $sin^2\theta_c$=0.2248(10). 

For GRV98  the charm quark distribution c=0. Noting that  $d^n=u^p$, $u^n=d^p$, $\bar{d}^n=\bar{u}^p$, and $\bar{u}^n=\bar{d}^p$)  we separate the distributions into non-charm production (ncp) and charm production (cp) terms.

For neutrino scattering on protons (here the items in parenthesis are explanations of the process)
\begin{eqnarray}
q^{\nu p} (W^+ncp)  &= & d_{v} cos^2\theta_c (d^p_{v}\rightarrow u)+s~sin^2\theta_c [s^p\rightarrow u) \nonumber\\
  &+& d_{sea} ~cos^2\theta_c (d^p_{sea}\rightarrow u) \nonumber\\
q^{\nu p} (W^+{cp})  &= & d_v sin^2\theta_c (d^p_{v}\rightarrow c)+s~cos^2\theta_c [s^p\rightarrow c) \nonumber\\
  &+& d_{sea} ~sin^2\theta_c (d^p_{sea}\rightarrow c) \nonumber\\
\bar{q}^{\nu p} (W^+ncp)&= &\bar{u}_{sea}(\bar{u}^p_{sea}\rightarrow [\bar{d}+\bar{s}]).
 \end{eqnarray}
 %
 For neutrino scattering on neutrons (here the items in parenthesis are explanations of the process)
  \begin{eqnarray}
q^{\nu n}(W^+ncp)  &= & u_v cos^2\theta_c (d^n_{v}\rightarrow u)+s~sin^2\theta_c [s^n\rightarrow u) \nonumber\\
 &+& u_{sea} cos^2\theta_c (d^n_{sea}\rightarrow u) \nonumber\\
q^{\nu n}(W^+{cp})  &= & u_v sin^2\theta_c (d^n_{sea}\rightarrow c)+s~cos^2\theta_c (s^n\rightarrow c) \nonumber\\
 &+ & u_{sea} ~sin^2\theta_c (d^n_{sea}\rightarrow c) \nonumber\\
\bar{q}^{\nu p} (W^+ncp)&= & \bar{d}_{sea}(\bar{u}^n_{sea}\rightarrow [\bar{d}+\bar{s}]).
 \end{eqnarray}
 %
For antineutrino scattering on protons
 \begin{eqnarray}
\bar{q}^{\bar \nu p}(W^-ncp)& =& \bar{d}_{sea} cos^2\theta_c(\bar{d}^p_{sea} \rightarrow \bar{u}) + \bar{s} ~sin^2\theta_c(\bar{s}^p \rightarrow \bar{u})    \nonumber \\
\bar{q}^{\bar \nu p}(W^-{cp})& =& \bar{d}_{sea} sin^2\theta_c(\bar{d}^p_{ses} \rightarrow \bar{c}) + \bar{s} ~cos^2\theta_c(\bar{s}^p \rightarrow \bar{c})    \nonumber \\
 q^{\bar{\nu} p}(W^-ncp)  &= & u_v (u^p_{v}\rightarrow [d+s]) \nonumber \\
  &+&u_{sea} (u^p_{sea} \rightarrow [d+s]) 
 \end{eqnarray}
 %
 For antineutrino scattering on neutrons
 \begin{eqnarray}
\bar{q}^{\bar \nu n}(W^-ncp)& =& \bar{u}_{sea} cos^2\theta_c(\bar{d}^n_{sea} \rightarrow \bar{u}) + \bar{s} ~sin^2\theta_c(\bar{s}^n \rightarrow \bar{u})    \nonumber \\
\bar{q}^{\bar \nu n}(W^-cp)& =& \bar{u}_{sea} sin^2\theta_c(\bar{d}^n_{sea} \rightarrow \bar{c}) + \bar{s} ~cos^2\theta_c(\bar{s}^n \rightarrow \bar{c})    \nonumber \\
 q^{\bar{\nu} n}(W^-ncp)  &= & d_v (u^n_{v} \rightarrow [d+s]) \nonumber \\
   &+& d_{sea} (u^n_{sea} \rightarrow [d+s]) 
 \end{eqnarray}

There are several  major difference between the case of  charged-lepton inelastic scattering and the case of neutrino scattering.  In the neutrino case we have one additional  structure functions  ${\cal F}_{3}^\nu(x,Q^{2})$.   In addition, at  low $Q^{2} $  there could be a difference between the vector and axial $K_{i}$ factors due a difference in the  non-perturbative  axial vector contributions. Unlike the vector ${\cal F}_2$ which must go to zero in  the $Q^2=0$ limit,  we expect  \cite{DL,kulagin}  that the axial part of ${\cal F}_2$ can be   non-zero in  the $Q^2=0$ limit. At $Q^2=0$, this non-zero axial contribution is purely longitudinal. This can be illustrated as follows. The neutrino structure functions must satisfy the following inequalities:
 \begin{eqnarray}
 0 &\le & \sqrt{1+\frac{Q^2}{\nu^2}} x{|{\cal F}_{3}|}\le 2x {\cal F}_1\le(1+\frac{Q^2}{\nu^2}) {\cal F}_2,\nonumber
 \end{eqnarray}
which indicates that only ${\cal F}_2$ can be non zero at $Q^2=0$.

We  already account for kinematic, dynamic  higher twist and higher order  QCD effects in ${\cal F}_{2}$  by fitting  the  parameters of the scaling variable $\xi_w$   (and the  $K$ factors)   to low $Q^2$ data for  ${\cal F}_{2}^{e\mu}(x,Q^{2})$.  These
 should also be valid for the vector part of ${\cal F}_2$  in neutrino scattering. 
 \begin{eqnarray}
{\cal F}_{2}^{\nu , vector}(x,Q^{2}) =
  \Sigma_i K_i^{vector}(Q^2) \xi_w q_i(\xi_w,Q^2)\nonumber \\
 +  \Sigma_j K_j^{vector}(Q^2) \xi_w \overline{q}_j(\xi_w,Q^2).
 \end{eqnarray}
 
 However, the higher order QCD effects in  ${\cal F}_{2}$ and $x{\cal F}_{3}$ are different.  We account for the different scaling violations in  ${\cal F}_{2}$ and $x{\cal F}_{3}$ (from higher order QCD terms)  by adding a  correction factor $H(x,Q^{2})$ as follows.
%
\begin{figure}[t]
\includegraphics[width=3.3in,height=4.3in]{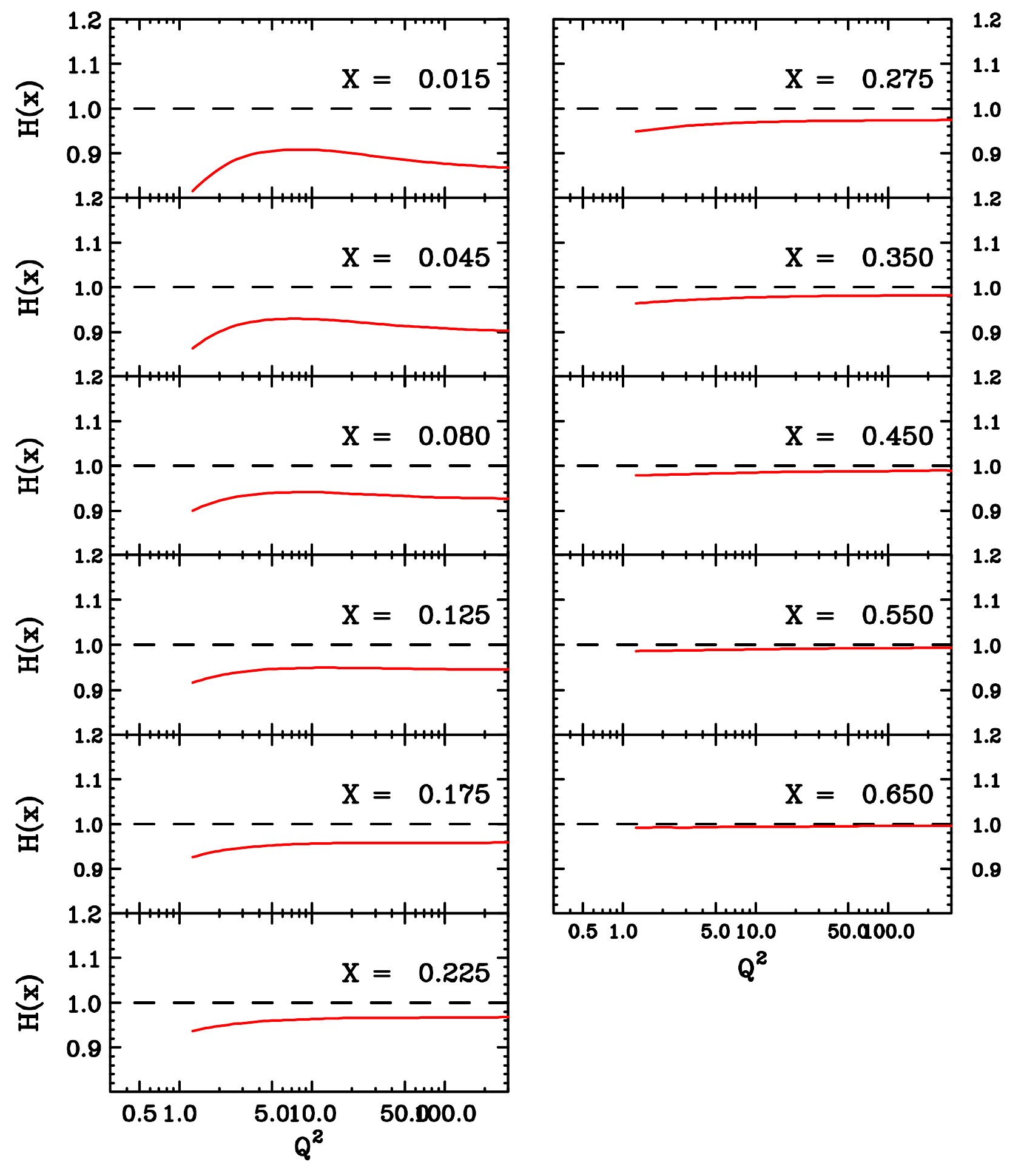}
\caption{The $x$ and $Q^2$ dependence of the factor $H(x,Q^2)$  that accounts for the difference in the QCD higher order corrections in
${\cal F}_2$ and  $x{\cal F}_3$ }
\label{fig:HxQ2}
\end{figure}
%
\begin{figure}[t]
\includegraphics[width=3.3in,height=3.0in]{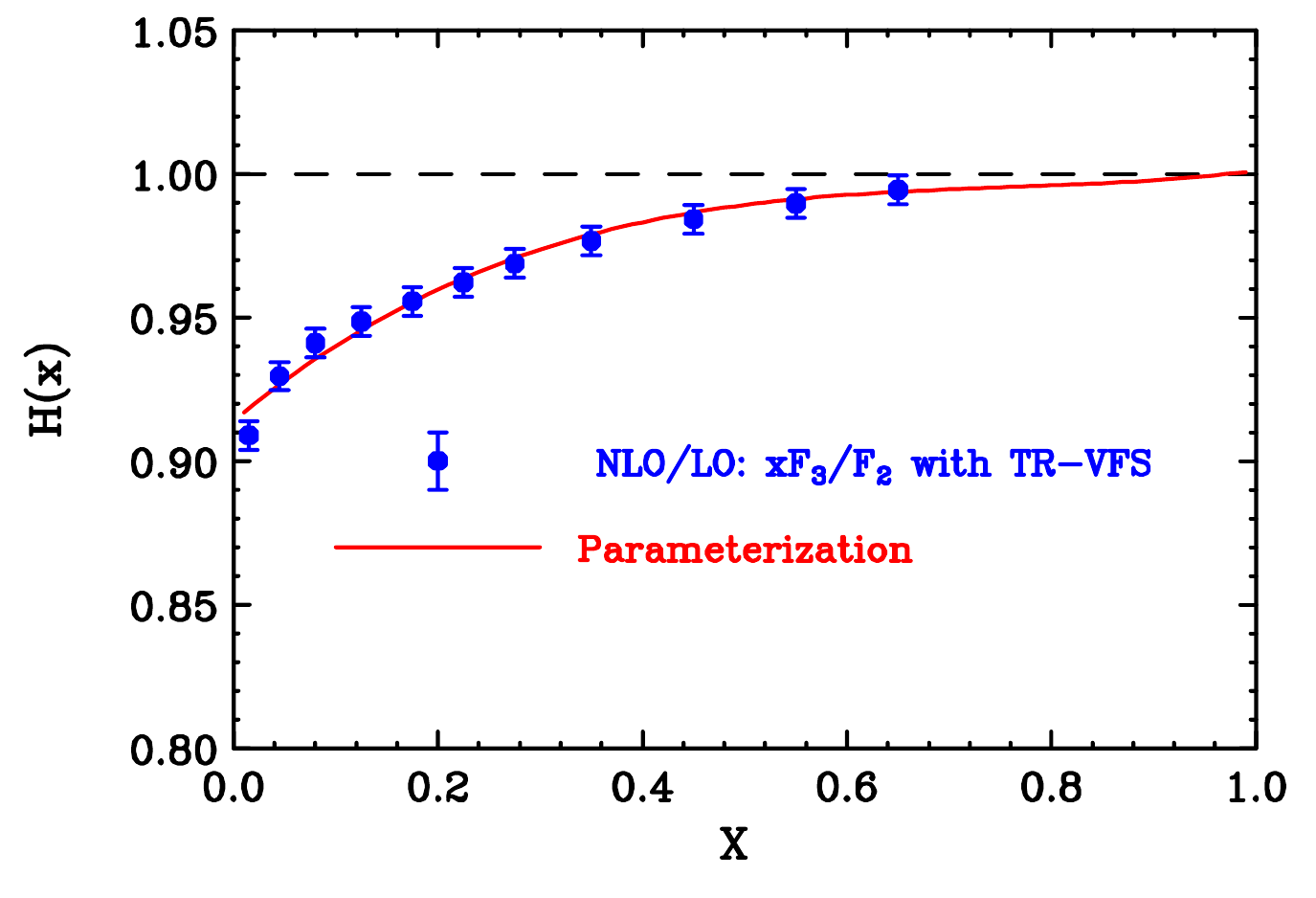}
\caption{A fit to the $x$ dependence of the factor  $H(x,Q^2)$  that accounts for the difference in the QCD higher order corrections in ${\cal F}_2$ and  $x{\cal F}_3$ (at $Q^2=8~GeV^2)$. }
\label{fig:H}
\end{figure}
\begin{eqnarray}
\label{xF3equation}
x{\cal F}_{3}^\nu(x,Q^{2} ) =  2  H(x,Q^{2}) \Biggl\{   \Sigma_i K^{vector}_i   \xi_w q_i(\xi_w,Q^2)  \nonumber \\
 -  \Sigma_j K^{vector}_j   \xi_w \overline{ q}_j(\xi_w,Q^2) \Biggr\}.
\end{eqnarray}
%
We obtain an approximate expression for $H(x,Q^{2})$ as the ratio of two ratios as follows:
\begin{eqnarray}
H(x,Q^{2})= D_{x{\cal F}_3}(x,Q^{2})/D_{{\cal F}_2}(x,Q^{2}),
\end{eqnarray}
where  
\begin{eqnarray}
 D_{xF3}(x,Q^{2}) &=& \frac{x{\cal F}_3^{nlo}(x,Q^2)}{x{\cal F}_{3}^{lo}(x,Q^2)} \nonumber  \\
 D_{F2}(x,Q^{2})  &=& \frac{{\cal F}_2^{nlo}(x,Q^2)} {{\cal F}_2^{lo}(x,Q^2)}.	
\end{eqnarray}
The double ratio $H(x,Q^{2})$ is calculated by the  TR-VFS scheme\cite{xf3calc}  with MRST991 NLO PDFs. This ratio turns out to be almost independent of $Q^2$.  The results of this calculation at $Q^{2}=8$ GeV$^{2}$, shown in Fig.~\ref{fig:H} are fitted with the following functional form:
\begin{eqnarray}
\label{Hequation}
 H(x,Q^2) &=& 0.914+0.296x  - 0.374x^{2}+0.165x^{3}.	
\end{eqnarray}
We use the above  approximation for $H(x,Q^2)$ for all values of $Q^{2}$.

In our previous~\cite{nuint01-2}  analysis we assumed  $H(x,Q^2)$=1,  and  $K_{i}^{axial}(Q^2)$= $K_{i}^{vector}(Q^2)$.  This assumption is  only valid for  at high  $Q^{2}$  ($Q^{2}>1~GeV^2$). Here,   we improve  on the previous analysis by  introducing  $K_{i}^{axial}(Q^2)$ factors which are different from  $K_{i}^{vector}(Q^2)$,  and include the $H(x,Q^2)$ correction for  $x{\cal F}_3$.  
%
\section{$ 2x{\cal F}_1$ and the longitudinal structure function}
\label{R-section}
In the extraction of the original GRV98 LO PDFs, no separate longitudinal contribution was included. The quark distributions were directly fit to ${\cal F}_2$ data.  A full modeling of electron and muon cross section requires also a description of $2x{\cal F}_1$.  In general, $2x{\cal F}_1^{e/\mu}$  and  ${\cal W}_1^{e/\mu}$  are given in
terms of  ${\cal F}_2$ and $ {\cal R}$  in   equation \ref{eq2xF1}.
For the vector contribution we  use a non-zero longitudinal $ {\cal R}$ in reconstructing $2x{\cal F}_1$ by using a fit of $ {\cal R}$ to measured data. 
%
%
The   $ {\cal R}_{1998}$ function\cite{R1998}  provides  a good description of the world's data  for $ {\cal R}$   in the $Q^2>0.3$ GeV$^2$ and $x>0.05$ region (where most of the $ {\cal R}$ data are available).
\begin{eqnarray}
 {\cal R}_{e/\mu} (x,Q^2>0.3) & =&  {\cal R}_{1998}(x,Q^2>0.3)  \nonumber 
 \label{eq:rmod}
\end{eqnarray}
However, the $ {\cal R}_{1998}$ function breaks down at low $Q^2$. Therefore, we freeze the function at $Q^2=0.3$ GeV$^2$ and introduce a  $K$ factor for  $ {\cal R}$ in the $Q^2<0.3$ GeV$^2$ region to make a smooth transition for $ {\cal R}_{e/\mu}$ from $Q^2=0.3$ GeV$^2$ down to $Q^2=0$ by forcing ${\cal R}_{vector}$ to approach zero at $Q^2=0$, as expected in the photoproduction limit. This procedure keeps a $1/Q^2$ behavior at large $Q^2$ and matches to $ {\cal R}_{1998}$ at  $Q^2=0.3$ GeV$^2$,

\begin{eqnarray}
 {\cal R}_{e/\mu}(x,Q^2<0.3) & = & 3.633 \frac {Q^2}{Q^4+1}\times  {\cal R}_{1998}(x,Q^2=0.3). \nonumber 
\end{eqnarray}

Using the above fits to $ {\cal R}$ as measured in electron/muon scattering we use the following expressions for the vector part of $2x{\cal F}_1$  neutrino scattering: 
\begin{eqnarray}
2x{\cal F}_1^{vector} (x,Q^{2}) &=& {\cal F}_2^{vector} (x,Q^{2})  \frac{1+4M^2x^2/Q^2}{1+ {\cal R}(x,Q^{2})}  \nonumber \\
   {\cal R}_{vector}(x,Q^2>0.3) & = &   {\cal R}_{1998} (x,Q^2>0.3) \nonumber \\
         {\cal R}_{vector}(x,Q^2<0.3) & = & {\cal R}_{e/\mu}(x,Q^2<0.3). \nonumber 
\end{eqnarray}

The above expressions have the correct limit for the vector contribution at $Q^2=0$.

A  recent fit to $ {\cal R}$  that includes updated $ {\cal R}$ measurements from Jefferson Lab (including 
resonance data) has been published by  M.E. Christy and P.E. Bosted\cite{jlabR}.  In the kinematic region of the fits the difference between the Christy-Bosted fit and the $ {\cal R}_{1998}$ fit  is small.

\subsection{ Nuclear Corrections to R}
\label{R-section_Fe}
In the analysis we use the $ {\cal R}_{1998}$ parametrization.    Preliminary results from the JUPITER Jefferson Lab collaboration indicates that   $ {\cal R}$  for heavy nucleus may be higher by about 0.1 than $R$ for deuterium.  Therefore, we  use   an error of $\pm$0.1 in R to estimate the systematic error in the cross sections from this source. 

%
%
\section{ Charm production in neutrino scattering}
%
Neutrino scattering is not as simple as the case of charged-lepton scattering because of the contribution from  charm production (cp). 
For  non-charm production (ncp) components  we use  the sum of the vector and axial contributions to  ${\cal F}_2^{ncp}(x,Q^{2})$,  and $2x{\cal F}_1^{ncp}(x,Q^{2}$) with $x{\cal F}_3^{ncp}(x,Q^{2})$  as described above. 

For the charm production components  of   ${\cal F}_2^{cp}(x,Q^{2})$,  $x{\cal F}_3^{cp}(x,Q^{2})$ and $2x{\cal F}_1^{cp}(x,Q^{2})$  the  variable $\xi_w$  is replaced with $\xi_w^{\prime}$   includes a non-zero final state quark mass  $M_{c}=1.32~GeV$.
\begin{eqnarray}
\label{eq:xiw-charm}
\xi_w ^{\prime}&=& \frac{Q^2+1.32^2+B}
        {M\nu [1+\sqrt{1+Q^2/\nu^2}]+A},
\end{eqnarray}

The target mass  calculations as discussed by  Barbieri et. al\cite{barb}   imply that ${\cal F}_2^{\nu-cp}$ is described by
 ${\cal F}_2^{\nu-cp} (\xi_w^\prime ,Q^{2})$, and   the other two structure functions are multiplied by the  factor $K_{charm} = \frac {Q^2}{Q^2+M_C^2}$. Consequently, we use the following expression for charm production processes:
\begin{eqnarray}
{\cal F}_{2}^{\nu, vector-cp}(x,Q^{2}) = \Sigma_i K_i^{vector}(Q^2) \nonumber  \\
\times  \left [\xi_w^\prime q_i(\xi_w^{\prime},Q^2) +\xi_w^{\prime} \overline{q}_i(\xi_w^{\prime},Q^2)  \right], \nonumber
\end{eqnarray}
\begin{eqnarray}
2x{\cal F}_1^{\nu,vector-cp}(x.Q^{2})& =& K_{charm}  \nonumber \\
&\times&    \frac{1+4M^2x^2/Q^2}{1+ {\cal R}(\xi_{w}^\prime,Q^{2})}
{\cal F}_{2}^{\nu, vector-cp}(x,Q^{2}),\nonumber \\
K_{charm}& =& \frac {Q^2}{Q^2+M_C^2},  \nonumber 
\end{eqnarray} 
and		
\begin{eqnarray}
&&x{\cal F}_{3}^{\nu,cp}(x,Q^{2} ) =   2  H(x,Q^{2}) K_{charm} \times \nonumber \\
&&\Biggl\{   \Sigma_i K^{vector}_i   \xi_w^{\prime} q_i(\xi_w^{\prime},Q^2) 
 - \Sigma_j K^{vector}_j   \xi_w^{\prime} \overline{ q}_j(\xi_w^{\prime},Q^2) \Biggr\}.\nonumber
\end{eqnarray}
%
\begin{figure}[t]
\includegraphics[width=3.3in,height=3.3in]{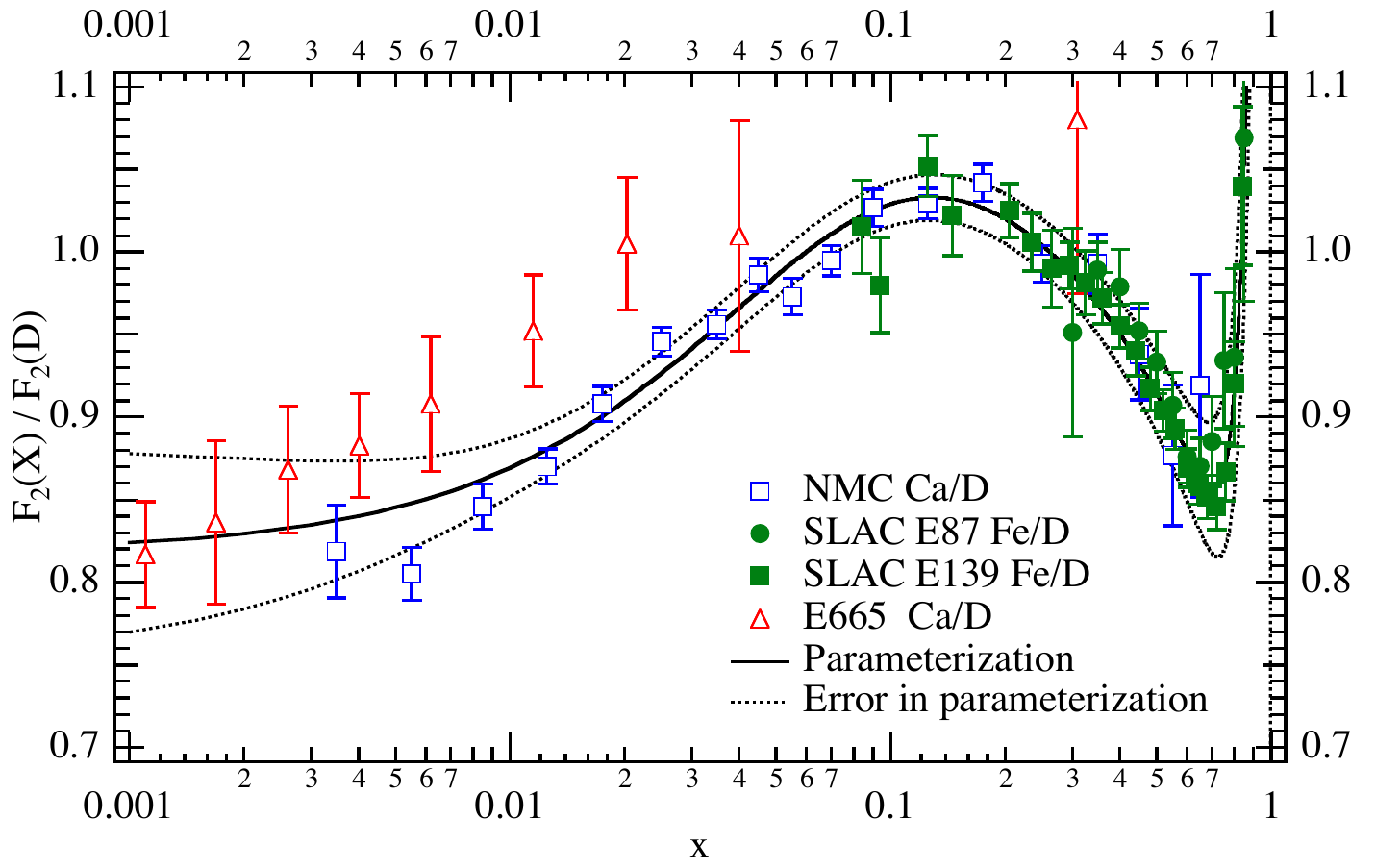}
\caption{The ratio of ${\cal F}_2$ data for heavy nuclear targets and deuterium as measured in charged-lepton scattering experiments(SLAC,NMC, E665). The band show the uncertainty of the parametrized curve (as a function of $x$)  from the statistical and systematic errors in the experimental data~\cite{selthesis}.We do not use this fit in our analysis. Instead we use another fit that includes only SLAC and Jefferson Lab data.}
\label{fig:nuclear_heavy}
\end{figure}
We  use  the $ {\cal R}_{1998}$ parametrization~\cite{R1998} for the vector part of  $ {\cal R}^{ncp}$ and $ {\cal R}^{cp}$. Because of the suppression of charm production at low $Q^2$ we assume that the vector and axial contributions to charm production are equal.

%
\begin{figure}[t]
\includegraphics[width=3.3in,height=3.3in]{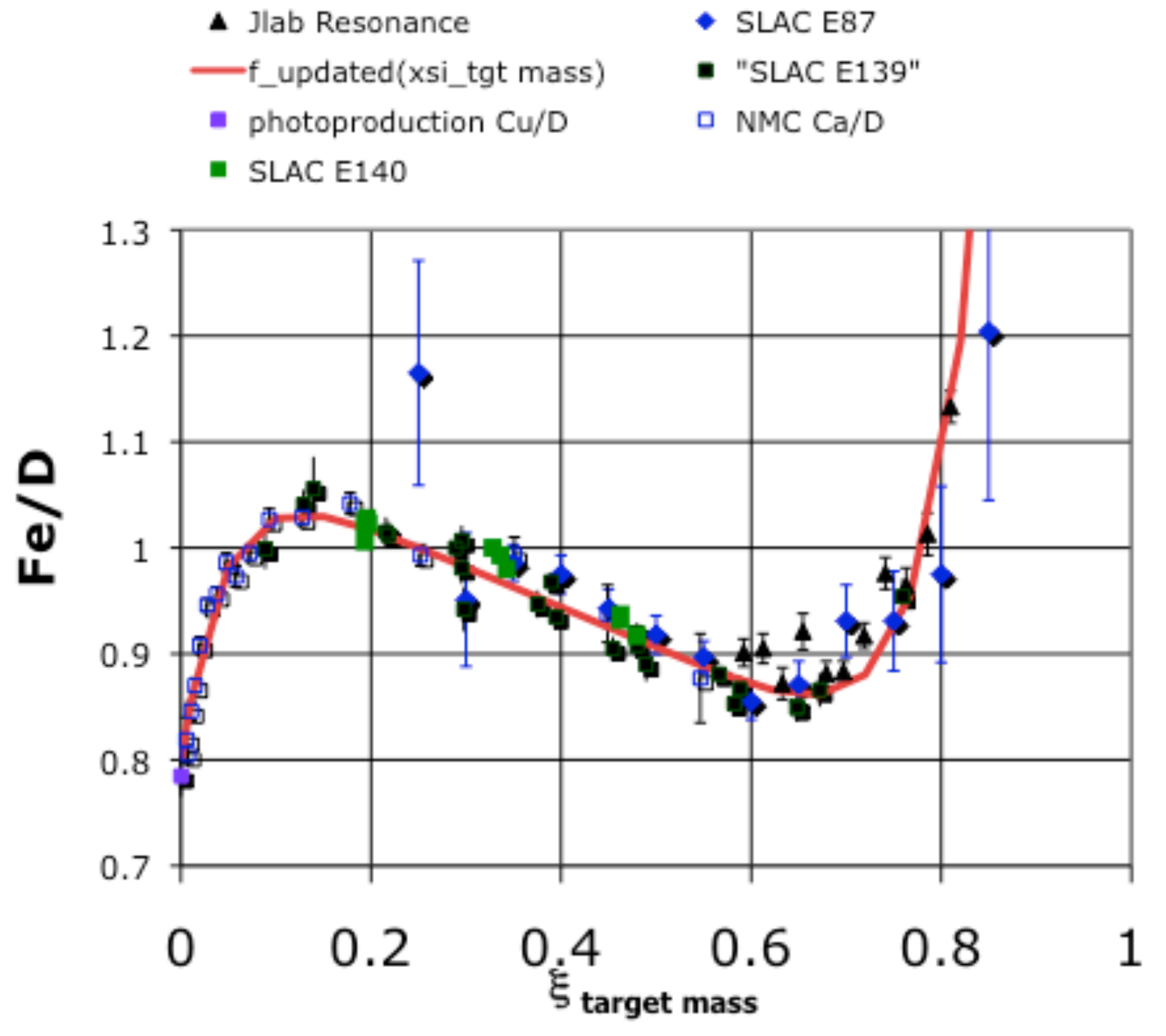}
\caption{A a comparison of Jefferson lab  measurements of   the ratio of electron scattering cross sections on an isoscalar iron target  to deuterium in the resonance region\cite{arrington}  to  data from SLAC E87\cite{e87},  SLAC E139\cite{e139},  SLAC E140\cite{e140} and NMC\cite{NMCnuc}  in the deep inelastic region.  The   data    plotted versus $\xi_{TM}$  are compared to the updated  fit function $f_{updated}^{Fe/D} (\xi_{TM})$.
  For comparison we also show the ratio as measured in photoproduction\cite{photo} at $\xi_{TM}=0$. We use this fit to cross section ratios in our analysis.
}
\label{fig:jlab2}
\end{figure}

%
\section{Nuclear corrections}
\label{nuclear}
%
In the comparison with neutrino charged-current differential cross section on an isoscalar iron target, a nuclear correction for iron targets should be applied. Previously, we used  the following parameterized function, $ f^{Fe/D}(x)$ (a fit to experimental electron and muon scattering data for the ratio of isoscalar iron to deuterium cross sections,  shown in Fig.~\ref{fig:nuclear_heavy}), to convert deuterium structure functions to (isoscalar) iron structure functions~\cite{selthesis};
\begin{eqnarray}
f^{Fe/D}(x) & = &   \frac{{\cal} F_2^{Fe}}{{\cal F}_2^{D}} = 1.096 -0.364~x  \nonumber \\
& - & 0.278~e^{-21.94~x }+   2.772~x^{14.417}.
\end{eqnarray}
However, in this publication we do not use the above nuclear corrections for iron since they are a function of $x$ and  include very high $Q^2$ data.  We find that  the ratio of iron to deuterium structure function measurements  at SLAC and Jefferson Lab are better described in terms of the target mass (or Nachtman) variable   $\xi_{TM}$.  If  $\xi_{TM}$ is used,  then the  function  that describes the iron to deuterium ratios in the  deep inelastic region is also valid in the  resonance region. 
In addition, since we are interested primarily in low energy neutrino cross sections we only include SLAC and Jefferson lab data in our fit.  We use the following  updated function  $f_{updated}^{Fe/D}(\xi_{TM})$.
\begin{eqnarray}
f_{updated}^{Fe/D} (\xi_{TM})& = &   \frac{{\cal} F_2^{Fe}}{{\cal F}_2^{D}}= 1.096 -0.38~\xi_{TM}  \nonumber \\
& - &
 0.3~e^{-23\xi_{TM}} +   8~\xi_{TM}^{15}.
\end{eqnarray}
Fig.~\ref{fig:jlab2} shows a comparison of Jefferson lab  measurements of   the ratio of electron scattering cross sections on iron to deuterium in the resonance region\cite{arrington}  to  data from SLAC E87\cite{e87},  SLAC E139\cite{e139}, SLAC E140\cite{e140} and NMC\cite{NMCnuc}  in the deep inelastic region.  The   data    plotted versus $\xi_{TM}$   are compared to the updated  fit ${\cal F}_{updated}(\xi_{TM})$.  For comparison we also show the ratios as measured in photoproduction\cite{photo} at $\xi_{TM}=0$.

\begin{figure}[t]
\includegraphics[width=3.3in,height=3.3in]{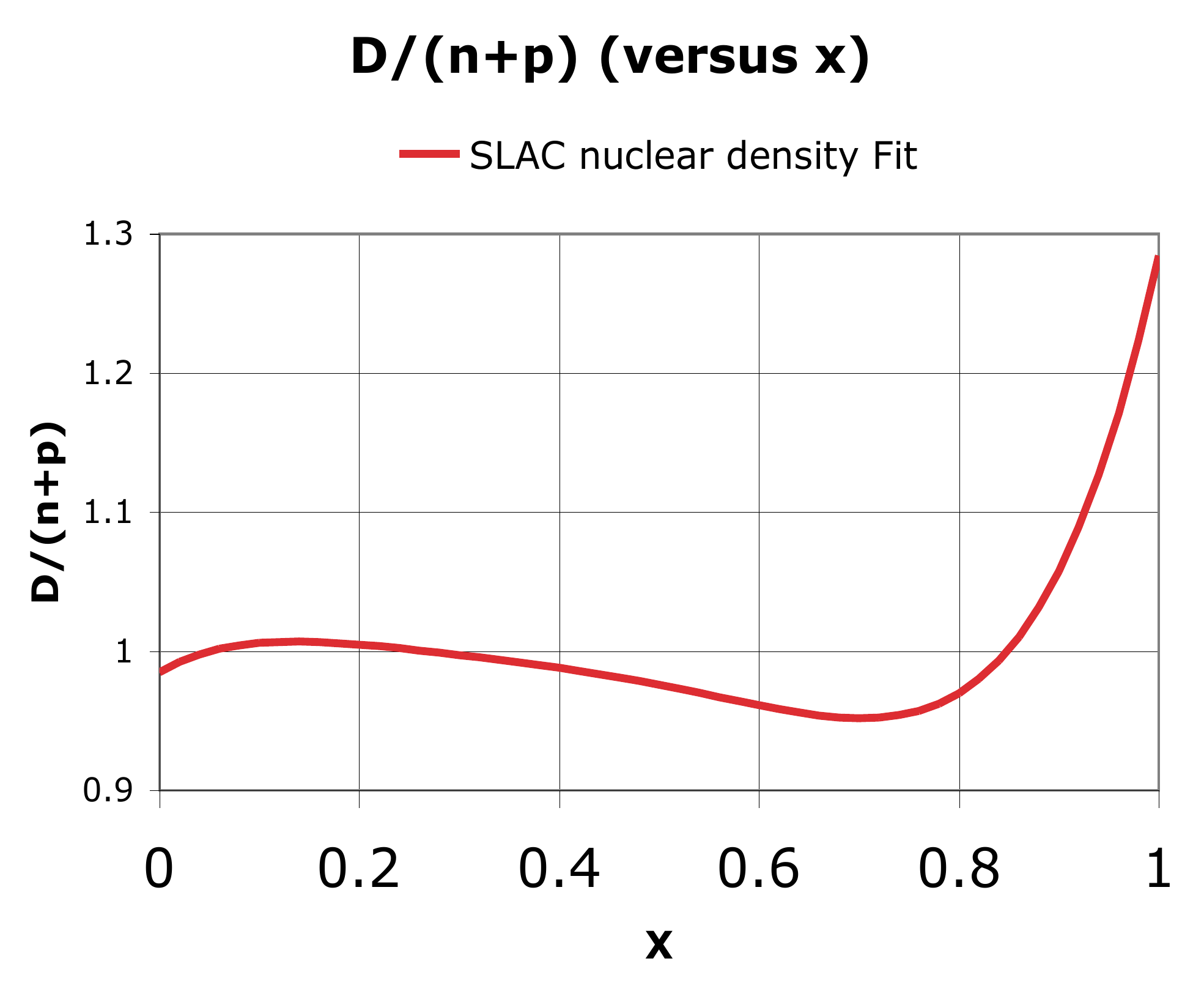}
\caption{The total correction for nuclear effects (binding and Fermi motion) in the deuteron,    $f^{D/(n+p)}(x) ={\cal F}_2^d/{\cal F}_2^{n+p}$, as a function of $x$, extracted from fits to the nuclear dependence of SLAC ${\cal F}_2$ electron scattering data. This correction   is only valid for $0.05<x<0.75$. }
\label{fig:f2dp}
\end{figure}

For the ratio of deuterium cross sections to cross sections on free nucleons we use the following function obtained from a fit to SLAC data on the nuclear dependence of electron scattering cross sections~\cite{yangthesis}.
\begin{eqnarray}
f^{D/(n+p)}(x) & = & 0.985 \times (1+0.422x -2.745x^2  \nonumber \\
         & +  &   7.570x^3  -10.335x^4+5.422x^5).
\label{eq:nucl-d}
\end{eqnarray}
This correction  shown in Fig.~\ref{fig:f2dp} is only valid in the $0.05<x<0.75$ region. 

\begin{figure}[t]
\includegraphics[width=3.3in,height=3.3in]{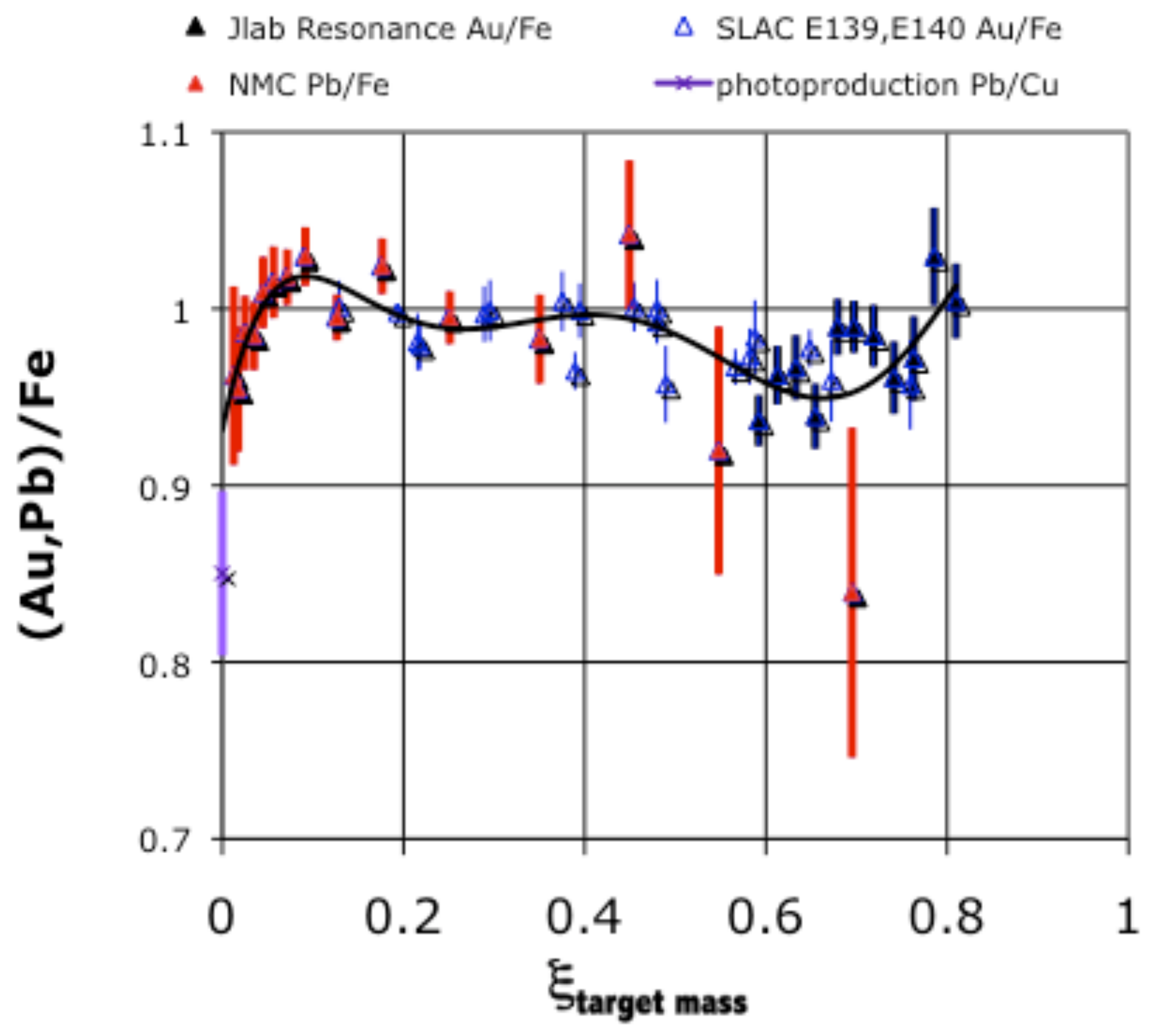}
\caption{The ratio of ${\cal F}_2$ data for gold (Au) to ${\cal F}_2$ data  for Iron (Fe)  as measured in charged-lepton scattering experiments in the deep inelastic region  (SLAC E139, SLAC E140) as compared to Jlab data in the resonance region versus the target mass  (or Nachtman) variable  $\xi_{TM} $. Also shown is the ratio of  ${\cal F}_2$ data for lead (Pb) to ${\cal F}_2$ data  for iron (Fe) from the NMC\cite{{NMCnuc}} collaboration. For comparison we also show the ratio of lead to copper cross sections (Pb/Cu)  as measured in photoproduction\cite{photo}. } 
\label{fig:jlab_Au}
\end{figure}

\begin{figure}[t]
\includegraphics[width=3.3in,height=3.3in]{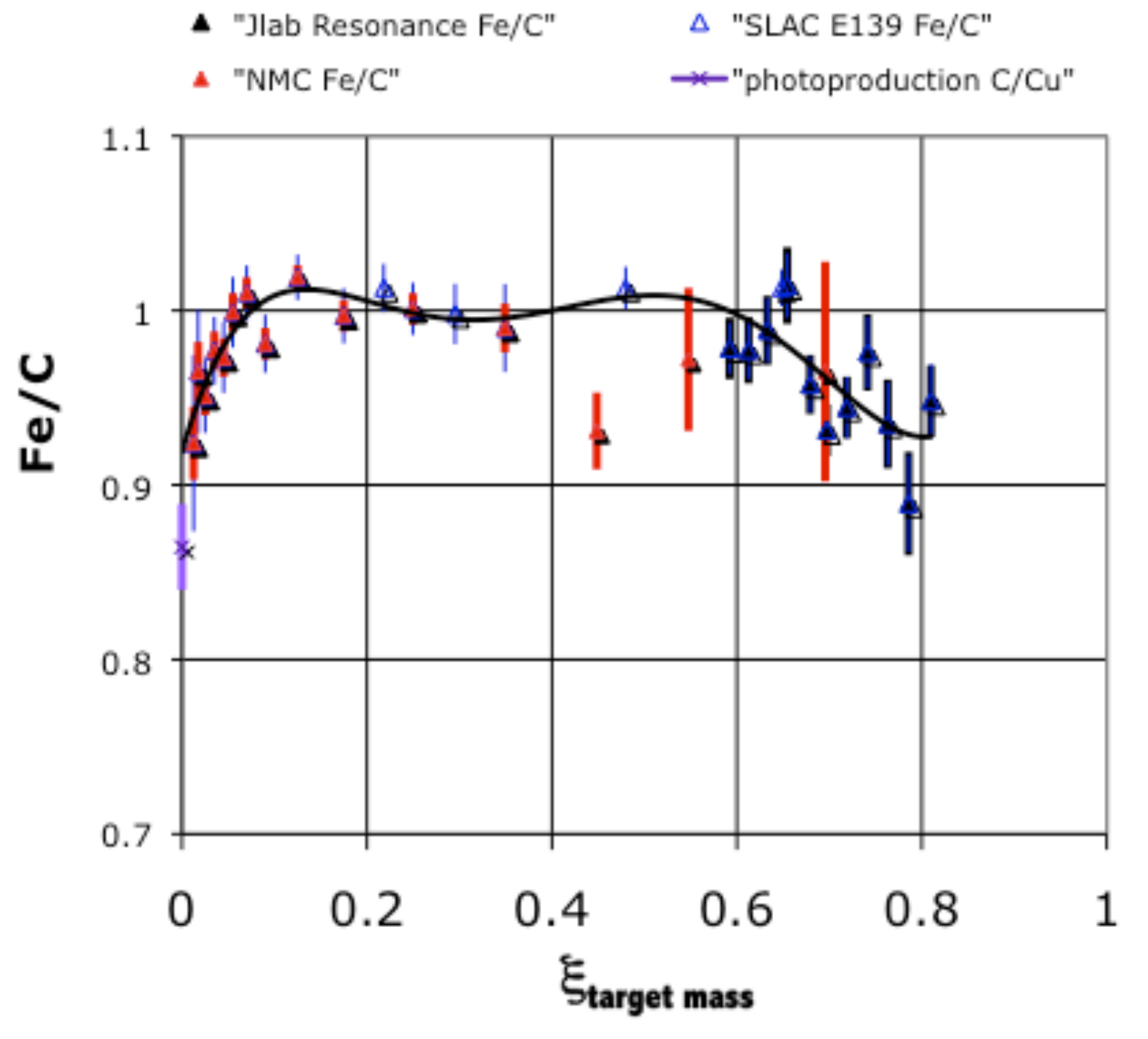}
\caption{The ratio of ${\cal F}_2$ data  for iron (Fe) to ${\cal F}_2$ data for  carbon (C)  to  as measured in charged-lepton scattering  in the deep inelastic region  (SLAC E139) as compared to Jlab data in the resonance region versus the target mass, or Nachtman variable\cite{Nachtman}  $\xi_{TM}$. Also shown is the ratio of  ${\cal F}_2$ data for carbon  to ${\cal F}_2$ data  for iron from the NMC\cite{NMCnuc} collaboration.  For comparison we also show the ratio  of of carbon to copper  cross sections (C/cu)  as measured in photoproduction\cite{photo}. } 
\label{fig:jlab_carbon}
\end{figure}

Figures~\ref{fig:jlab_Au} show the measured ratio of structure functions for gold (Au)\cite{e140}  or lead (Pb)\cite{NMCnuc}  to the structure functions for iron (Fe) versus  $\xi_{TM}$.  Fig.~\ref{fig:jlab_carbon}  shows the ratio of the structure functions for iron  to the structure functions for carbon versus $\xi_{TM}$. 

 The gold (and lead)   data are described by the function 
 \begin{eqnarray}
  &&  \frac{{\cal F}_2^{Au,Pb}}{{\cal F}_2^{Fe}}(\xi_{TM})=
 0.932+2.461\xi -24.23\xi_{TM}^2  \\
& +&101.03\xi_{TM}^3  
  -203.47~\xi_{TM}^4 +193.85~\xi_{TM}^5-69.82~\xi_{TM}^6. \nonumber
\end{eqnarray}

The carbon data\cite{e140,jlabC} are described by the function
 \begin{eqnarray}
   \frac{{\cal} F_2^{Fe}}{{\cal F}_2^{C}}(\xi_{TM}) &=& 0.919+1.844\xi_{TM}  -12.73\xi_{TM}^2 \nonumber \\  &+&36.89\xi_{TM}^3 
   -46.77\xi_{TM}^4 +21.22\xi_{TM}^5. 
   \end{eqnarray} 
    All of these ratios are for structure functions which have been corrected for the neutron excess in the nucleus and therefore account for nuclear effects only.

In neutrino scattering, we assume that the nuclear correction factors for ${\cal F}_{2}$, $x{\cal F}_{3}$ and $2x{\cal F}_{1}$ are the same.  This is a source of systematic error because the nuclear shadowing corrections at low $x$ can be different for the vector and axial structure functions. This difference can be accounted for by assuming a  specific theoretical model\cite{kulagin}.
\subsection{ Avoiding double counting of Fermi motion}
Note that when the model implemented in neutrino Monte Carlo generators we must be careful not to double count the effect of Fermi motion.  The above fits include the effect of Fermi motion at high $\xi_{TM}$.  If Fermi motion is applied to the structure functions, than it is better to assume that  the ratio of iron to deuterium without Fermi motion for  $\xi_{TM}>0.65$ is equal to the ratio at $\xi_{TM}=0.65$.
%
%
%

\section{ d/u correction}
\label{dovu}
%
The $d/u$ correction for the GRV98 LO PDFs is obtained from the NMC data for ${\cal F}_2^D/{\cal F}_2^P$. Here, Eq.~\ref{eq:nucl-d} is used to remove nuclear binding effects in the NMC deuterium ${\cal F}_2$ data. The correction term, $\delta (d/u)$ is obtained by keeping the total valence and sea quarks the same.
\begin{eqnarray}
\delta (d/u)(x) = -0.00817 + 0.0506x + 0.0798x^2,
\end{eqnarray}
where the corrected $d/u$ ratio is $(d/u)'=(d/u)+\delta (d/u)$. Thus, the modified $u$ and $d$ valence distributions are given by
\begin{eqnarray}
u_v' = \frac{u_v}{1+\delta (d/u) \frac{u_v}{u_v+d_v}} \\
d_v' = \frac{d_v+u_v \delta (d/u)}{1+\delta (d/u) \frac{u_v}{u_v+d_v}}.
\end{eqnarray}
The same formalism is applied to the modified $u$ and $d$ sea distributions. We find that the modified $u$ and $d$ sea distributions (based on NMC data) also agree with the NUSEA data in the range of $x$ between 0.1 and 0.4. Thus, we find that  corrections to $u$ and $d$ sea distributions  are not necessary.
%
 \section{Axial structure functions ${\cal F}_2$, and  $2x{\cal F}_1$}
 \label{axial-section}
 %
 At $Q^2=0$ the vector structure function ${\cal F}_2^{\nu -vector}$ is required to go to zero.  In contrast, the axial structure function ${\cal F}_2^{\nu- axial} $  is not constrained to go to zero at $Q^2=0$. At  higher  $Q^2$  ($>$1.5 GeV$^2$) the vector and axial structure functions should be equal. 
 Since the contribution of the structure function  $2x{\cal F}_1$ to the  cross section near $Q^2=0$ is very small we set
 \begin{eqnarray}
{2x\cal F}_{1}^{\nu- axial}(x,Q^{2}) &= &{2x \cal F}_{1}^{\nu -vector}(x,Q^{2}).
\end{eqnarray}
The axial contribution at small $Q^2$ is primarily longitudinal and only contributes to ${\cal F}_{2}$. 

 We compare neutrino data to  two  versions of the model as shown below.
%
\subsection {Effective LO PDFs Model Type I (axial=vector) }
The first version of the model (which we refer to as Type I(A=V)) assumes that the vector and axial components of the structure function   ${\cal F}_2^{\nu} $ are equal at all values of $Q^2$. i.e. 
  \begin{eqnarray}
{\cal F}_{2}^{\nu- axial}(x,Q^{2}) = {\cal F}_{2}^{\nu -vector}(x,Q^{2})~(Type~I). 
\end{eqnarray}
This is the assumption that has been made in previous implementations of our model.   This assumption underestimates the neutrino cross section at low $Q^2$. In contrast, the assumption that the K axial factors are 1.0  overestimates the cross section.  To  properly model neutrino interactions propose the  Type II (A$>$V) model described below.
\subsection {Effective LO PDFs model Type II (A$>$V) (a better model)}
%
In this version of the model we account for the fact that the axial and vector structure functions are not equal at $Q^2$=0 as follows:
 \begin{eqnarray}
{\cal F}_{2}^{\nu -axial}(x,Q^{2}) = \nonumber
 \Sigma_iK_i^{axial}(Q^2) \xi_w q_i(\xi_w,Q^2)\nonumber \\
 +      \Sigma_j K_j^{axial}(Q^2)   \xi_w \overline{q}_j(\xi_w,Q^2). 
\end{eqnarray}
%
\subsubsection{Axial sea}
%
For sea quarks, use use the same axial $K$ factor for all types of quarks.	
\begin{eqnarray}		
\label{eq:kfac-axial}	     
 K_{sea}^{axial}(Q^2) &=& \frac{Q^2+ P_{sea}^{axial}C_{sea}^{axial} }{Q^2 + C_{sea}^{axial}} \nonumber
 \end{eqnarray}	 
 where $P_{sea}^{axial} =0.55\pm0.05$, and $C_{sea}^{axial}= 0.75\pm0.25$ yielding
 \begin{eqnarray}	     
 K_{sea}^{axial}(Q^2) &=&  \frac{Q^2+ 0.41}{Q^2 + 0.75} ~(Type~II).
\end{eqnarray}
We use 30\% of the difference  between the cross section predictions of the Type I (A=V)  and Type II (A$>$V)  models as an estimate of the uncertainty in the axial $K$ factors. 
\subsubsection{Explanation of the origin of the axial sea}
We refer to the non-zero value of the $K_{sea}^{axial}$ at $Q^2$=0 as the PCAC term in ${\cal F}_2$.  We obtain the parameters  $P_{sea}^{axial} =0.55\pm0.05$, 
and $C_{sea}^{axial}= 0.75\pm0.25$ using the following relation,
\begin{eqnarray}		
	{\cal F}_2^{axial-sea} = {\cal F}_2^{vector-sea} +{\cal F}_2^{PCAC}
\end{eqnarray}
where  ${\cal F}_2^{PCAC}$ is from the model of Kulagin and Peti\cite{kulagin}.
%

As a check, we note that the CCFR\cite{bonnie} collaboration has reported on a measurement  (via an extrapolation)  of  ${\cal F}_2^{Fe}$  at $Q^2=0$.  The CCFR value for an iron target (per nucleon)  ${\cal F}_2^{Fe} (Q^2=0)$=0.210$\pm$0.02, is in agreement with  our model prediction for  ${\cal F}_2^{axial,Fe}(x=10^{-5}, Q^2=0)$=0.251$\pm$0.025. Our value is obtained using $P_{sea}^{axial} =0.55\pm0.05$ in conjunction with  ${\cal F}_2^{GRV98}$=0.57 for $Q^2=0.8$ GeV${^2}$ and $x=10^{-5}$ (assuming a nuclear shadowing ratio ${\cal F}_2^{Fe}/{\cal F}_2^D=0.8$.)
\subsubsection{Axial valence}
For  the  valence quarks, we note that the following is a good approximation to the vector  $K$ factor.
 \begin{eqnarray}
   K^{vector}_{valence}(Q^2) \approx  [1-G_D^2(Q^2)]  \approx  \frac {Q^2}{Q^2+0.18}. 
\end{eqnarray}
Where $Q^2$ is in units of  GeV$^2$.
We use a similar form for the axial $K$ factor for valence quarks. 
 \begin{eqnarray}
 K^{axial}_{valence}(Q^2)= \frac {Q^2 +P_{valence}^{axial}\times0.18}{Q^2+ 0.18} ~ (Type~II).
\end{eqnarray}
Where $P_{valence}^{axial}=0.3$ is chosen to get better agreement with measured high energy neutrino and antineutrino total cross sections. In summary, 
 \begin{eqnarray}
          K^{axial}_{valence}(Q^2)= \frac {Q^2+0.054\pm0.009}{Q^2+ 0.18} ~ (Type~II),
\end{eqnarray}
which implies that the axial $K$ factor for the valence quarks at $Q^2=0$ is 0.3.  We use the same axial $K$ factor for the $u$ and $d$  valence quarks.  As mentioned earlier, we assume $2x{\cal F}_1^{axial}=  2x {\cal F}_1^{vector}$.   This is because the non-zero PCAC component of  ${\cal F}_2^{axial}$ at low $Q^2$  is purely longitudinal and therefore does not contribute to $2x {\cal F}_1^{axial}$ which is purely transverse.
%
  \begin{figure}
\includegraphics[width=3.6in,height=3.3in]{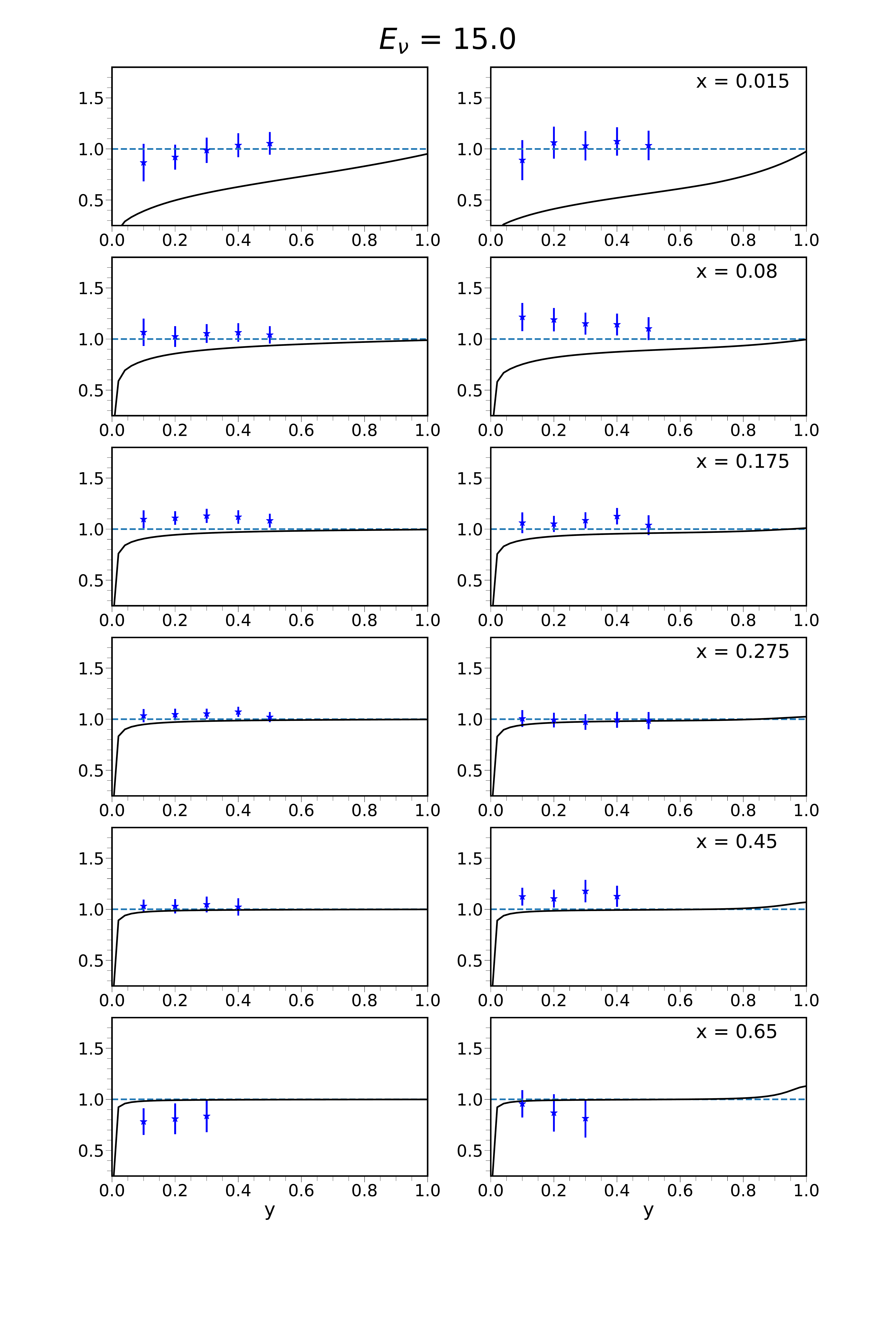}
\includegraphics[width=3.6in,height=3.3in]{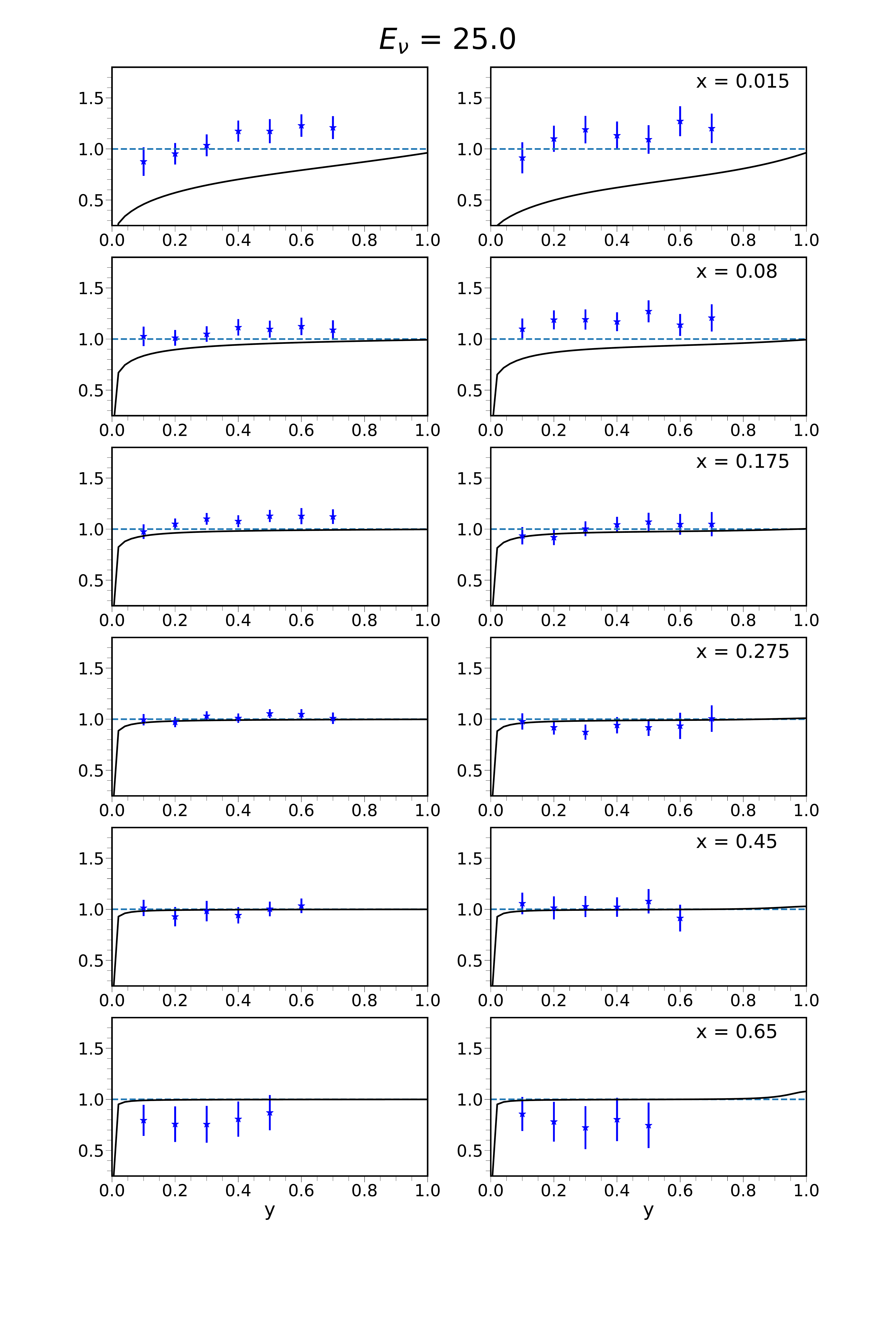}
\caption  {Same as Fig.\ref{fig:neutrinoD3} for energies of 15 and 25 $GeV$ (CCFR data only). } 
\label{fig:neutrinoD1}
\end{figure}
%
  \begin{figure}
\includegraphics[width=3.6in,height=3.3in]{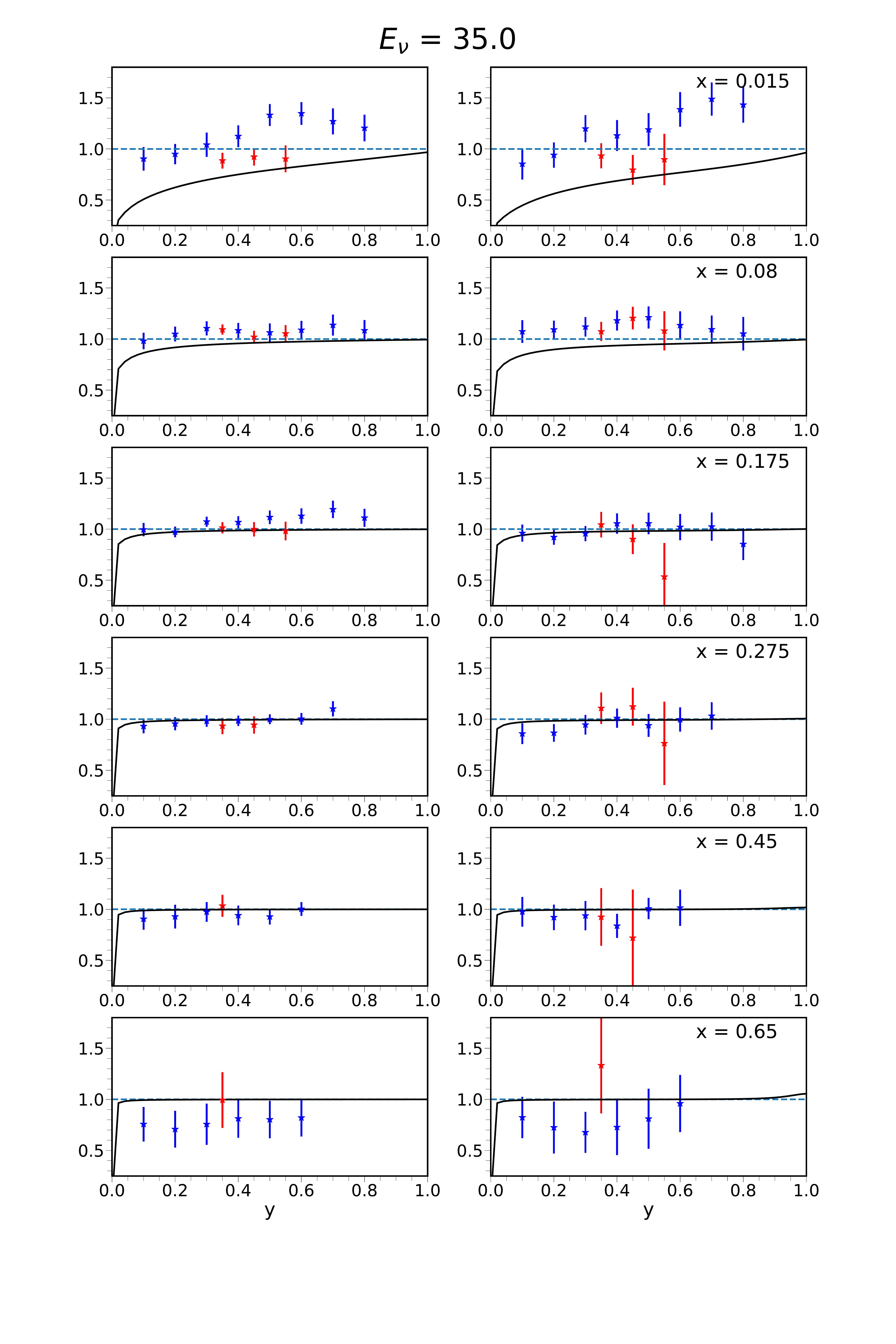}
\includegraphics[width=3.6in,height=3.3in]{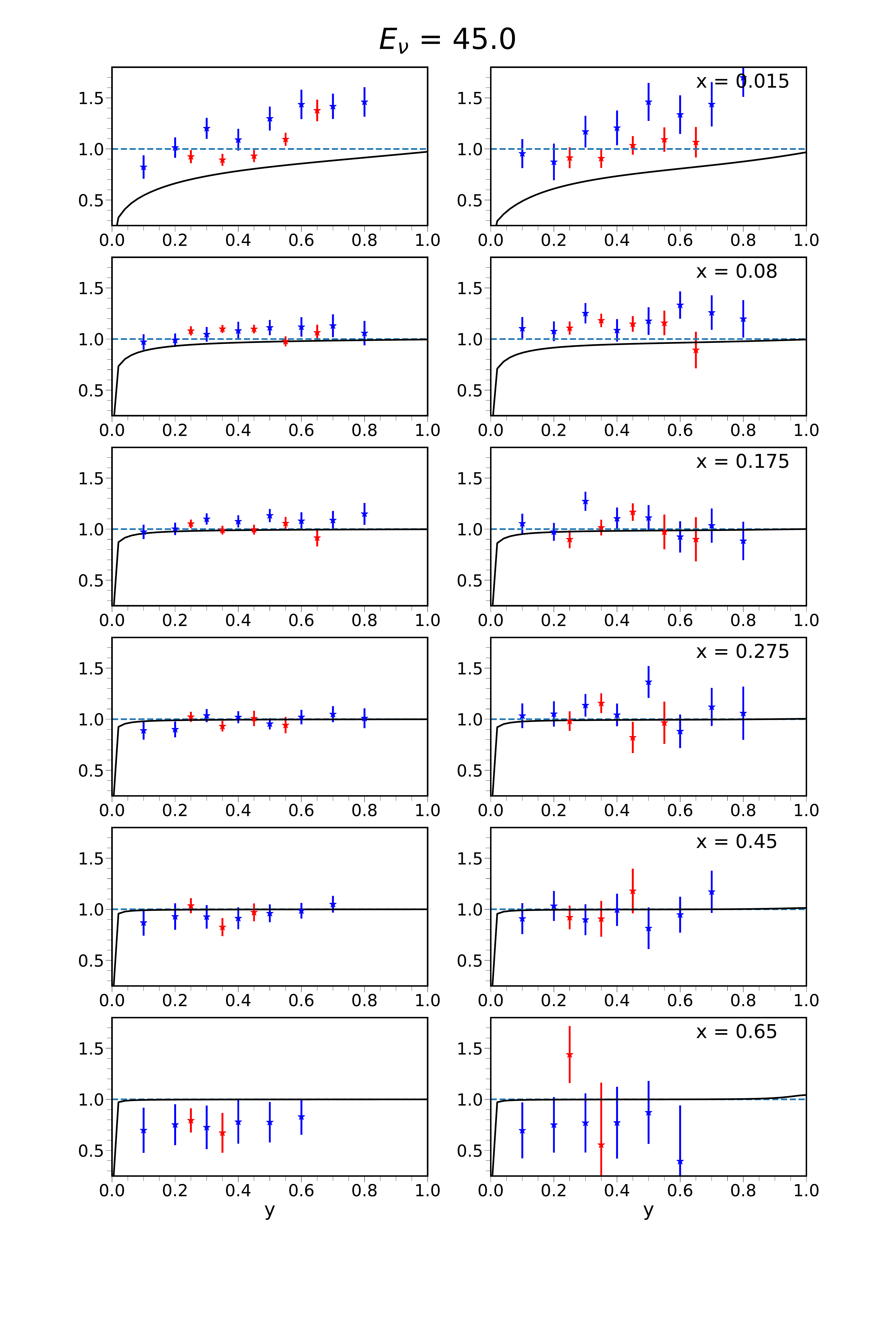}
\caption { The ratio of  charged-current neutrino and antineutrino 
differential cross sections $d^2\sigma/dxdy$ on lead from CHORUS~\cite{chorus} (blue points) and CCFR cross sections (red points) on iron~\cite{yangthesis,rccfr} to  Type II (A$>$V) default model. The ratios are shown for  energies of 35 and 45 $GeV$.   On the left side we show the comparison for neutrino cross sections and on the right side we show the comparsons for antineutrinos.   The black line is the ratio of the predictions of the  Type I (A=V) model for which the  axial structure functions are set equal to the vector structure functions,  to the predictions of the  Type II (A$>$V)  default model. The CHORUS and CCFR data favor the  Type II (A$>$V)  model. }
\label{fig:neutrinoD3}
\end{figure}
%
 \begin{figure}
\includegraphics[width=3.6in,height=3.3in]{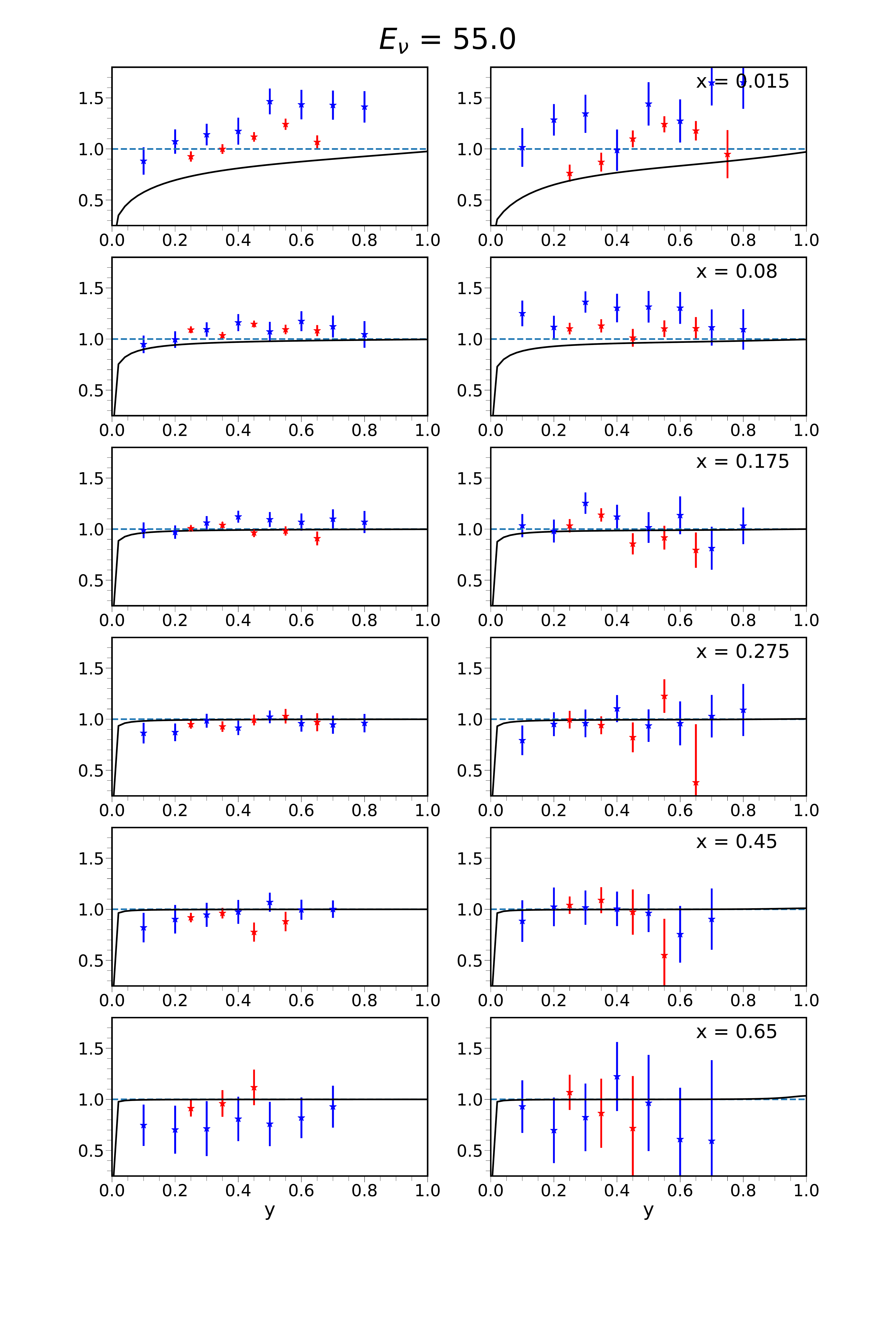}
\includegraphics[width=3.6in,height=3.3in]{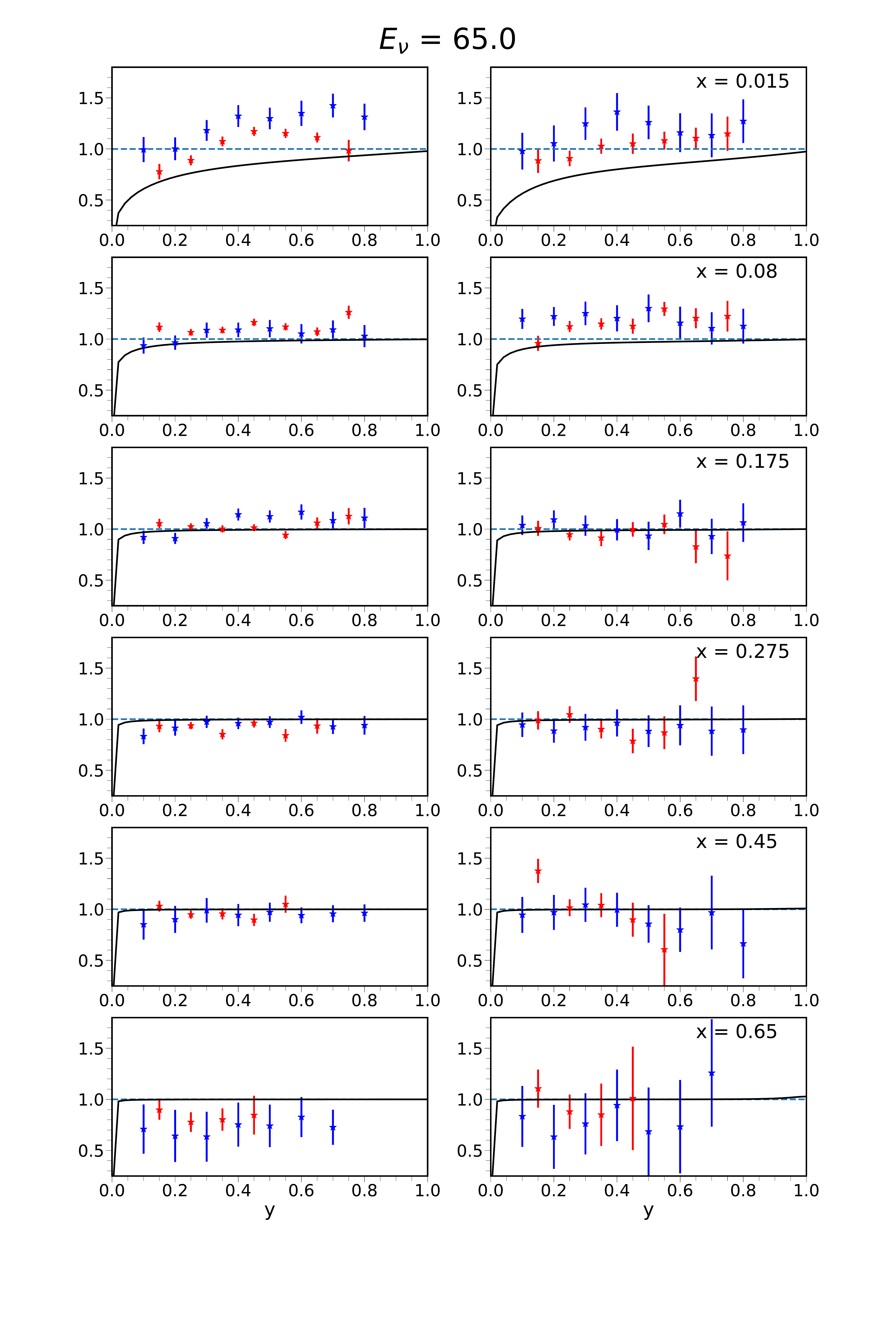}
\caption {Same as Fig.\ref{fig:neutrinoD3} for energies of 55 and 65 $GeV$.  }
\label{fig:neutrinoD4}
\end{figure}%
%
 %
 \section{Comparison to inelastic $\nu$ and \nubar~cross sections on nuclear targets }
 \label{nuclear-targets}
\subsection {$\nu$ and \nubar ~differential cross section data}
We compare the model predictions to neutrino differential cross sections ($\frac{d^2\sigma^{\nu(\overline{\nu})}}{dxdy}$) on lead (CHORUS~\cite{chorus}) and iron (CCFR~\cite{yangthesis,rccfr}).  We multiply the CHORUS cross sections by  ratio of the nuclear corrections for iron divided by the nuclear correction for lead, such that both differential cross sections can be shown on the same plot and compare to the predictions for the neutrino differential cross sections on iron.
The neutrino (antineutrino) differential  cross section is given by:
\begin{eqnarray}
\label{eqn:disxsec}
\frac{d^2\sigma^{\nu(\overline{\nu})}}{dxdy} &=& \frac{G_F^2 M  E}{\pi} {\Big(} \Big[1-y(1+\frac{Mx}{2E})\Big]  {\cal F}_{2}
+\frac{y^2}{2} (2x {\cal F}_{1})   \nonumber \\
&\pm&  \Big[y-\frac{y^2}{2}\Big]x{\cal F}_{3}{\Big)}.
\end{eqnarray}
%
%
where $ {\cal F}_{2}={\cal F}_{2}^{vector}+{\cal F}_{2}^{axial}$,  $ {\cal F}_{1}={\cal F}_{1}^{vector}+{\cal F}_{1}^{axial}$, and
  $G_F /(\hbar c)^3 = 1.1663787(6)\times10^{-5}$~GeV$^{-2}$ is the Fermi coupling constant and $ (\hbar c)^2$= 0.389 379 3656(48) GeV$^2$ mbarn.
  
In the comparison we assume that the ratio of neutrino  structure functions for nucleons bound in a nucleus to neutrino  structure functions  free nucleons for neutrinos is equal to the ratio measured in electron/muon scattering for ${\cal F}_{2}$. We also assume that the nuclear corrections  are  the same for the axial and vector part of the structure functions. This is a source of systematic error because the nuclear shadowing corrections at low $x$ can be different for the vector and axial terms (this difference can be accounted for by assuming a  specific theoretical model\cite{kulagin}). 

The published  CHORUS and CCFR  differential cross sections have been corrected for radiative corrections. In addition, the  CHORUS  and CCFR data are corrected for the  neutron excess in lead and iron.  Consequently, we  compare the  CHORUS data to the model prediction for  isoscalar  (i.e. equal number of neutrons and protons) lead and iron targets .

Figures~\ref{fig:neutrinoD1}-\ref{fig:neutrinoD4} show the ratio of  charged-current neutrino and antineutrino  differential cross sections $d^2\sigma/dxdy$ on lead from CHORUS (blue points)  and CCFR cross sections on iron (red points), to the  Type II (A$>$V)  default model.   The ratios are shown for neutrino energies of 15, 25, 35, 45, 55 and 65 $GeV.$  On the left side we show the comparison for neutrinos  and on the right side we show the comparison for antineutrinos.  The black line is the ratio of the predictions of the  Type I (A=V)  model for which the  axial structure functions are set equal to the vector structure functions, to the predictions of the Type II (A$>$V) default model. The CHORUS and CCFR data  favor the  Type II (A$>$V) model. 
%
\begin{table}[ht]
    \begin{center}
\begin{tabular}{|l|l|l|l|}
\hline
 $P_{sea}^{axial}$ & $C_{sea}^{axial}$ &  $P_{sea}^{valence}$ \\
$0.55$  &$0.75$ &   $0.3$   \\
  \hline
  \hline
  $C^{low-\nu (\nu)}_{axial}$  &  $C^{low-\nu (\overline{\nu})}_{axial}$&    \\
 $0.436$ & $ 0.654$ &\\       
 \hline
 \hline
 \end{tabular}
\caption{ A summary of the axial parameters for the Type II (A$>$V)  default model. All parameters are in units of GeV$^2$.}
\label{Table-axial}
  \end{center}
\end{table}
%
%
%
\subsection{Modeling $\nu$ and \nubar~cross sections in  the resonance region}
\label{duality}
As mentioned earlier, the  $K^{LW}_{vector}$ factor should be included for a better description of electron scattering and photo-production  $average$ cross sections in in the resonance region.  For the vector structure functions $K^{LW}_{vector}=(\nu^2 + C^{low-\nu}_{vector})/\nu^2$ (where $C^{low-\nu}_{vector}$=0.218~GeV$^2$ ) as shown in equation \ref{eq:kfac2}.

In order to better describe the low energy neutrino and antineutrino total cross sections (as discussed below) we find that the 
$K^{LW}$ factor  for the $axial$ part of the  cross section in the resonance region is larger and is different for neutrinos and antineutrinos, 
  $$K^{LW- (\nu,\overline{\nu})}_{axial}=\frac{\nu^2 + C_{axial}^{low-\nu(\nu, \overline{\nu})} }{\nu^2}$$
  For neutrinos:
$$C^{low-\nu (\nu)}_{axial}=0.436~GeV^2$$
For antineutrinos
  $$C^{low-\nu (\overline{\nu})}_{axial}=0.654~GeV^2.$$
 
 \noindent A summary of the axial parameters is given in Table \ref{Table-axial}.
 \subsection{Comparisons to $\nu$ and \nubar~ total cross section measurements for $E_\nu>30$ GeV}
To test the validity of the model we compare the model predictions for the   $\nu$ and \nubar~total cross sections to measurements.
In the calculation of the neutrino total cross sections includes the following components.
\begin{enumerate}
\item The contributions of the quasielastic (QE) cross section and the  cross section for the $\Delta$($W<1.4~GeV$) resonance region are extracted from measurements as  described in section \ref{all_cross}.
\item The contribution of the higher resonances  $1.4<W<1.8~GeV$ is calculated using our model.
\item The contribution of the inelastic $W>1.8$ GeV continuum is calculated using our model.
\end{enumerate}
\begin{table}[ht]
    \begin{center}
\begin{tabular}{|l|l|l|l|l|}
\hline            
$E_\nu$ & $QE$ &$W<1.4$ & $1.4<W<1.8$& $W>1.8$  \\
GeV& & GeV & GeV & GeV \\
 & &$\Delta$(1238) & resonances& inelastic  \\
\hline
 3  & 23.8\% &  19.7\%&  31.3\% & 25.2\%\\
  \hline
 5 & 16.2\% &  12.5\%&  22.2\% & 48.1\%\\
  \hline
10  & 7.2\%&  6.5\%  & 13.4\% & 72.8\%\\
\hline
40   & 1.5\%  & 1.6\%  & 6.5\%  & 90.4\% \\
 \hline
 \hline
 $E_{\overline{\nu}}$ & $QE$ & $W<1.4 $ & $1.4<W<1.8$& $W>1.8$ \\
 GeV& & GeV & GeV & GeV \\
  & &$\Delta$(1238) & resonances& inelastic  \\
\hline 
3  & 40.7\% &  27.1\%&  25.6\%&  6.6\% \\
  \hline
 5  & 27.8\% &  20.8\%&  33.5\%&  17.9\% \\
  \hline
10  & 15.0\%&  11.7\%  & 25.2\%& 48.1\% \\
\hline
40  & 3.1\%  & 3.4\%  & 7.1\% &86.4\%  \\
 \hline
 \hline
 \end{tabular}
\caption{ Percent contributions to the total cross section of   QE,  $\Delta (W<1.4)$ GeV,  higher resonances $1.4<W<1.8$ GeV 
and inelastic continuum $W>1.8$ GeV.}
\label{Table-percent}
  \end{center}
\end{table}
%
\begin{table}[ht]
\caption{ \label{high_energy}
  Model predictions for  $\sigma_{\nu}$/E per nucleon (for nucleons bound in an isoscalar iron target)  in units of $10^{-38}~cm^2/GeV$,  $\sigma_{\bar \nu}$/E per nucleon  in units of $10^{-38}~cm^2/GeV$, and the ratio $\sigma_{\bar \nu}$/$\sigma_{ \nu}$ for an average neutrino energy of 40 GeV. The  predictions 
  for both the Type I (A=V) and  the Type II (A$>$V)   default models 
  are compared to the averages \cite{MINOS2} of all of the world's data for energies between 30 and 50 GeV.
  The axial parameters for the   Type II (A$>$V) model were tuned to agree with data.}
  \begin{tabular}{|c|c|c|c|} \hline \hline
 &  Type I (A=V) &  Type II (A$>$V)&  World~Average \\ \hline
  $\sigma_\nu$ /E &0.656 $\pm$0.024 & 0.674 $\pm$0.024 & 0.675 $\pm$0.006 \\
   ${\sigma_{\bar\nu}}$/E &0.311 $\pm$ 0.016 &0.327 $\pm$ 0.016 & 0.329 $\pm$ 0.011\\
    ${\sigma_{\bar\nu}}/{\sigma_{\nu}}$ & 0.474 $\pm$ 0.012 &  0.487 $\pm$ 0.012 &  0.485 $\pm$ 0.005 \\     
    \hline \hline
\end{tabular}
\end{table}
The fractional contributions to the total $\nu$ and \nubar~cross section of  the QE,  $\Delta (W<1.4)$, and higher resonance $1.4<W<1.8$ regions are shown in Table \ref{Table-percent}.
For neutrino and antineutrino energies of 40 GeV  the contributions from  the QE,  $\Delta (W<1.4)$, and higher resonance $1.4<W<1.8$ regions small and the cross sections are dominated  by inelastic  $W>1.8$ GeV continuum.
  Consequently, comparisons of our predictions  to total cross section measurements at 40 GeV provide a good test of the  modeling of the inelastic continuum.

Table~\ref{high_energy} shows  comparisons of the Type (A=V) and Type II  (A$>$V) model  predictions for  $\sigma_{\nu}$/E per nucleon (in an isoscalar iron nucleus)  at average neutrino energy of 40 GeV to the averages \cite{MINOS2} of all of the world's data.   The axial parameters for the  Type II (A$>$V) model were tuned to agree with the high energy total cross section  measurements.   
 %
 \begin{figure}
\includegraphics[width=3.5in,height=2.1in]{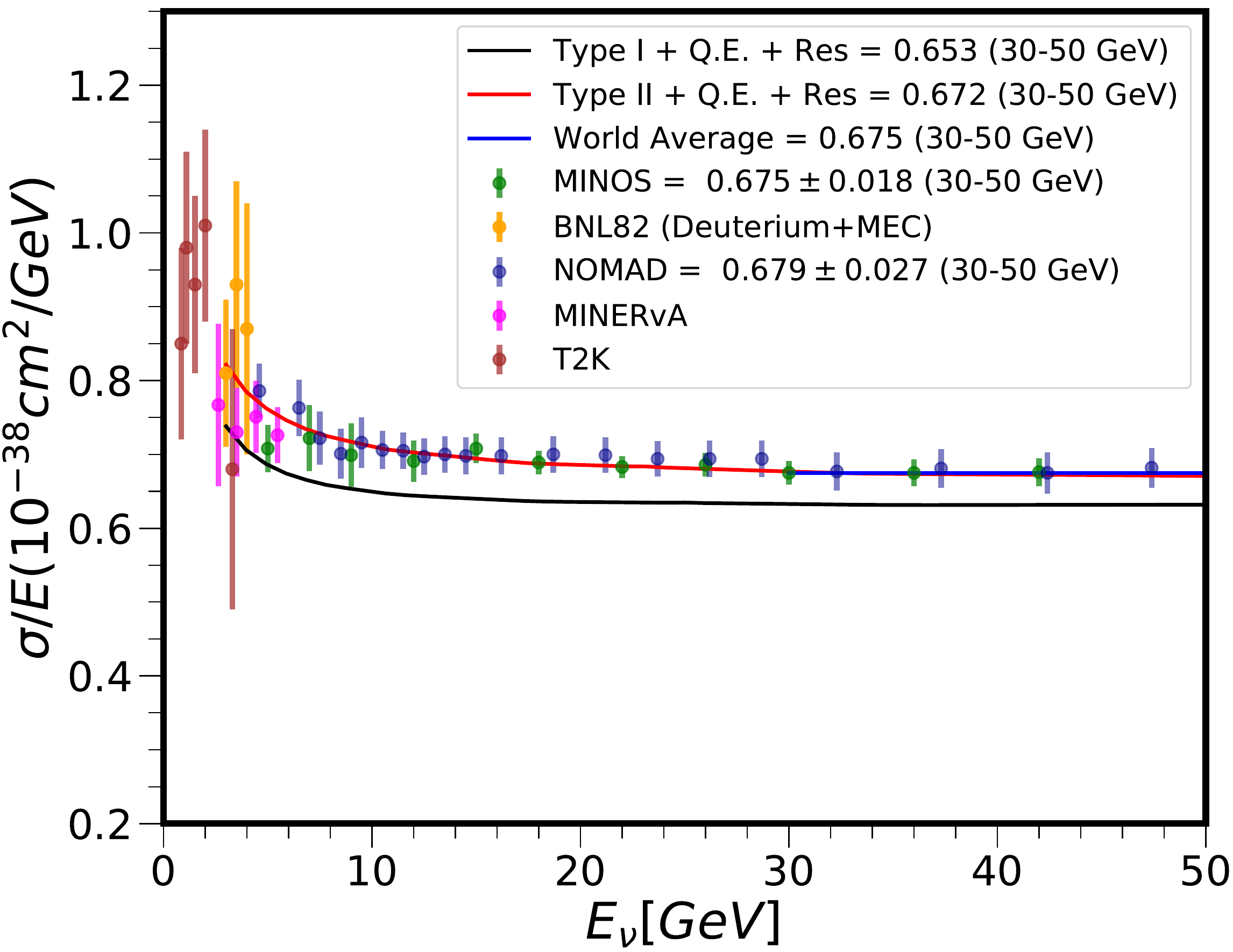}
\includegraphics[width=3.5in,height=2.1in]{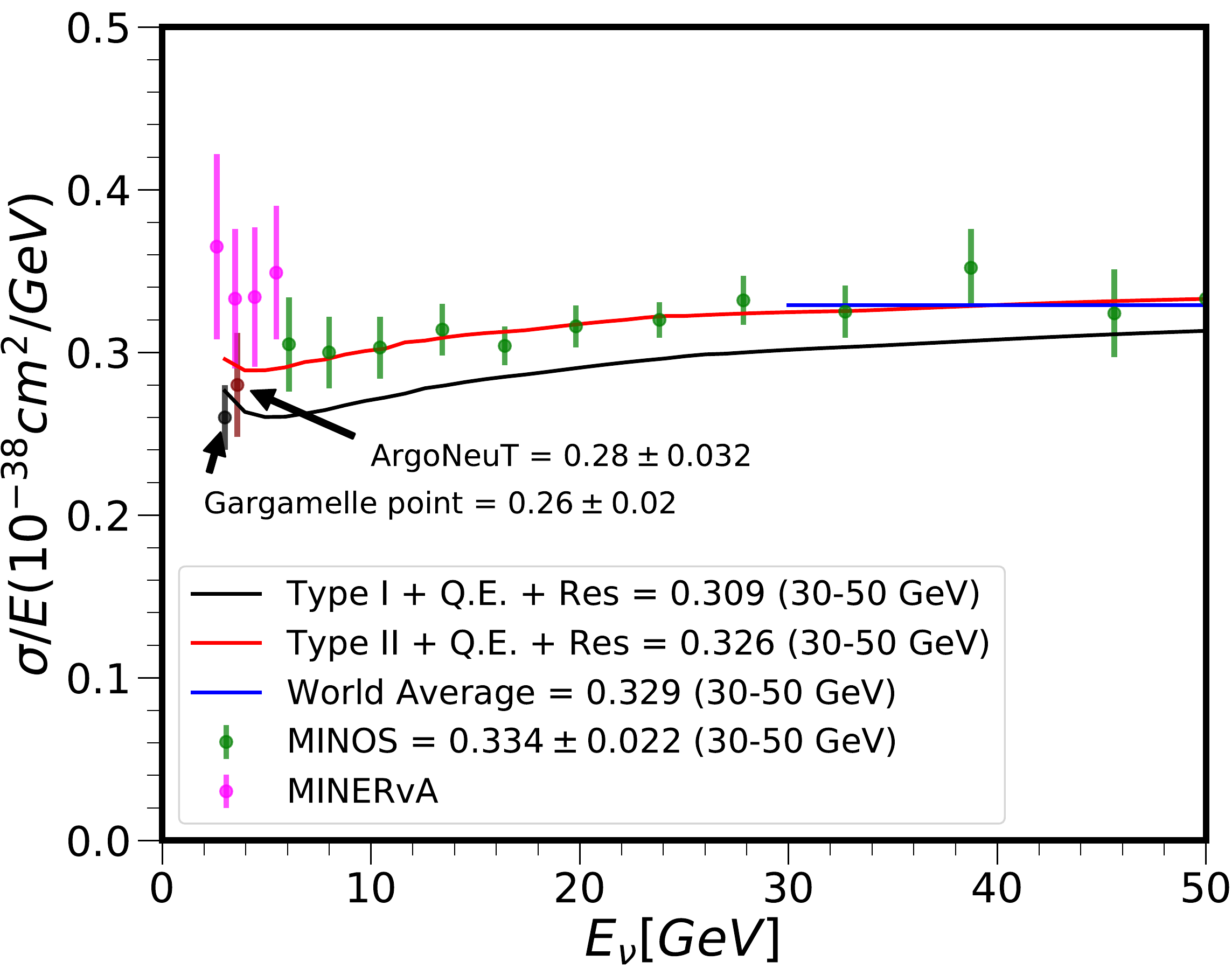}
\includegraphics[width=3.5in,height=2.1in]{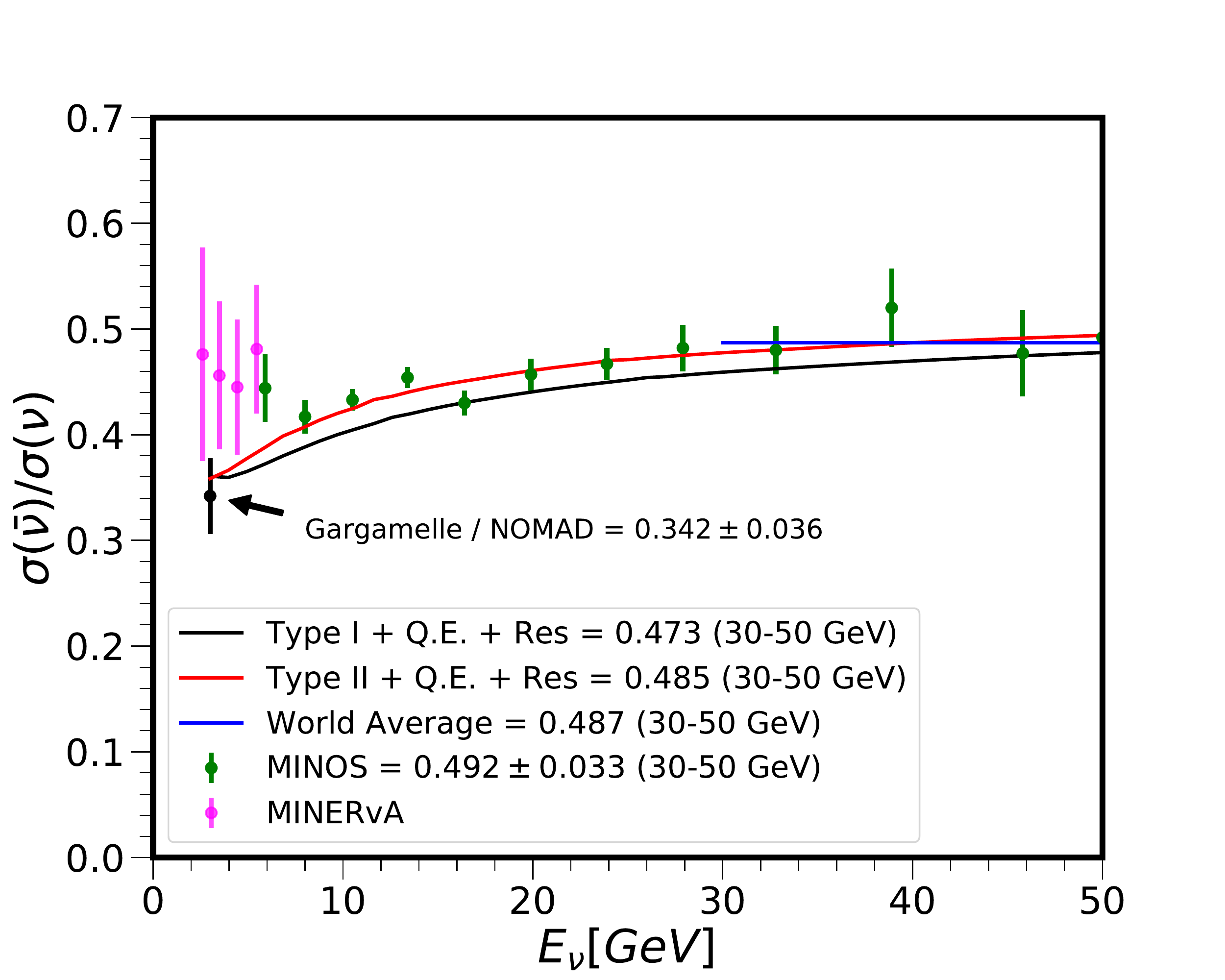}
\caption {Model  predictions (per nucleon) for an isoscalar iron target  compared to measurements. The top part of the figure is for $\sigma_{\nu}$/E,  the middle  panel is  for  $\sigma_{\bar \nu}$/E, and  the bottom panel is for the ratio $\sigma_{\bar \nu}$/$\sigma_{ \nu}$  as a function of energy.
The green points are MINOS\cite{MINOS2} data, and the blue points are NOMAD\cite{NOMAD,New_data_set} $\sigma_{\nu}$/E measurments. The yellow crosses are  BNL82\cite{BNL82} data as discussed in section \ref{all_cross}.  The MINERvA\cite{minerva1} and T2K\cite{T2K} are shown in purple and brown, respectively.
The Gargamelle\cite{Zeller,GGM,New_data_set}  and ArgoNeut\cite{argoneut} of  $\sigma_{\bar \nu}$/E per nucleon are identified.  On the ratio plot we also show the Gargamelle antineutrino $\sigma_{\bar \nu}$/E at 3 GeV divided by the NOMAD neutrino $\sigma_{\nu}$/E at 4.6 GeV. The axial parameters of the Type II (A$>$V) model were tuned to agree with the total cross section data,
}
\label{fig:neutrinoD12}
\end{figure}
\subsection{Comparisons to  $\nu$ and  \nubar ~total cross section measurements for $E_\nu < $ 10 GeV}
 Because of quark hadron  duality and the tuning of parameter described in section \ref{duality} the model also describes the  $average$ cross section in the  $1.4<W<1.8~GeV$  resonance region. As shown in Table \ref{Table-percent}, 
 at an incident energy of 5 GeV,  the contribution of the $1.4<W<1.8~GeV$  resonance region is significant. Therefore, comparison of our model predictions to low energy neutrino cross sections is a test of our modeling of the cross section in this resonance region.

Fig.~\ref{fig:neutrinoD12} shows model  predictions (per nucleon) for an isoscalar iron target that contains and equal number of protons and neutrons compared to measurements. The top panel is for $\sigma_{\nu}$/E,  the middle  panel is  for  $\sigma_{\bar \nu}$/E, and  the bottom panel is for  the ratio $\sigma_{\bar \nu}$/$\sigma_{ \nu}$  as a function of energy.

The green points are MINOS\cite{MINOS2} data, and the blue points are NOMAD\cite{NOMAD,New_data_set} $\sigma_{\nu}$/E measurements. The yellow crosses are BNL82\cite{BNL82} data as discussed in section \ref{all_cross}.  The MINERvA\cite{minerva1} and T2K\cite{T2K} are shown in purple and brown, respectively.
The Gargamelle\cite{Zeller,GGM,New_data_set}  and ArgoNeut\cite{argoneut} of  $\sigma_{\bar \nu}$/E per nucleon are identified.  On the ratio plot we also show the Gargamelle antineutrino $\sigma_{\bar \nu}$/E at 3 GeV divided by the NOMAD neutrino $\sigma_{\nu}$/E at 4.6 GeV.

The Type II  (A$>$V) red lines are the prediction of  the Type II default model.  The black lines are the prediction of the  Type (A=V) model for which the  axial structure functions are set equal to the vector structure functions. The blue lines  above 30 GeV are the averages \cite{MINOS2} of all of the world's data (on isoscalar iron)  for energies between 30 and 50 GeV. The axial parameters of the  Type II  (A$>$V) default model were tuned to agree with the total cross section measurements.

For $\nu$ and \nubar~scattering, the Type II  (A$>$V) default model describes the  $1.4<W<1.8~GeV$ resonance region on  $average$.  However if a better resonance model  is available, we suggest that it be used and smoothly matched to our model at $W>1.8$ GeV.  Since our model also describes  $average$ cross sections in the  $1.4<W<1.8~GeV$ region, this matching should be continuous.  In addition, comparisons of other resonance modesl predictions to our model in the  $1.4<W<1.8~GeV$ region provides an estimate of the systematic errors associated with the modeling of the resonances.
%
  \section{Systematic errors in the application of the model}
  \label{systematic-section}
 %
 The model predicts neutrino cross sections at the Born  level.  Therefore, radiative corrections must be applied to the model if it is  compared  to  non-radiatively corrected neutrino or charged-lepton  scattering data.  In general, all published charged-lepton scattering data are radiatively corrected. Similarly,   published neutrino differential cross sections (e.g. CCFR, CDHSW, CHORUS, NuTeV) are radiatively corrected,  and therefore can be directly compared to the model.
  
 The model describes all inelastic charged-lepton scattering data and photoproduction   on hydrogen and deuterium  for $W>1.8~GeV$ at all values of $Q^2$ (and gives a reasonable $average$ cross section in the resonance region for  $W>1.4~GeV$).  Therefore, under the assumption of CVC, the model describes the vector part of the cross section in neutrino scattering well.  The axial parameters of the  Type II  (A$>$V) default model were tuned to agree with the total cross section measurements.

Estimates of the  systematic error in the total cross sections in Table \ref{table6} were obtained by varying the parameters in the model within our estimated uncertainties.  in addition, a rough estimated  of the uncertainties can be obtained by
 writing the differential cross sections in terms of quark and antiquark distributions. Within the naive quark parton model, the vector and axial structure functions are the same i.e. Type I (A=V) and the structure functions are related to the quark distribution by the following expressions: 
 $${\cal F}_2=2x[q(x,Q^2)+\overline {q}(x,Q^2)]$$
   $$x{\cal F}_3=2[xq(x,Q^2)-\overline {q}(x,Q^2)]$$
We define   $Q_{dist}(x,Q^2)=2xq(x,Q^2)$  and $\overline{Q}_{dist}(x,Q^2)=2x\overline{q}(x,Q^2)$. We define  
$$Q_T=\int_{0}^{1}2xq(x,Q^2)dx$$ 
$$\overline{Q}_T=\int_{0}^{1}2x\overline{q}(x,Q^2)dx$$. 

The neutrino (antineutrino) differential  cross section are then given by :
\begin{eqnarray}
\frac{d^2\sigma^{\nu(\overline{\nu})}}{dxdy}& = &\frac{G_F^2 M  E}{\pi}  \nonumber \\
& \times&  {\Big(} \Big[1-y(1+\frac{Mx}{2E}) +\frac{y^2}{2} \frac{1+Q^2/\nu^2}{1+ {\cal R} (x,Q^2)}\Big] {\cal F}_{2} 
\nonumber \\
&\pm&  \Big[y-\frac{y^2}{2}\Big]x{\cal F}_{3}{\Big)}.
\end{eqnarray}
or 
\begin{eqnarray}
\frac{d^2\sigma^{\nu}}{dxdy}& = &\frac{G_F^2 M  E}{\pi}  {\Big(} Q_{dist}(x,Q^2) + (1-y)^2\overline {Q}_{dist}(x,Q^2)  \nonumber \\
&-& \frac{y^2}{2} \frac{ {\cal R} (x,Q^2) }{1+ {\cal R} (x,Q^2)}(Q_{dist}+\overline{Q}_{dist}) \\
& + &  \Big[-\frac{Mxy}{2E}+ \frac {Mxy/E}{1+{\cal R} (x,Q^2)}\Big] (Q_{dist}+\overline{Q}_{dist}){\Big)}.\nonumber
\end{eqnarray}
and
\begin{eqnarray}
\frac{d^2\sigma^{\overline \nu}}{dxdy}& = &\frac{G_F^2 M  E}{\pi}  {\Big(} \overline Q_{dist}(x,Q^2) + (1-y)^2 Q_{dist}(x,Q^2)  \nonumber \\
&-& \frac{y^2}{2} \frac{ {\cal R} (x,Q^2) }{1+ {\cal R} (x,Q^2)}(Q_{dist}+\overline{Q}_{dist})  \\
& + &  \Big[-\frac{Mxy}{2E}+ \frac {Mxy/E}{1+{\cal R} (x,Q^2)}\Big](Q_{dist}+\overline{Q}_{dist}){\Big)}.
\nonumber
\end{eqnarray}
%
%
Integrating over x and y, the  cross sections for neutrino (anti-neutrino) (at high energy)  can then be approximately expressed in terms of (on average)  the  fraction antiquarks  $f_{ \overline {Q}}=\overline{Q}_T/(Q_T+\overline{Q}_T$)  in the nucleon, and (on average) the ratio of longitudinal to transverse cross sections   ${\cal R}$ as follows:
\begin{equation}
\sigma(\nu N) \approx \frac{G_{F}^{2}ME}{\pi}(Q+\overline{Q})\Big[(1- f_{\overline {Q} })+\frac{1}{3} f_{\overline {Q}}
-\frac{1}{6}\frac {{\cal R}}{(1+{\cal R)}} \Big],\label{nu-approx}\end{equation}
and
\begin{equation}
\sigma(\overline{\nu}N \approx
\frac{G_{F}^{2}ME}{\pi}(Q+\overline{Q})\Big[\frac{1}{3}(1- f_{\overline {Q}})+ f_{\overline {Q}}
-\frac{1}{6}\frac {{\cal R}}{(1+{\cal R)}}  \Big].\label{nub-approx}
\end{equation}
With   $\langle{\cal R}\rangle=0.3$ at low $Q^2$  and  $\langle f_{\overline{Q}}\rangle=0.175$,  we obtain  $\langle {\sigma_{\bar\nu}}/{\sigma_{\nu}}\rangle=0.487$, which is the world's experimental average value in the 30-50 GeV energy range. The above expressions are only approximate.  We use the exact expressions to estimate the systematic errors in the modeling the cross section originating from uncertainties in  ${\cal R}$, uncertainties in  $f_{\overline {q}}$, uncertainties in the  axial K factors, and overall normalization. These are  summarized in Table \ref{table6}.
 
\begin{table}[ht]
\caption{ \label{table6} Sources of systematic error in the predicted inelastic contribution to the  total cross section on iron.  The  change (positive or negative)  in the neutrino, antineutrino and the ${\sigma_{\bar\nu}}/{\sigma_{\nu}}$ ratio that originate from a plus one standard deviation change in the  ratio of transverse to longitudinal structure functions (R), the fraction of antiquarks ($f_{\overline {q}}$), the difference between axial and vector K factors, and the overall normalization of the structure functions (N).} 
  \begin{tabular}{|c|c|c|c|c|} \hline \hline
source &  change  &  change & change&  change  \\
& (error)    &  in $\sigma_\nu$  & in ${\sigma_{\bar\nu}}$ & in ${\sigma_{\bar\nu}}/{\sigma_{\nu}}$ \\  \hline
 {\cal R}    &   +0.1   &  -1.3\%  & -2.7\%& -1.4\% \\ 
$f_{\overline {Q}}$
    &   +5\%   &  -0.4\% & +0.9\% & +1.4\% \\ 
$K^{axial}-K^{vector}$
    &  -30\%   &  -0.8\% &  -1.5\% & -0.7\%\\ 
    \hline
    Subtotal& & $\pm1.6\%$& $\pm3.2\%$&  $\pm2.1\%$\\
N  &   +3\%   &   +3\% & +3\%&  0 \\  \hline
    Total& & $\pm3.4\%$& $\pm4.4\%$&  $\pm2.1\%$
     \\\hline \\ \hline
        Experimental & & & &  \\
         uncertainties& & & &  \\
       in Total $\sigma$ & & & &  \\
     measurements  & & $\pm0.9\%$& $\pm3.4\%$&  $\pm1.0\%$
     \\\hline \hline
\end{tabular}
\end{table}

We estimate the total systematic error in  the modeling of the cross sections on iron for the  $W>1.8~GeV$ region to be  $\pm3.4 \%$ for neutrinos,  $\pm4.4\%$ for antineutrinos, and  $\pm 2.1\% $  in the  ${\sigma_{\bar\nu}}/{\sigma_{\nu}}$ ratio (for neutrino energies below 50 GeV).  The errors are dominated by the PDF normalization errors of $\pm$3\%.  However, since the axial parameters were tuned to agree with the world's total cross sections measurement, the smaller experimental uncertainties in the total cross section measurements shown in Table \ref{table6} may  be taken as a lower limits of the systematic errors. 
 
 The following sources contribute to the systematic error.
 \begin{enumerate}
 \item  Longitudinal structure function:  In the analysis we use the $ {\cal R}_{1998}$ parametrization.    Preliminary results from the JUPITER Jefferson Lab collaboration indicates that   $ {\cal R}$  for heavy nucleus may be higher by about 0.1 than $R$ for deuterium.  Therefore, we  use   an error of $\pm$0.1 in R to estimate the systematic error in the cross sections from this source. 
 \item The antiquark fraction in the nucleon ($f_{\overline {q}}$).  We estimate an uncertainty of  $\pm  5\%$ in the fraction of  the sea quarks at low $Q^2$. 
  \item We assign a $\pm3\%$  error in the overall normalization of the structure functions (N) on iron, partly from the error in normalization of the SLAC data on deuterium and partly from   the level of consistency of the $Fe/D$ cross section  ratio among the various measurement as seen in Fig.\ref{fig:jlab2}.
  \item Axial $K$ factors for sea  and valence quarks:   We use 30\% of the difference
  between the cross section predictions of the Type I (A=V) and  the Type II  (A$>$V) models as an estimate of the uncertainty in the axial $K$ factors. 
  \item Charm sea: Since the GRV98 PDFs do not include a charm sea, the charm sea contribution must be added  separately. This can be implemented either by using a boson-gluon fusion model, or by incorporating a charm sea   from another set of PDFs.      We modeled  the contribution of the charm sea using a photon-gluon fusion model when we compared the  predictions to photo-production data at HERA.    If   the charm sea contribution is neglected,  the  model  underestimates  the cross section at very high neutrino  energies in the  low $x$  and large $\nu$ region. For  neutrino energies less than 50 GeV,  the charm sea contribution is very small and can be neglected.
\end{enumerate}
The following are additional sources of systematic errors which which  are not included in Table \ref{table6}.
 \begin{itemize}
  \item Nuclear corrections:    The model is primarily a model for the structure functions of free nucleons.  Only hydrogen and deuterium data are included in the fits. 
   
\item   However, electron scattering data indicate that nuclear effects change the shape of the $x$ and $Q^2$ dependence of the structure functions of bound nucleons.  Therefore in order to predict differential neutrino cross sections on  heavy targets, we assume  that the  nuclear corrections  are the same for the three structure functions. We also assume that the   corrections are the same for  the   axial  and vector contributions (and  are equal to the nuclear corrections  for ${\cal F}_2$ as measured  in charged-lepton scattering), and  that the nuclear  corrections are  only a function of   $\xi_{TM}$  and are independent of $Q^2$.    In general, nuclear corrections can be different for sea and valence quarks,  and  also for the longitudinal and transverse structure functions. Some of the systematic  error  in the modeling of the scattering from nuclear targets can be reduced when Jefferson Lab data on the nuclear dependence of $ {\cal R}=\sigma_L/\sigma_T$ are published.   Other systematic  errors in the nuclear corrections can be reduced by assuming specific theoretical models\cite{kulagin} to account for the differences in the nuclear  corrections  between neutrino and charged-lepton scattering  (as a function of $Q^2$ and $x$ for various nuclear targets).
  \end{itemize}           
  \begin{table}[ht]
    \begin{center}
\begin{tabular}{|l|l|l|l|l}
\hline            
$A$ & $B$ & $C_{v2d}$ & $C_{v2u}$  \\
$0.538$ & $0.305$ & $0.255$ & $0.189$   \\
\hline
\hline
 $C_{sea}^{down}$ & $C_{sea}^{up}$ &  $C_{v1d}$&  $C_{v1u}$ \\
$0.621$  &$0.363$ &   $0.202$  & $0.291$  \\
  \hline
  \hline
  $C_{sea}^{strange}$  & $$&  ${\cal F}_{valence}$   & $N$ \\
 $0.621$ & $$   & $[1-G_D^2(Q^2)]$   & $1.015$   \\
 \hline
 \hline
 \end{tabular}
\caption{ Parameters from  NUINT04 version of the model which are  currently implemented (with GRV98  PDFs) in the GENIE neutrino event generator.  When applicable, all parameters are in units of GeV$^2$.}
\label{Table-nuint04}
  \end{center}
\end{table}
\section{Updating the model  in neutrino MC generators}
\label{updates}
%
The current (2016) version of the GENIE\cite{GENIE} neutrino generator  is using the NUINT04\cite{nuint04} version of the model. This early version of the model assumes that the axial structure functions are the same as the vector structure  functions. As noted earlier, in this update, we refine the model and also account for the difference  between the axial and vector structure functions at low values of $Q^2$. Table \ref{Table-nuint04} shows the vector parameters of the NUINT04 version.  Implementation of the 2021 Type II  (A$>$V) default in neutrino MC generators can be done by updating the NUINT04 model as follows:
 \begin{enumerate}
\item The  vector parameters in Table \ref{Table-nuint04}  should be  replaced by the vector parameters in table \ref{iteration2} (see equation \ref{eq:kfac2}).
\item The axial K factors as described in section \ref{axial-section} should be used for the axial structure functions.
\item
Note that when the model implemented in neutrino Monte Carlo generators we must be careful not to double count the effect of Fermi motion.  The above fits include the effect of Fermi motion at high $\xi_{TM}$.  If Fermi motion is applied to the structure functions, than it is better to assume that  the ratio of iron to deuterium without Fermi motion for  $\xi_{TM}>0.65$ is equal to the ratio at $\xi_{TM}=0.65$.
 \item The  $K^{LW}_{vector}$ factor should be included for a better description in the resonance region. Here, $K^{LW}_{vector}=(\nu^2 + C^{low-\nu}_{vector})/\nu^2$ (where $C^{low-\nu}_{vector}$=0.218) as shown in equation \ref{eq:kfac2}.
  \item The  $K^{LW-(\nu, \overline{\nu})}_{axial}$ factor should be included for a better description in the resonance region. Here, 
  $$K^{LW- (\nu,\overline{\nu})}_{axial}=\frac{\nu^2 + C_{axial}^{low-\nu(\nu, \overline{\nu})} }{\nu^2}$$
  For neutrinos:
$$C^{low-\nu (\nu)}_{axial}=0.436.$$
For antineutrinos
  $$C^{low-\nu (\overline{\nu})}_{axial}=0.654.$$
 \item The structure function  $x{\cal F}_{3}$  should be multiplied by the  $H(,x,Q^2)$ as described in  equations \ref{xF3equation} and  \ref{Hequation}.
 \item The sea quark and antiquark contributions should be increased by 5\% as shown in equation \ref{sea_part}.   
%
\end{enumerate}

\section{Tests of duality in the for QE and $\Delta$(1238) production}

\begin{figure}
\includegraphics[width=3.7in,height=2.9in]{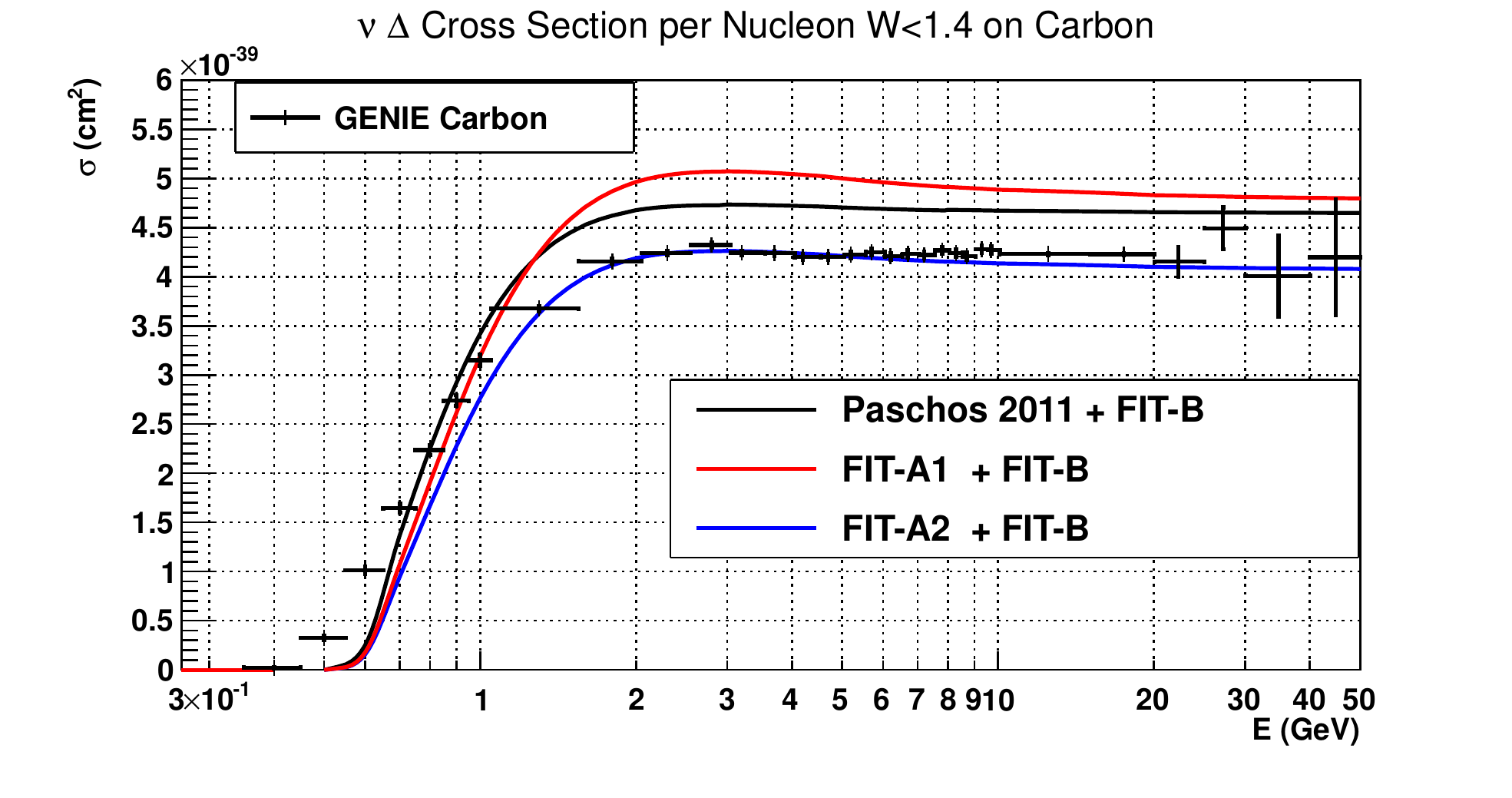}
\includegraphics[width=3.7in,height=2.9in]{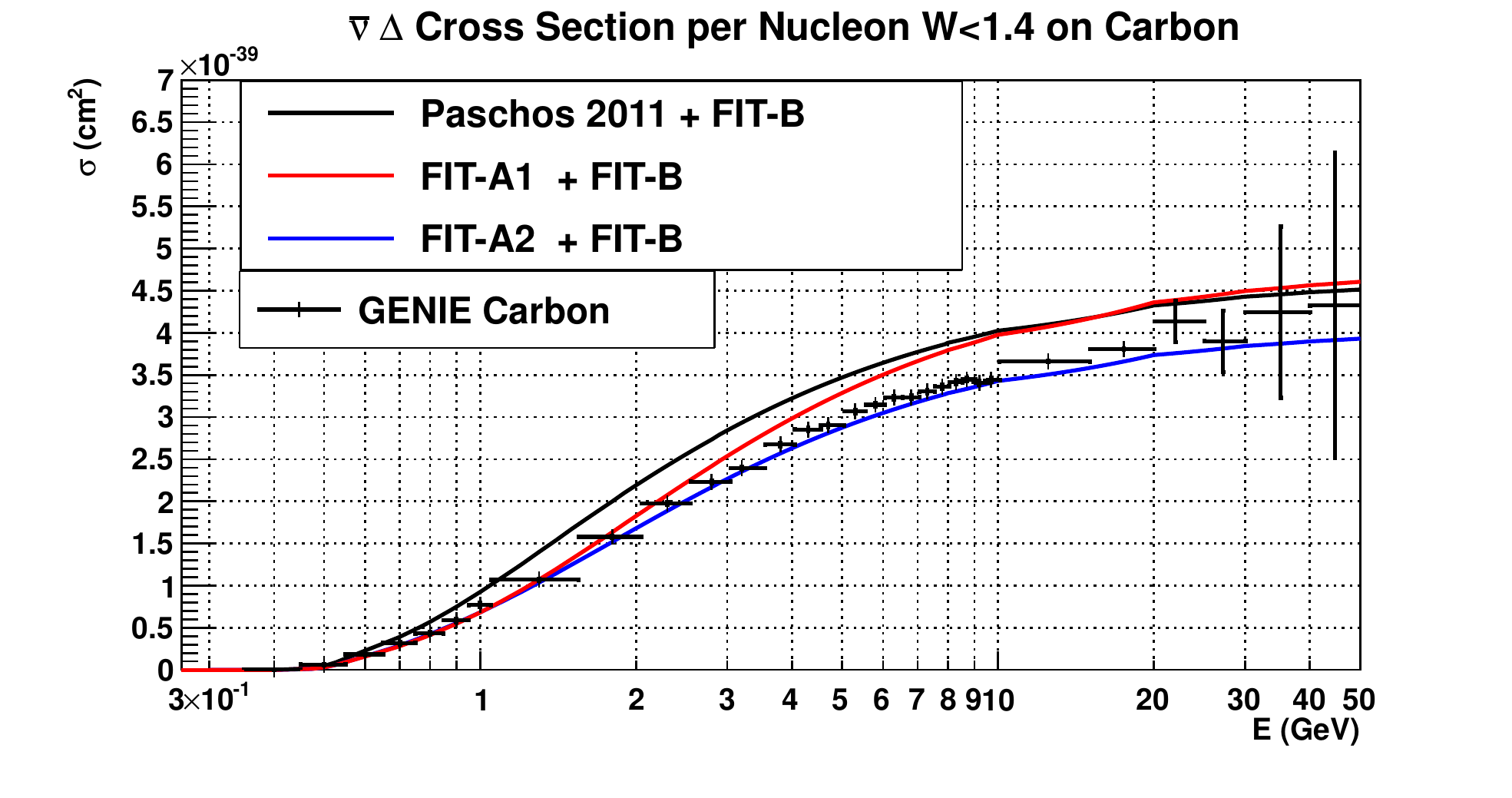}
\vspace{-0.3in} 
\caption{  Figure from reference \cite{lownu}. The total cross sections on carbon (per nucleon) predicted by GENIE for $W<1.4$ GeV (black points with MC statistical errors) for $\nu_\mu C \to (\mu^- \Delta^{++}$ or  $\Delta^{+}$) are shown on the top panel, and for $\nub_\mu C \to \mu^+  (\Delta^0$ or $\Delta^{-}$) are shown on the bottom panel.  The cross sections include the inelastic continuum for  $W<1.4$ GeV.  The red line and the green line span the range of  experimental measurements of the cross sections for this region, as investigated in reference \cite{lownu}.  We take the midpoint between the red and green line as the best estimate of the cross sections for $W<1.4$ GeV  (color online).}
\label{deltan-carbon}
\end{figure}

\begin{figure}
\includegraphics[width=3.5 in,height=2.9in]{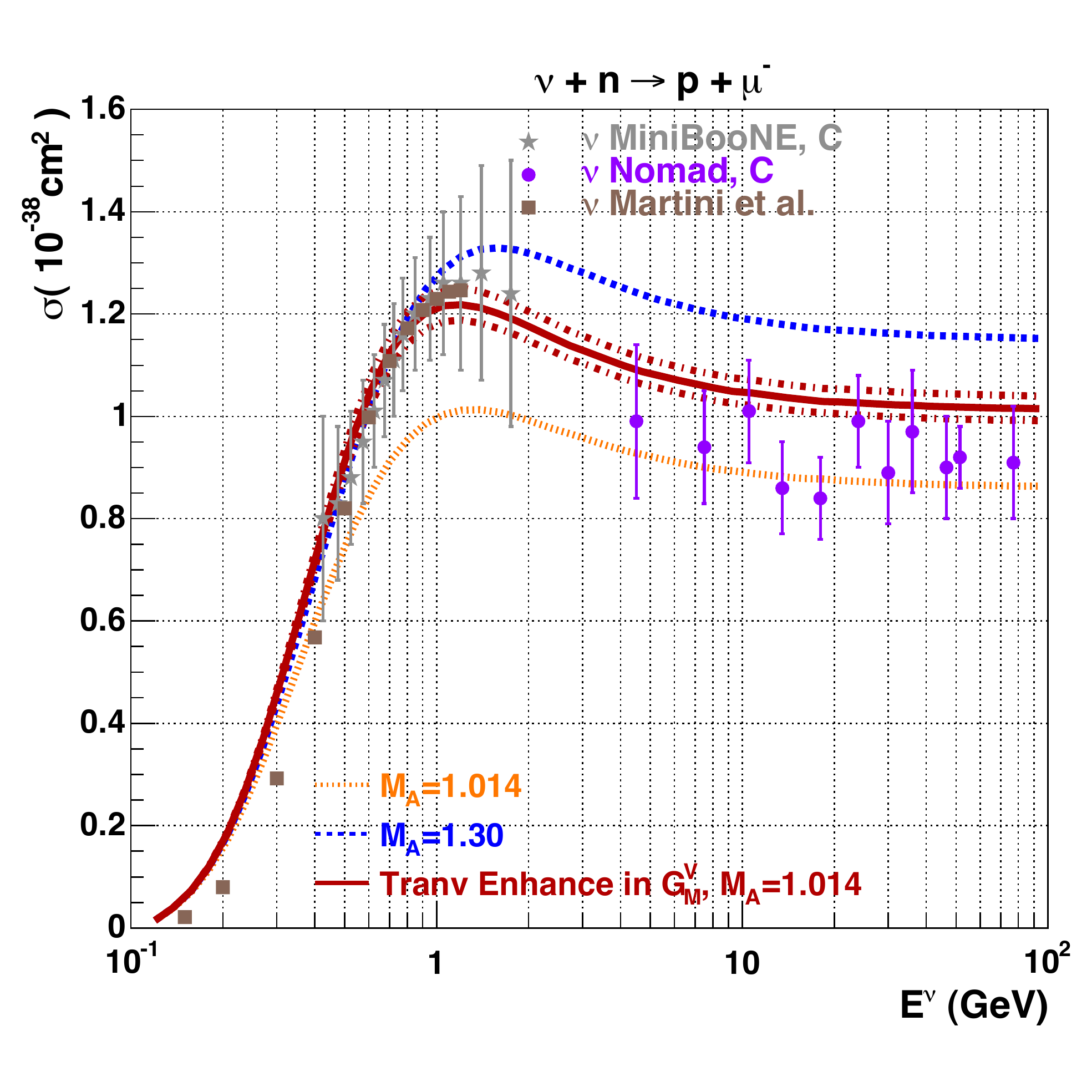}
 \includegraphics[width=3.5 in,height=2.9in]{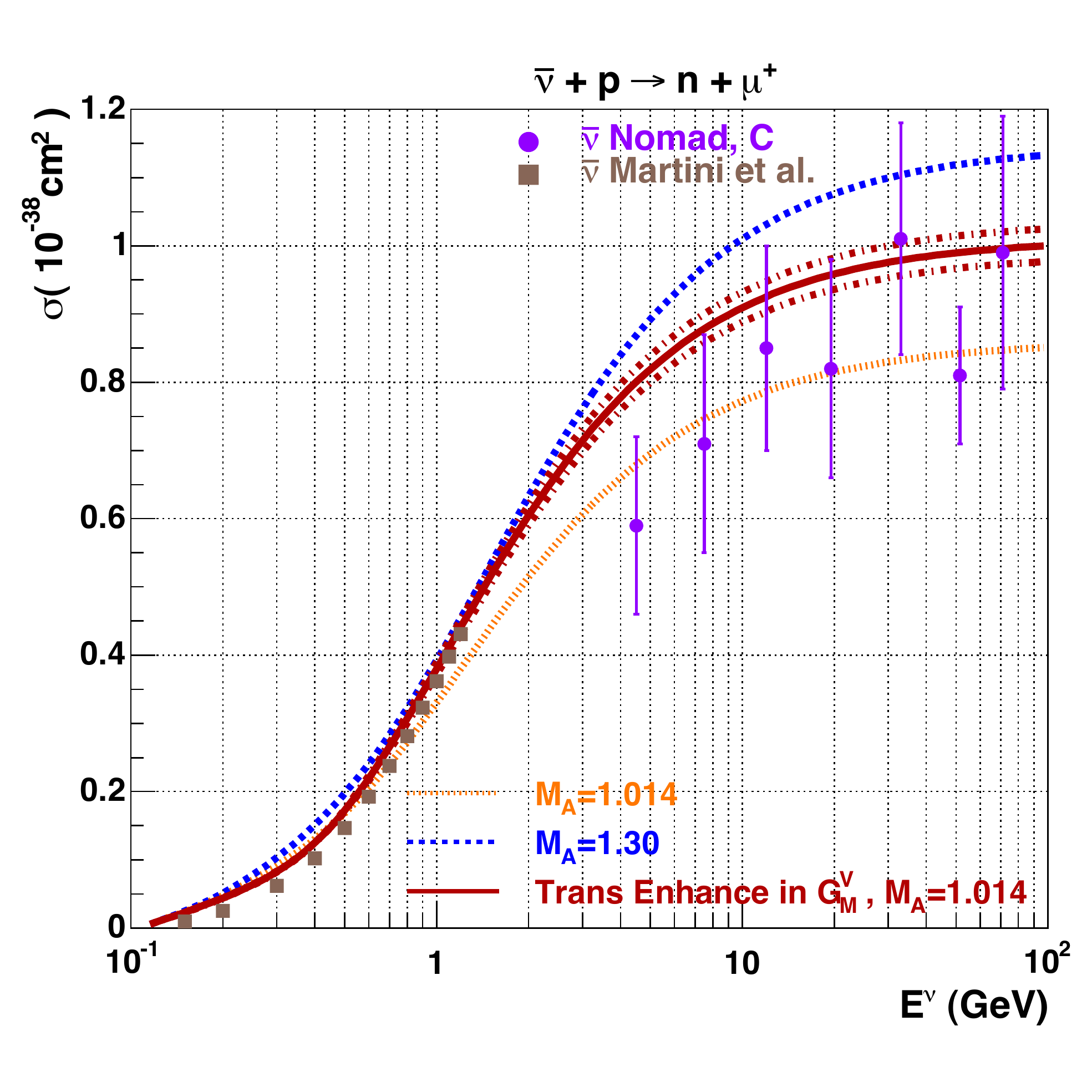}
\caption{Figure from reference \cite{lownu}. Comparison of predictions for the   $\nu_{\mu}$, $\bar{\nu}_\mu$  total QE cross section sections from  the nominal  TE model \cite{lownu},
the "Independent Nucleon (MA=1.014)"  model, the 
"Larger $M_A$  ($M_A$=1.3) model",   and  the 
  "QE+np-nh RPA"  MEC model of Martini et al.\cite{MEC5}.  We use the TE model to estimate the QE cross section.
  The data points are the   measurements of MiniBooNE\cite{MiniBooNE} (gray stars)  and NOMAD\cite{NOMAD} (purple circles) (color online).}
\label{totalMEClog}
\end{figure}

Table \ref{Duality} shows a comparison of the sum of the measured $\sigma/E$ (in units of $10^{-39}~cm^2/GeV$) for  QE and $\Delta (W<1.4)$ GeV,
   to the prediction of the Type II ($A>V$)(=BY II) model for  $1.08<W<1.4$ GeV. The experimental  errors for the QE and $\Delta$ cross sections are assumed to be 10\%.
  The experimental cross sections are taken from  Figures \ref{deltan-carbon} and  \ref{totalMEClog}. The model predictions for the integrated cross section in the  $1.08<W<1.4$ GeV region appears describe the $sum$ of the QE and
 the $\Delta (W<1.4)$ GeV measured cross sections.
%
%
\begin{table}[ht]
    \begin{center}
\begin{tabular}{||clc|cl|c||c|l}
\hline            
$E_\nu$ & $QE$ &$W<1.4$ & $1.08<W<1.4$&  \\
GeV& & GeV &  GeV& Ratio \\
 & & $ \rm^{\Delta(1238)} $ &$\rm ^{BY~II~model}$&${ \rm  ^{BY~II/(QE+\Delta)}}$   \\
\hline
 3  & 1.83 &  1.57&  3.72 & 1.09$\pm$0.15\\
  \hline
 5 & 1.10 &  0.92 &  2.25 & 1.1$\pm$0.16\\
  \hline
10  & 0.53&  0.45 & 1.13 & 1.16$\pm$0.16\\
\hline
40   & 0.13  &0.11  & 0.30  &1.21$\pm$0.17 \\
 \hline
 \hline
 $E_{\overline{\nu}}$  & $QE$ &$W<1.4$ & $1.08<W<1.4$&   \\
GeV& & GeV &  GeV& Ratio \\
 & & $ \rm^{\Delta(1238)} $ &$\rm ^{BY~II~model}$&${ \rm  ^{BY~II/(QE+\Delta)}}$   \\\hline 
3  & 1.20 &  0.80&  2.15 & 1.08$\pm$0.15 \\
  \hline
 5  & 0.81 &  0.60&  1.56& 1.11  $\pm$0.16 \\
  \hline
10  & 0.46&  10.35  & 0.90 & 1.11$\pm$0.16 \\
\hline
40  & 0.13  & 0.10  &0.27 &1.20$\pm$0.17  \\
 \hline
 \hline
 \end{tabular}
\caption{ Test of duality: Comparison of the sum of the measured $\sigma/E$ (in units of $10^{-39}~cm^2/GeV$) for  QE and $\Delta (W<1.4)$ GeV,
   to the prediction of the Type II ($A>V$)(=BY II) model for  $1.08<W<1.4$ GeV. The experimental  errors for the QE and $\Delta$ cross sections are assumed to be 10\%.
 The model predictions for the integrated cross section in the  $1.08<W<1.4$ GeV region appears describe the $sum$ of the QE and
 the $\Delta (W<1.4)$ GeV measured cross sections.}
\label{Duality}
  \end{center}
\end{table}
%
%
%
%
\section{QE cross sections and cross section in the region of the  $\Delta$ resonance ($W<1.4$ GeV)}
\label{all_cross}
\subsection{cross section in the region of the  $\Delta$ resonance ($W<1.4$ GeV)}
Figure \ref{deltan-carbon} is taken from reference \cite{lownu}. The total cross sections on carbon (per nucleon) predicted by GENIE for $W<1.4$ GeV (black points with MC statistical errors) for $\nu_\mu C \to (\mu^- \Delta^{++}$ or  $\Delta^{+}$) are shown on the top panel, and for $\nub_\mu C \to \mu^+  (\Delta^0$ or $\Delta^{-}$) are shown on the bottom panel.  The cross sections include the inelastic continuum for  $W<1.4$ GeV.  The red line and the green line  span the range of  experimental measurements of the cross sections for this region, as investigated in reference \cite{lownu}.  We take the midpoint between the red and green line as the best estimate of the cross sections for $W<1.4$ GeV.

\subsection{Neutrino and antineutrino  quasielastic cross sections on nuclei}

Figure \ref{totalMEClog} is taken from reference \cite{lownu}. Shown are comparisons of predictions for the   $\nu_{\mu}$, $\bar{\nu}_\mu$  total QE cross section sections from  the nominal  TE model\cite{lownu},
the "Independent Nucleon (MA=1.014)"  model, the 
"Larger $M_A$  ($M_A$=1.3) model",   and  the 
  "QE+np-nh RPA"  MEC model of Martini et al.\cite{MEC5}. 
  The data points are the   measurements of MiniBooNE\cite{MiniBooNE} (gray stars)  and NOMAD\cite{NOMAD} (purple circles). We use the TE model to estimate the QE cross section.
  %
\section{Appendix: The Adler sum rule }
\label{adler}
%
The Adler sum rules are derived from current algebra and are therefore valid at all values of $Q^{2}$.   The equations below
are for $strangeness~conserving (sc)$ processes.  These are related to the PDFs by a factor of $cos^2\theta_c$. 

The Adler sum rules for the vector  part of the structure function ${\cal W}_{2}^{\nu-vector}$ is given by:
\begin{eqnarray}
&&|F_V(Q^2)|^{2}+  \int_{\nu_{0}}^{\infty}{\cal W}_{2n-sc}^{\nu  -vector}(\nu,Q^2)  d\nu \nonumber \\
 &-&  \int_{\nu_{0}}^{\infty}{\cal W}_{2p-sc}^{\nu-vector}(\nu,Q^2)  d\nu = 1,   
\end{eqnarray}
where the limits of the integrals are from pion threshold $\nu_{0}$ where $W= M_{\pi}+M_P$ to $\nu=\infty$.
At  $Q^2=0$, the inelastic part of ${\cal W}_{2}^{\nu-vector}$ goes to zero,  and the sum rule is saturated by the quasielastic contribution $ |F_V(Q^2)|^{2}$.  Here  $=Q^2/(4M^2)$, and 
 $$ |F_V(Q^2)|^{2}- \frac{[G_E^V(Q^2)]^2+ \tau [G_M^V(Q^2)]^2}{1+\tau}.$$
 In the dipole approximation we have
 \begin{eqnarray}
   G_E^{V}(Q^2) &=&  G_E^P(Q^2)-G_E^N(Q^2) \approx G_D(Q^2)\\
  G_M^{V}(Q^2)& =& G_M^P(Q^2) - G_M^N(Q^2) \approx 4.706~G_D(Q^2)\\
  G_D &=& 1/(1+Q^2/M_V^2)^2 ,
  \end{eqnarray}
 where $M_V^2=0.71~GeV^2$.  Note that in all  of the calculations, we do not use the dipole approximation  (we use BBBA2008~~\cite{quasi}  vector and axial form factors).

The Adler sum rule for ${\cal W}_{2}^{\nu-axial}$ is given by:
\begin{eqnarray}
&& |{\cal F}_A(Q^2)|^2 +   \int_{\nu_{0}}^{\infty} {\cal W}_{2n-sc}^{\nu  -axial}(\nu,Q^2) d\nu \nonumber \\
 &-&  \int_{\nu_{0}}^{\infty}{\cal W}_{2p-sc}^{\nu-axial}(\nu,Q^2) d\nu = 1,\nonumber    
\end{eqnarray}
where in the dipole approximation 
$${\cal F}_A \approx -1.267/(1+Q^2/M_A^2)^2$$
and $M_A= 1.014~GeV$ from  reference\cite{quasi}.

The Adler sum rule for   ${\cal W}_{1}^{\nu-vector}$ is given by:
\begin{eqnarray}
&& \tau |G_{M}^{V}(Q^2)|^{2}+  \int_{\nu_{0}}^{\infty}{\cal W}_{1n}^{\nu  -vector}(\nu,Q^2) d\nu \nonumber \\
 &-&  \int_{\nu_{0}}^{\infty} {\cal W}_{1p}^{\nu-vector}(\nu,Q^2)  d\nu = 1.
\end{eqnarray}

The Adler sum rule for ${\cal W}_1^{\nu-axial}$ is given by:
\begin{eqnarray}
&&(1+\tau)|{\cal F}_A(Q^2)|^2+  \int_{\nu_{0}}^{\infty}{\cal W}_{1n-sc}^{\nu  -axial}(\nu,Q^2) d\nu \nonumber \\
 &-&  \int_{\nu_{0}}^{\infty} {\cal W}_{1p-sc}^{\nu-axial}(\nu,Q^2)  d\nu = 1.  
\end{eqnarray}

The Adler sum rule for ${\cal W}_3^{\nu}$ is given by:
\begin{eqnarray}
&&2{\cal F}_A(Q^2)G_{M}^{V}(Q^2)+  \int_{\nu_{0}}^{\infty}{\cal W}_{3n-sc}^{\nu} (\nu,Q^2) d\nu \nonumber \\
 &-&  \int_{\nu_{0}}^{\infty} {\cal W}_{3p-sc}^{\nu}(\nu,Q^2)  d\nu = 0. 
\end{eqnarray}

We use the Alder sum rule for ${\cal W}_{2}^{\nu-vector}$ to constrain the form of the $K_{valence}^{vector}(Q^2) $ factor for ${\cal W}_{2}^{\nu-vector}$.  At low $Q^2$  we  approximate  $|F_V(Q^2)|^{2}$
by $G_D^{2}(Q^2)$, and  use the following $K$ factors for ${\cal W} _{2}^{\nu-vector}$. 
\begin{eqnarray}	
 K_{valence}^{vector}(Q^2) &=&[1-G_D^2(Q^2)] 
  \left(\frac{Q^2+C_{v2}} 
{Q^{2} +C_{v1}}\right), 
\end{eqnarray}
where the values of the parameters $C_{v2d}$, $C_{v1d}$, $C_{v2u}$ and $C_{v1u}$  are obtained from  a fit to the charged-lepton scattering and photoproduction data as discussed in section \ref{section3}.

 With this $K_{valence}^{vector}(Q^2) $ factor, the Adler sum rule for ${\cal W} _{2}^{\nu-vector}$  is then  approximately satisfied.  At  $Q^2=0$, the inelastic part of ${\cal W} _{2}^{\nu-vector}$ goes to zero,  and the sum rule is saturated by the quasielastic contribution.   Note that the contribution of the $\Delta(1232)$ resonance to the Adler sum rule is negative. Near  $Q^2=0$  the $\Delta(1232)$ contribution is small in the vector case (since it must be zero at $Q^2=0$) and can be neglected.    However,  for  the axial case the contribution of the $\Delta(1232)$at $Q^2=0$ is large and negative and cannot be neglected. 
 %
%
 
 \section{Appendix -Results with GRV94 PDFs }
\label{GRV94-section}
For completeness we describe the early NUINT01  analysis~\cite{nuint01-2} in which  we used another  modified scaling variable ~\cite{omegaw} $x_w$ with   GRV94 PDFs(instead of GRV98)  and simplified $K$ factors.  In that analysis  we modified the   leading order GRV94 PDFs  as follows:
\begin{enumerate}
\item  We increased the $d/u$ ratio at high $x$ as  described in section \ref{dovu} (and reference \cite{highx}).
\item  Instead of the scaling variable $x$ we used the scaling variable $x_w = (Q^2+B)/(2M\nu+A)$ (or =$x(Q^2 +B)/(Q^2+Ax)$).
This modification was used in early fits to SLAC data~\cite{bodek}. The parameter A provides for an approximate way to include $both$ target mass and higher twist effects at high $x$, and the parameter B allows the fit to be used all the way down to the photoproduction limit ($Q^{2}$=0).
\item  In addition as was done in earlier non-QCD based fits~\cite{DL,bonnie}  to low energy data, we multiplied all PDFs by a factor $K$=$Q^{2}$ / ($Q^{2}$ +C). This was done in order for the fits to describe low $Q^2$  data in the photoproduction limit, where  ${\cal F}_{2}$ is related to the photoproduction cross section.
\item Finally,  we froze  the evolution of the GRV94 PDFs at a value of $Q^{2}=0.24$ (for $Q^{2}<0.24$), because GRV94 PDFs are only valid down to $Q^{2}=0.23~GeV^2$.
\end{enumerate}
\begin{table}[ht]
    \begin{center}
\begin{tabular}{|l|l|l|l|l}
\hline            
$A$ & $B$ & $C$ & $\chi^2/ndf$  \\
$1.735$ & $0.624$ & $0.188$ & $1351/958$   \\
\hline
\hline
 \end{tabular}
\caption{ Parameters from  NUINT01 version of the model (with GRV94  PDFs).  When applicable, all parameters are in units of 
GeV$^2$. }
\label{Table-nuint01}
  \end{center}
\end{table}
In the  GRV94 analysis, the measured structure functions were also corrected for the BCDMS magnetic field systematic error shift\cite{bcdms}  and for the relative normalizations between  SLAC, BCDMS and NMC data~\cite{highx,nnlo}. The deuterium data were corrected for nuclear binding effects~\cite{highx,nnlo}.  A simultaneous fit  to both proton and deuteron SLAC, NMC and BCDMS data (for $x>0.07$) yields the following values  A=1.735, B=0.624 and C=0.188 GeV$^2$) with GRV94 LO PDFs ($\chi^{2}=$ 1351/958 DOF).  These parameters are summarized in Table~\ref{Table-nuint01}. Note that for $x_w$ the parameter A  accounts for $both$ target mass and higher twist effects.

In our studies with GRV94 PDFs we used the earlier   $ {\cal R}_{world}$ fit~\cite{slac}   for  $ {\cal R}^{ncp}$
and $ {\cal R}^{cp}$. ${\cal R}_{world}$ is parameterized by:
\begin{eqnarray}
 {\cal R}_{world}(x,Q^2>0.35) & = & \frac{0.0635}{ln(Q^2/0.04)} \theta(x,Q^2) \nonumber \\
  & +  &  \frac{0.5747}{Q^2}-\frac{0.3534}{Q^4+0.09},	        
\end{eqnarray}
where $\theta = 1. + \frac{12 Q^2}{Q^2+1.0} \times \frac{0.125^2}{0.125^2 + x^2}$. The  $ {\cal R}_{world}$ function  provided a good description of the world's data for $ {\cal R}$  at that time  in the $Q^2>0.35$ $GeV^{2}$ and $x>0.05$ region (where most of the $ {\cal R}$ data are available). However, for electron and muon scattering and for the  vector part of neutrino scattering   the $ {\cal R}_{world}$ function breaks down below $Q^2=0.35$ $GeV^{2}$. 

Here,  we freeze the function at $Q^2=0.35$ GeV$^2$. For electron and muon scattering and for the  vector part of  ${\cal F}_{1}$ we introduce a  $K$ factor for  $ {\cal R}$ in the $Q^2<0.35$ GeV$^2$ region. The $K$ factor provides a smooth transition for  the vector  $ {\cal R}$ (we use $ {\cal R}_{vector}$=$ {\cal R}_{e/\mu}$) from $Q^2=0.35$ GeV$^2$ down to $Q^2=0$ by forcing $ {\cal R}_{vector}$ to approach zero at $Q^2=0$ as expected in the photoproduction limit (while keeping a $1/Q^2$ behavior at large $Q^2$ and matching to $ {\cal R}_{world}$ at  $Q^2=0.35$ GeV$^2$).
\begin{eqnarray}
 {\cal R}_{vector}(x,Q^2<0.35) & = & 3.207 \times \frac {Q^2}{Q^4+1} \nonumber \\
           & \times   &  {\cal R}_{world}(x,Q^2=0.35).\nonumber
\end{eqnarray}
\section{Acknowlegements}
Research supported by the U.S. Department of Energy under grant number DE-SC0008475
and Promising-Pioneering Researcher Program  through Seoul National University.f


\begin{thebibliography}{9}
\bibitem{ATM}  S. Fukuda {\em et al.}, Phys.
Rev. Lett. {\bf 85}, 3999 (2000); T. Toshito,   hep-ex/0105023.
\bibitem {MINOS}  D.G. Michael {\em et al.}(MINOS), Phys. Rev. Lett. {\bf 97}, 191801 (2006);  
http://www-numi.fnal.gov/Minos/
\bibitem {MINOS2} P. Adamson {\em et al.}(MINOS), Phys. Rev. D {\bf 81}, 072002 (2010).
\bibitem {NOVA}P. Adamson {\em et al.}(NOVA,   Phys. Rev.  D {\bf 93} 051104 (2016 )( http://www-nova.fnal.gov/)
\bibitem {K2K} M. H. Ahn {\em et al.}(K2K), Phys. Rev. D {\bf 74}, 072003 (2006);
 http://neutrino.kek.jp/
\bibitem {superK}   Y.  Ashie  {\em et al.}(SuperK), Phys. Rev. D {\bf 71}, 112005 (2005);
\bibitem {T2K}  K. Abe et al. (T2K), Phys. Rev. D 87, 092003 (2013); K. Abe et al. (T2K), Phys. Rev. D 90, 052010 (2014); K. Abe et al. (T2K), Phys. Rev. D 93, 072002 (2016). We applied Isoscalar correction from ref. \cite{minerva1} to T2K  total cross sections.
\bibitem {MiniBooNE}   A. A.   Aguilar-Arevalo  {\em et al.}(MiniBooNE), Phys. Rev. Lett {\bf 98}, 231801(2007)
\bibitem {DUNE}  The DUNE Collaboration, B. Abi et al. "The DUNE Far Detector Interim Design Report Volume 1: Physics, Technology and Strategies"   arXiv:1807.10334 [physics.ins-det]
\bibitem {sciboone}   Y. Nakjima  {\em et al.}, (SciBoonE) arXiv:hep-ex/1011.213
\bibitem {MicroBooNE} MicroBooNE collaboration, ?Design and Construction of the MicroBooNE Detector?, arXiv:1612.05824, JINST 12, P02017 (2017)
\bibitem {argoneut} R. Acciarri et al. (ArgoNeuT), Phys. Rev. D 89, 112003 (2014).
\bibitem{minerva1} J. DeVan et. al. (MINERvA) Phys. Rev. D 94, 112007 (2016) arXiv:1610.04746	
\bibitem {ICARUS} ICARUS at Fermilab, https://icarus.fnal.gov/
\bibitem{nuint01-2} A. Bodek and U.K. Yang (NUINT01),Nucl. Phys. B  Proc. Suppl.{\bf 112}, 70 (2002), arXiv:hep-ex/0203009; A. Bodek and U. K. Yang  (NUINT02), arXiv:hep-ex/0308007.
\bibitem{nuint04} A. Bodek , Ikyu Park and U. K. Yang, (NUINT04) Nucl. Phys. B Proc. Suppl.{\bf 139}, 113 (2005), arXiv:hep-ph/0411202.
\bibitem{NEUT} Y. Hayato, Nucl Phys. Proc. Suppl.. {\bf 112}, 171 (2002)
\bibitem{GENIE} C.Andreopoulos (GENIE),  Nucl. Instrum. Meth. A614, 87 (2010)
\bibitem{NEUGEN} H. Gallagher (NEUGEN), Nucl. Phys. Proc. Suppl. 112 (2002)
\bibitem{NUANCE} D. Casper (NUANCE) , Nucl. Phys. Proc. Suppl. 112, 161 (2002);  http://nuint.ps.uci.edu/nuance/
\bibitem{highx} U. K. Yang and A. Bodek, Phys. Rev. Lett. {\bf82}, 2467 (1999)
\bibitem{nnlo} U. K. Yang and A. Bodek, Eur. Phys. J. C{\bf13}, 241 (2000)
\bibitem{yangthesis} U. K. Yang, Ph.D. thesis, Univ. of Rochester (2001), 
FERMILAB-THESIS-2001-09, available at $http://inspirehep.net/record/567983$.
\bibitem{slac} L.W. Whitlow, E.M. Riordan, S. Dasu, S. Rock, A. Bodek  (SLAC-MIT), Phys. Lett. B{\bf 282}, 433 (1995);
L.W. Whitlow, PhD thesis, Stanford University, SLAC Report 357 (1990)
\bibitem{bcdms} A. C. Benvenuti {\it et al.} (BCDMS), Phys. Lett. B{\bf237}, 592 (1990);   M. Virchaux and A. Milsztajn, Phys. Lett. B 274, 221 (1992)
 \bibitem{nmcdata}   M. Arneodo {\it et al.}  (NMC), Nucl. Phys. B{\bf 483}, 3 (1997)
\bibitem{barb} R. Barbieri {\it et al.},  Phys. Lett. B{\bf 64}, 171 (1976), and  Nucl. Phys. B{\bf117}, 50 (1976)
\bibitem{Nachtman}O. Nachtmann, Nucl. Phys. B63 (1973) 237; O. Nachtmann, Nucl. Phys. B78 (1974) 455;  O. W. Greenberg and D. Bhaumik, Phys. Rev. D4 (1971) 2048; H. Georgi and H. D. Politzer, Phys. Rev. D{\bf 14}, 1829 (1976); J. Pestieau and J. Urias, Phys.Rev.D{\bf 8}, 1552 (1973)
\bibitem{grv98} M. Gluck, E. Reya, A. Vogt, Eur. Phys. J {\bf C5}, 461 (1998).
\bibitem{DL} A. Donnachie and P. V. Landshoff, Z. Phys. C {\bf61}, 139 (1994).
\bibitem{bonnie} B. T. Fleming {\it et al.}(CCFR), Phys. Rev. Lett. {\bf 86}, 5430 (2001).
\bibitem{omegaw}F. W. Brasse  {\it et al.},  Nucl. Phys. B {\bf 839}, 421 (1972). 
\bibitem{bodek} A. Bodek {\it et al.}, Phys. Rev. D{\bf 20}, 1471 (1979).
\bibitem{close} S. Stein {\it et al.}, Phys. Rev. D{\bf 12}, 1884 (1975);  K. Gottfried,  Phys. Rev. Lett. {\bf18}, 1174 (1967).
\bibitem{adler} S. Adler, Phys. Rev. {\bf 143}, 1144 (1966); F. Gillman, Phys. Rev. {\bf 167}, 1365 (1968).
\bibitem{adler2} O. Lalakulich, W. Melnitchouk, and  E. A. Paschos, Phys. Rev. C {\bf 75},015202 (2007).
 \bibitem{h1data} C. Adloff {\it et al.}     (H1) , Eur Phys J C30, 32 (2003);  http://www-h1.desy.de/
\bibitem{photo} Photoproduction:  David O. Caldwell, {\it et al.}   Phys. Rev. Lett.  25, 609 (1970); 
T.A. Armstrong {\it et al.}   Nucl. Phys. B41, 445 (1972);   T.A. Armstrong {\it et al.}   Phys. Rev. D5, 1640 (1972);  
David O. Caldwell, {\it et al.}   Phys. Rev. D7, 1362 (1973) (nuclear targets); David O. Caldwell {\it et al.}    Phys. Rev. Lett. 40, 1222, (1978);   S. Chekanov {\it et al.} (ZEUS)  Nucl. Phys. B627, 3 (2002); T. Ahmed {\it et al.} (H1)  Phys. Lett. B299 374 (1993).
\bibitem{jlab}
C. Keppel, Proc. of the Workshop on Exclusive Processes at High $P_T$, Newport News, VA, May (2002).
\bibitem{bloom}E. D. Bloom and F. J. Gilman, Phys. Rev. Lett. {\bf 25}, 1140 (1970).
\bibitem{Liang:2004tj}
  Y.~Liang {\it et al.}(E94-110),  arXiv:nucl-ex/0410027.
\bibitem{R1998}K. Abe {\it et al.},  Phys. Lett. B{\bf 452}, 194 (1999)
\bibitem{xf3calc}  R.S. Thorne and R.G. Roberts, Phys. Lett. B 421, 303 (1998); Eur. Phys. J. C 19, 339 (2001).
\bibitem{kulagin}  S. A. Kulagin and R. Pett, Phys. Rev. D{\bf 76}, 094023 (2007), ibid Nucl. Phys. A765, 26 (2006).  
 \bibitem{jlabR} P.E. Bosted and M.E. Christy, Phys. Rev. C{\bf 77}, 065206 (2008);  M.E. Christy and P.E. Bosted,
  	Phys. Rev. C{\bf81}, 055213 (2010),  arXiv:0712.3731.  Fortran program for $R$ ia  available at $http://www.jlab.org/~christy/cs_fits/F1F209.f$ .
\bibitem{selthesis}
W. G. Seligman, Ph.D. thesis, (CCFR) Columbia Univ., Nevis reports 292 (1997).
 \bibitem{arrington} J. Arrington et al (Jefferson Lab),  Phys.Rev. C{\bf 73}, 035205 (2006). 
\bibitem{e87} A. Bodek   {\it et al.}  (E87), Phys. Rev. Lett. 50, 1431 (1983).
\bibitem{e139} J. Gomez {\it et al.} (E139, Phys. Rev. D49, 4348 (1994). 
\bibitem{e140} S. Dasu {\it et al.} (E140) , Phys. Rev. Lett. 60, 2591 (1988); S. Dasu {\it et al.} (E140) Phys. Rev. D 49, 5641 (1994).
\bibitem{NMCnuc} M. Arneodo (NMC){\it et al.},  Nucl. Phys. B 481, 3 (1966). 
\bibitem{jlabC} R. Seely  {\it et al.} (Jefferson Lab data on Carbon), Phys. Rev. Lett. 103, 202301 (2009).
\bibitem{chorus} R. Oldeman, Proc. of  30th International Conference on High-Energy Physics (ICHEP 2000),  Osaka, Japan, 2000; R. G. C. Oldeman, Ph.D. thesis, University of Amsterdam, 2000; G. Onengut et al. (CHORUS ) Phys.Lett. B632, 65 (2006). http://choruswww.cern.ch/Publications/DIS-data
\bibitem{rccfr} U. K. Yang {\it et al.}(CCFR), Phys. Rev. Lett.  {\bf 87}, 251802 (2001).
 \bibitem{cdhsw} P. Berge {\em et al.} (CDHSW), Zeit. Phys. {\bf C49}, 607 (1991).
\bibitem{quasi}  A. Bodek, S. Avvakumov, R. Bradford, and  H. Budd, Eur. Phys. J. C{\bf 53}, 349 (2008). 
 \bibitem{lownu}  A. Bodek , U. Sarica, D. Naples  and L. Ren, Eur.Phys.J.C 72 (2012) 1973
\bibitem{NOMAD}  
  V.~Lyubushkin  et al.  (NOMAD Collaboration),
  Eur.\ Phys.\ J.\  C {\bf 63}, 355 (2009);  Q. Wu {\it et al.}(NOMAD Collaboration), Phys. Lett. 
{\bf B60}, 19 (2008).
\bibitem{Serp96} V.B. Anikeev, et al. (Serpukhov) Z. Phys. C {\bf 70}, 39 (1996)
\bibitem{BNL82} N. J. Baker et al. (BNL)  Phys. Rev. D {\bf 25}, 617 (1982).
\bibitem{MEC5}
M. Martini, M. Ericson, G. Chanfray, and J. Marteau, Phys. Rev. C 80: 065501, 2009; ibid
 Phys. Rev. C 81: 045502, 2010. 
 \bibitem{Zeller} J.A. Formaggio, G.P. Zeller, 	Rev. Mod. Phys. 84, 1307 (2012) (arXiv:1305.7513 [hep-ex])
 \bibitem{GGM}  Ciampolillo, S., et al. (Gargamelle Neutrino Propane Collaboration, Aachen-Brussels-CERN-Ecole Poly-Orsay-Padua
Collaboration), 1979, Phys.Lett. B84, 281.	
\bibitem{New_data_set}    M.R. Whalley,  Nucl.Phys.B Proc.Suppl. 139 (2005) 241 (hep-ph/0410399)
 
$http://durpdg.dur.ac.uk/hepdata/online/neutrino/$

%
\end{thebibliography}
\end{document}